\documentclass[%
 reprint,
 preprintnumber,
 amsmath,amssymb,
 aps,prd,
 nofootinbib,
]{revtex4-1}

\pdfoutput=1

\usepackage{bbold}
\usepackage{mathtools}
\usepackage{color}
\usepackage{graphics}
\usepackage[caption=false]{subfig}
\usepackage{xcolor,colortbl}
\usepackage{slashed}
\usepackage{hyperref}

\allowdisplaybreaks
\title{}
\date{}

\def\eq#1{Eq.~(\ref{#1})}
\def\Eq#1{Eq.~(\ref{#1})}
\newcommand{\fig}[1]{Fig.~\ref{fig:#1}}
\definecolor{prd_blue}{RGB}{41, 41, 133}

\hypersetup{
    colorlinks=true,
    linkcolor=prd_blue,
    filecolor=prd_blue,      
    urlcolor=prd_blue,
    citecolor=prd_blue
}

\newcommand{\nn}{\nonumber}

\newcommand\Eqns[2]    {Eqs.\,(\ref{#1}) and~(\ref{#2})}

\newcommand\eqn[1]     {eq.\,(\ref{#1})}

\newcommand\sect[1]    {section~{\ref{#1}}}

\newcommand\appx[1]     {appendix~\ref{#1}}

\newcommand{\be}{\begin{equation}}
\newcommand{\ee}{\end{equation}}
\newcommand{\beq}{\begin{equation}}
\newcommand{\eeq}{\end{equation}}
\newcommand{\bea}{\begin{eqnarray}}
\newcommand{\eea}{\end{eqnarray}}

\newcommand{\eps}{\epsilon}
\newcommand \slsh [1] {\not\!{#1}}
\newcommand{\ord}{{\cal O}}

\newcommand{\LP}{\mathrm{LP}}
\newcommand{\NLP}{\mathrm{NLP}}
\newcommand{\PSell}{\int\! \frac{d^{4-2\eps}\ell}{(2\pi)^{4-2\eps}}\,}
\newcommand{\PSellRed}{\int\!\frac{d\ell^- d^{2-2\eps}\ell_\perp}{(2\pi)^{4-2\eps}}\,}
\newcommand{\omx}{(1\!-\!x)}
\newcommand{\ome}{1\!-\!\eps}
\newcommand{\h}{\hspace{.25pt}}

\begin{document}

\preprint{NIKHEF 2020-019}

\title{Towards all-order factorization of QED amplitudes at next-to-leading power}

\author{E.~Laenen$^{1,2,3}$}
\author{J.~Sinninghe Damst\'{e}$^{1,2}$}
\author{L.~Vernazza$^{4}$}
\author{W.~Waalewijn$^{1,2}$}
\author{L.~Zoppi$^{1,2}$}

\affiliation{$^{1}$Institute for Theoretical Physics Amsterdam and Delta Institute for Theoretical Physics, University of Amsterdam, Science Park 904, 1098 XH Amsterdam, The Netherlands}

\affiliation{$^{2}$Nikhef, Theory Group, Science Park 105, 1098 XG, Amsterdam, The Netherlands}
\affiliation{$^{3}$ITF, Utrecht University, Leuvenlaan 4, 3584 CE Utrecht, The Netherlands}
\affiliation{$^{4}$Dipartimento di Fisica Teorica, Universit\`a di Torino and INFN, Sezione di Torino, Via P. Giuria 1, I-10125 Torino, Italy}
\date{\today}

\begin{abstract}
    We generalise the factorization of abelian gauge theory amplitudes to next-to-leading power (NLP) in a soft scale expansion, following a recent generalisation for Yukawa theory. From an all-order power counting analysis of leading and next-to-leading regions, we infer the factorized structure for both a parametrically small and zero fermion mass. This requires the introduction of new universal jet functions, for non-radiative and single-radiative QED amplitudes, which we compute at one-loop order. We show that our factorization formula reproduces the relevant regions in one- and two-loop scattering amplitudes, appropriately addressing endpoint divergences. It provides a description of virtual collinear modes and accounts for non-trivial hard-collinear interplay present beyond the one-loop level, making this a first step towards a complete all-order factorization framework for gauge-theory amplitudes at NLP.
\end{abstract}

\maketitle

\section{Introduction}

\begin{figure}[t]
\begin{center}
	\includegraphics[width = .22\textwidth]{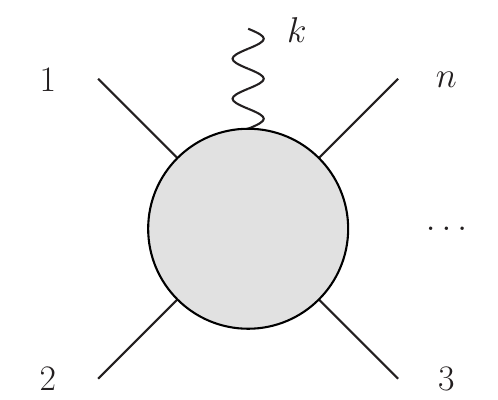}
	\caption{$n$-particle scattering with the emission of an additional soft photon.}
	\label{fig:n_partons}
\end{center}
\end{figure}

Deepening our understanding of gauge theory scattering 
amplitudes in the limit where radiation is soft has important phenomenological benefits as well as 
significant intrinsic value. For $n$-particle scattering processes in QED with the emission of an additional soft photon with momentum 
$k$ (as in \fig{n_partons}), the scattering amplitude 
${\cal M}_{n+1}$ can be expressed as a power expansion 
in the energy $E = k^0$ of the photon,
\be \label{softexpansion}
{\cal M}_{n+1} = {\cal M}_{n+1}^{\rm LP}
+ {\cal M}_{n+1}^{\rm NLP} + \ord(E)\,,
\ee
where the \emph{leading power} (LP) term has scaling 
${\cal M}_{n+1}^{\rm LP} \sim 1/E$, and the \emph{next-to-leading
power} (NLP) contribution is of order ${\cal M}_{n+1}^{\rm NLP}\sim 
E^0$. Crucially, the coefficients in this expansion
can be expressed in terms of simpler objects, which 
relate the radiative amplitude to the non-radiative or 
\emph{elastic} amplitude ${\cal M}_{n}$. Such relations 
go under the name of \emph{factorization} or \emph{soft} 
theorems. Their physical interpretation rests upon the long 
wavelength of soft radiation not being able to resolve the 
hard scattering. However, at each subsequent power in 
\eqref{softexpansion}  more is revealed. In this paper we investigate aspects of factorization at NLP, and focus in particular on the objects that can appear 
at higher orders in perturbation theory.
The simpler objects that enter in the coefficients of 
\eqref{softexpansion} describe soft and collinear dynamics 
and are universal, i.e.~they do not depend on the particular 
scattering process. For instance, it is well known 
that the LP term in \eqn{softexpansion} takes the universal 
form~\cite{Yennie:1961ad,Grammer:1973db}
\begin{align}\label{factLP}
{\cal M}_{n+1}^{\rm LP}(\{p_i\},k) &= S_n^{(0)} \,{\cal M}_n(\{p_i\})\,, \nn\\
 S_n^{(0)} &= e \sum_{i=1}^n q_i\, \frac{p_i^{\mu} \varepsilon_{\mu}(k)}{p_i \cdot k}\,,
\end{align}
where $p_i^\mu$ and $q_i$ denote the momentum and 
electric charge (in units of the elementary charge $e$) 
of the $i$-th hard particle, and $\varepsilon_{\mu}(k)$ 
is the polarisation vector of the soft photon. The soft 
function $S_n^{(0)}$ describes a set of eikonal 
interactions between the external particles and the emitted 
soft photon; in other words, soft radiation at LP is 
sensitive only to the direction and charge of the emitting particle. (The expression in \eq{factLP} is at lowest 
order in the coupling $e$, as indicated by the superscript 
$(0)$, and receives loop corrections.) For multiple 
soft photons the function $S_n$ can be 
calculated as the vacuum expectation value of a set of Wilson lines, one 
for each hard emitting particle, expanded to the appropriate order in the coupling.  

The factorization in \eq{factLP} is not only of theoretical 
interest, but also relevant for phenomenology. The $1/E$ singularity 
in the soft limit enhances soft radiation in scattering processes. 
Measurements that are sensitive to soft radiation involve
a small scale, and the corresponding cross section contains 
large logarithms of the ratio of this small scale and the scale 
of the hard scattering. Such large logarithms potentially 
spoil the convergence of the expansion in the coupling $e$, 
a problem that can be addressed by resummation. The development of theorems such as the one in \eqn{factLP} led to the proofs of factorization~\cite{Bodwin:1984hc,Collins:1985ue,Collins:1988ig}, but also 
constitutes the first step towards resummation, as it allows 
one to decompose a multi-scale amplitude (or cross section) 
into the product of simpler single-scale functions.

The resummation of large logarithms in QCD 
originating from the LP term in the equivalent 
expansion of \eqn{softexpansion} has been an 
active research topic for many years. Resummation
of soft gluon radiation at LP has been systematically 
applied to most processes of interest at lepton and 
hadron colliders. In the seminal papers~\cite{Sterman:1986aj,Catani:1989ne}
soft gluon resummation in Drell-Yan and DIS was 
achieved by means of diagrammatic techniques, 
and in~\cite{Korchemsky:1992xv,Korchemsky:1993uz}
it was shown that soft radiation can be described 
in terms of Wilson lines, whose exponentiation 
properties are at the basis of resummation. Later, 
by means of similar diagrammatic techniques, 
resummation was extended to more processes, 
including those with coloured particles in the final 
state, see e.g.~\cite{Contopanagos:1996nh,Catani:1996yz,Kidonakis:1997gm,Kidonakis:1998nf,Laenen:1998qw,Catani:2003zt}. 
Soft gluon resummation by means of renormalisation-group 
techniques was first studied in \cite{Forte:2002ni} and in a 
different method in \cite{Ravindran:2005vv,Ravindran:2006cg}, and 
this has more recently also been accomplished using 
effective field theory techniques, see e.g.~\cite{Bauer:2002nz,Manohar:2003vb,Idilbi:2005ky,Chay:2005rz,Becher:2006nr,Becher:2006mr,Becher:2007ty,Ahrens:2010zv}.

By contrast, the factorization and resummation of 
the NLP contribution in \eq{softexpansion} is still  under much investigation. This NLP term exhibits a more involved structure: emitted 
(next-to-)soft radiation becomes sensitive to the spin 
of the hard particles, and starts to reveal details of the 
internal structure of the hard interaction. The first insight 
into the structure of ${\cal M}_{n+1}^{\rm NLP}$ was 
already achieved a long time ago in papers by 
Low, Burnett and Kroll~\cite{Low:1958sn,Burnett:1967km},
who realised that, in the case of massive emitting particles, 
the structure of the NLP term is dictated by gauge invariance, 
by means of Ward identities. This early formulation, now 
known as the ``LBK'' theorem, was proven~\cite{DelDuca:1990gz} 
to hold only in the region $k^0 \ll m^2/Q$, with $Q$ the 
centre of mass energy. 
For $m^2/Q < k^0 <m$, the LBK theorem must be extended 
to account for NLP contributions arising from soft photons 
emitted from loops in which the exchanged virtual particles 
have momenta collinear to the external particles (i.e.~having 
a small virtuality, while retaining momentum components 
which are large compared to the soft radiation),
which can be taken into account by a 
\emph{radiative jet}~\cite{DelDuca:1990gz}.

The factorization proposed in \cite{DelDuca:1990gz} was 
later confirmed by some of us in \cite{Bonocore:2015esa}, 
and extended to non-abelian theories in \cite{Bonocore:2016awd}, unveiling many features of soft radiation at NLP. 
For instance, NLP radiation at next-to-leading order (NLO)
in perturbation theory has a universal structure, where the 
radiative matrix element squared can be expressed as a 
reweighing of a kinematics-shifted non-radiative matrix element squared.
This reweighing factor is universal, in the sense that it only 
depends on the (colour) charge of the particles participating 
in the scattering~\cite{DelDuca:2017twk,vanBeekveld:2019prq}. 
These developments enabled the soft gluon resummation 
at NLP at leading logarithmic (LL) accuracy for 
Drell-Yan~\cite{Bahjat-Abbas:2019fqa} (which has 
also been achieved using effective field theory 
methods~\cite{Beneke:2018gvs}, discussed more 
below), proving earlier conjectures~\cite{Kramer:1996iq,Catani:2001ic,Laenen:2008ux}.
Other methods, based on evolution equations, have been used in \cite{Moch:2009hr,deFlorian:2014vta,Presti:2014lqa,Ajjath:2020ulr,Ajjath:2020sjk,Grunberg:2009yi}.
A phenomenological 
study of NLP resummation at LL accuracy for prompt 
photon production was carried out in~\cite{vanBeekveld:2019cks}. 

While the radiative jet function significantly extends 
the applicability of the factorization beyond the original 
LBK theorem, there are additional types of radiative 
jets beyond one loop. These describe for example a 
soft emission from \emph{multiple} highly energetic particles in 
the \emph{same} collinear direction emanating from 
the hard scattering, which has been studied for 
Yukawa theory in \cite{Gervais:2017yxv}. 
The extension of this analysis to QED, classifying 
all types of radiative jets, is necessary to describe soft 
radiation in QED at NLP to all orders in perturbation theory, and is addressed in this paper.

Factorization theorems for amplitudes or cross sections 
involving soft emissions have also been studied using 
an effective field theory approach, specifically within 
Soft-Collinear Effective Theory (SCET)~\cite{Bauer:2000ew, Bauer:2000yr, Bauer:2001ct,Bauer:2001yt,Beneke:2002ph}.
SCET describes the soft and collinear limits of QCD as 
separate degrees of freedom, each with their own Lagrangian.
The elastic amplitude ${\cal M}_n$ in the factorization in 
\eqref{factLP} is encoded in terms of effective $n$-jet 
operators and their corresponding short-distance 
coefficients, which capture the contribution from 
hard loops. At LP, soft emissions from hard 
particles can be described by Wilson lines, as in the 
diagrammatic picture. Beyond LP, these soft emissions 
follow from time-ordered non-local operators made out 
of soft and collinear fields,
where the power suppression follows either from additional insertions of the 
power-suppressed soft and collinear Lagrangian, or from subleading operators describing the hard scattering.
Within SCET it is possible to define matrix elements 
which are equivalent to the radiative jets of the
diagrammatic approach~\cite{Larkoski:2014bxa,Moult:2019mog,Beneke:2019oqx}.
Several investigations have been conducted within 
SCET, including but not limited to soft gluon corrections, such as studies of the anomalous
dimension of power-suppressed operators~\cite{Beneke:2017ztn, Beneke:2018rbh,Beneke:2019kgv}, 
the basis of power-suppressed hard-scattering 
operators for several processes~\cite{Moult:2017rpl,Feige:2017zci,Chang:2017atu}, 
the application to subtractions \cite{Moult:2016fqy,Boughezal:2016zws,Moult:2017jsg,Boughezal:2018mvf,Ebert:2018lzn,Boughezal:2019ggi,Ebert:2018gsn} 
and resummation of NLP LLs in a variety of processes~\cite{Moult:2018jjd,Beneke:2018gvs, Beneke:2019mua,Moult:2019uhz,Liu:2019oav,Wang:2019mym,Liu:2020ydl,Liu:2020eqe}. 

SCET provides a systematic approach, as each operator 
and Lagrangian term have by construction a definite power 
counting. When all operators are included that are consistent
with symmetries up to the desired power, this 
completeness ensures that the resulting factorization
is valid to all orders in the coupling constant. Consequently, factorization can, 
at least formally, be extended beyond NLP, by simply 
adding more power-suppressed operators. On the other 
hand, the diagrammatic approach is often more direct 
compared to the full effective field theory treatment, 
and may also offer a way (as we will see in this paper) to address so-called endpoint 
singularities in convolution integrals. These convolutions 
between ingredients in the factorization are only well defined 
in dimensional regularisation, thus posing a challenge for 
SCET, where one first renormalises each ingredient in the 
factorization theorem to derive the renormalisation group 
equations needed for resummation, causing these 
convolution integrals to become divergent. (However 
see \cite{Liu:2019oav, Beneke:2020ibj,Liu:2020tzd} for recent progress in addressing this issue.) One may hope that the diagrammatic 
approach will provide an easier path to resummation, 
as resummation exploits exponentiation properties of 
soft radiation and the replica trick~\cite{Laenen:2008gt,Gardi:2010rn}, 
which can be carried out within dimensional regularisation.
\begin{figure}[t]
\begin{center}
	\includegraphics[width = .44\textwidth]{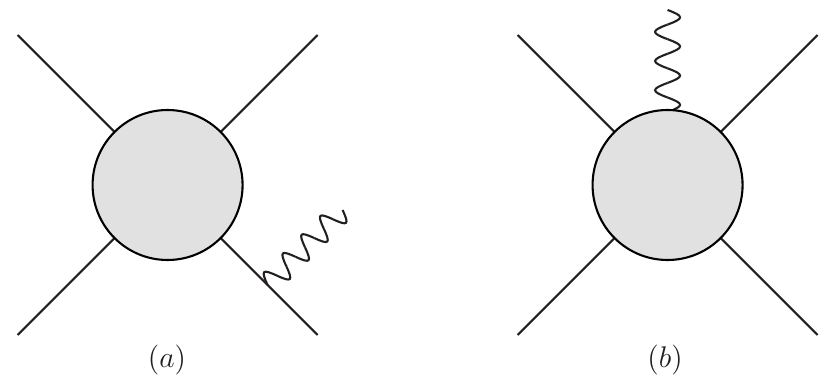}
	\caption{Decomposition of the radiative amplitude into 
	${\cal M}_{n+1}^{\rm ext}$ (a) and ${\cal M}_{n+1}^{\rm int}$ (b), 
	as defined in \eq{LBKeq}.}
	\label{fig:LBK}
\end{center}
\end{figure}
In order to set the stage for the development of 
a factorization theorem for the radiative amplitude
${\cal M}_{n+1}^{\rm NLP}$ in \eq{softexpansion}, 
let us recall in more detail where the original LBK 
theorem fails. Within the latter treatment, one separates the 
radiative amplitude in two contributions: one in which 
the radiation is emitted from the external legs, plus 
another in which the radiation is emitted from a particle 
within the hard scattering kernel. Schematically this 
is written as 
\be\label{LBKeq}
{\cal M}_{n+1} = {\cal M}_{n+1}^{\rm ext} 
+ {\cal M}_{n+1}^{\rm int}\,,
\ee
where the two terms are represented in figure 
\ref{fig:LBK}. The amplitude ${\cal M}_{n+1}^{\rm int}$ 
can be obtained by means of the Ward identity from 
${\cal M}_{n+1}^{\rm ext}$. Concerning
the amplitude involving an emission from the external legs, 
consider as an example the emission from an outgoing
fermion $i$: in this case ${\cal M}_{n+1}^{\rm ext}$ takes 
the form 
\begin{align}\label{LBKwrong}
{\cal M}_{n+1}^{\rm ext} &= \bar u(p_i) (i e q_i \gamma^{\mu})
\, \frac{i(\slashed{p}_i+\slashed{k}+m)}{(p_i+k)^2 - m^2} 
\nn\\ &\qquad\times {\cal M}_n(p_1,\ldots, p_i+k, \ldots p_n)\,,
\end{align}
where ${\cal M}_n$ represents the elastic amplitude
(stripped off the spinor $\bar{u}(p_i)$). Within the LBK
theorem, one expands the amplitude in the soft 
momentum $k$. As discussed, this expansion 
gives correct results only in the regime 
$p_i\cdot k/Q \ll m^2/Q$. But for parametrically small masses 
$m^2/Q < p_i\cdot k/Q <m$, naively expanding the elastic 
amplitude in the soft momentum $k$ misses a contribution 
in which the soft photon is emitted from internal particles
collinear to the external leg, which is thus included
neither in ${\cal M}_{n+1}^{\rm ext}$, nor in 
${\cal M}_{n+1}^{\rm int}$.

From a practical perspective, it is known that a correct 
expansion of ${\cal M}_{n+1} $ can be obtained by means 
of the method of regions~\cite{Beneke:1997zp,Smirnov:2002pj}: 
one splits a given loop integral into the sum of several integrals, 
in which the loop momentum is assumed to take different scalings, 
related to the scaling of the momenta of the external particles. 
Within each region one is allowed to expand the integrand in 
the small scales in that region, and then the sum over 
all regions is expected to provide the correct expansion of 
the original integral. In particular, using this method it is 
possible to check that the missing contribution in 
\eq{LBKwrong} indeed arises from the region where 
the loop momentum has scaling collinear to the external 
particles \cite{Bonocore:2014wua,Bahjat-Abbas:2018hpv}.

From the point of view of factorization, \eq{LBKwrong}
tells us that in order to understand the factorization properties 
of the radiative amplitude ${\cal M}_{n+1}$ we also need
to obtain the correct factorization structure of the elastic 
amplitude ${\cal M}_{n}$, in presence of a small off-shellness 
$p_i\cdot k \sim m^2 \ll Q^2$. At leading power the factorization
structure is known, see for instance~\cite{Collins:1989bt,Dixon:2008gr}, 
and it takes the following schematic form:
\be\label{factLP_nonrad}
{\cal M}_n = H_n\times S_n\times\prod_{i=1}^n \, \frac{J_{i}}{{\cal J}_i}\,.
\ee
In this equation the jet and soft functions $J_j$ and $S$
describe long-distance collinear and soft virtual radiation in 
${\cal M}_n$. These functions are universal, i.e. they depend 
only on the colour and spin quantum numbers of the external 
states, and determine also the structure of collinear and soft 
singularities of the elastic amplitude. Note that one must divide 
each jet by its eikonal counterpart ${\cal J}_i$, to avoid the double counting of soft and collinear divergences.

Given this premise, our first task is thus to determine the 
analogue of \eq{factLP_nonrad} at NLP.
In the absence of soft radiation, the elastic amplitude would 
only depend on hard scales. Following \cite{Gervais:2017yxv},
we will consider the external fermions to have a parametrically 
small mass $m$, providing us with a variable for the power 
expansion. We derive the power counting, which we then 
generalise to the case of massless particles, and obtain an 
all-order NLP factorization formula. In either case new, 
universal jet functions are required at the NLP level, consisting 
of multiple collinear particles along the same direction, probing 
the hard scattering process.
We restrict ourselves to the first non-trivial jet function, which we calculate for a parametrically small fermion mass up to NLP. Subsequently, we perform checks at one- and two-loop level that validate the obtained jet function and the corresponding part of the factorization 
formula. We carry out a similar analysis for single-radiative 
amplitudes in the massless fermion scenario.
We stress that our study is exploratory in nature, a full characterisation 
of the radiative jet functions is left to future work. 
Having in hand the factorization for both massless and massive fermions would allow the application of our results to a larger class of scattering processes of interest 
at the LHC, including the production of heavy 
coloured particles, such as top quarks, or scalar quarks and 
gluinos in supersymmetry. 

The outline of the paper is as follows: in section \ref{sec:power counting} we carry out the power counting analysis that underpins the (non-radiative) all-order NLP factorization formula presented in that section for two scenarios: one with a parametrically small fermion mass $m$ and one for massless fermions. We focus on the massive case in section \ref{sec:MassiveCase}, computing the first non-trivial jet function and performing checks at the  one- and two-loop level.  In section \ref{sec:rad_massless}, we consider single-radiative amplitudes in the massless fermion scenario, again performing one- and two-loop checks. We conclude in section \ref{sec:conclusions}, while certain technical aspects are relegated to appendices.

\section{From power counting to factorization} \label{sec:power counting}

The factorization of an $n$-particle scattering amplitude with emission of a soft 
photon ${\cal M}_{n+1}$ crucially depends on the factorization 
properties of the corresponding elastic amplitude ${\cal M}_{n}$. 
Therefore we start our analysis by extending \eq{factLP_nonrad} 
to NLP. Specifically, we set out to obtain a classification of the jet-like structures, 
consisting of virtual radiation collinear to any of the $n$ external 
hard particles, contributing at subleading power. 
Phrased differently,
we wish to derive which jet functions contribute up to NLP in a 
parametrically small scale, corresponding to a fermion mass or a soft external 
momentum. 

In the following we will distinguish two fermion mass scenarios. One of which is the truly massless theory ($m=0$), the standard approximation in high energy calculations, where it is well understood that (virtual) collinear effects beyond the LBK theorem play an important role. In the other scenario, we consider fermion masses to be non-zero but parametrically small, and in fact comparable to the scale associated to soft emissions: we assume $m \!\sim\! \lambda Q$, such that $p_i\cdot k \!\sim\! m^2 \!\sim \!\lambda^2 Q^2 \!\ll\! Q^2$, where $k$ is a soft momentum. This more intricate small-mass approximation could be of phenomenological importance if soft gluons are emitted from particles with an intermediate-size mass, as mass effects may be comparable in size to the aforementioned collinear effects. Resummation of resulting NLP threshold logarithms (beyond LL accuracy) would, in that case, require a proper understanding of massive radiative jet functions. In addition, this second scenario may prove useful for the resummation of logarithmic mass terms, $\log(m/Q)$, even in \emph{non-radiative} processes, where the small fermion mass $m$ and the hard scale $Q$ are the only scales in the problem.  

\begin{figure}[t]
\begin{center}
\includegraphics[width = .3\textwidth]{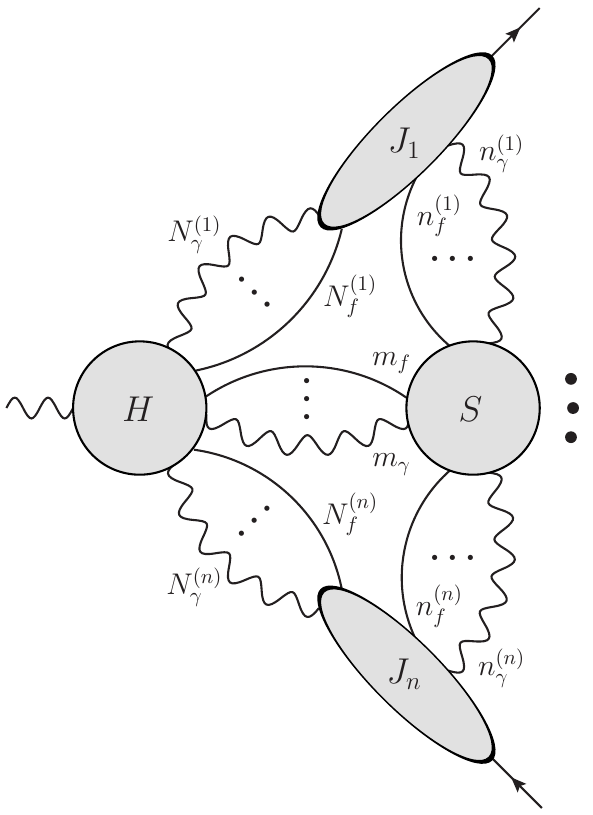}
\caption{The reduced diagram for a general, vector boson 
induced QED process with $n$ well-separated hard particles in the final state. 
Ellipses denote the presence of an arbitrary number of 
photon/(anti-)fermion lines.}
\label{fig:reduced_diagram}
\end{center}
\end{figure}

We derive our results by power counting the 
\textit{pinch surfaces}, that underlie the collinear 
(and soft) contributions we wish to describe in terms 
of jet (and soft) functions, for a general QED scattering 
amplitude. This was done recently for Yukawa theory in \cite{Gervais:2017yxv} 
for the same two mass scenarios. The pinch surfaces are 
the solutions of the Landau equations~\cite{Landau:1959fi} 
and are represented by reduced diagrams in
the Coleman-Norton picture~\cite{Coleman:1965xm}. 
In these diagrams, all off-shell lines are shrunk 
to a point, while the on-shell lines are kept and may be 
organised according to the nature of the singularity they 
embody, be it soft or collinear. This results in the general 
reduced diagram of \fig{reduced_diagram}, in which one 
distinguishes a soft ``blob" containing all lines carrying solely 
soft momentum, $n$ jets $J_i$ comprised of lines with 
momenta collinear to the respective external parton and 
lastly, a hard blob $H$ collecting all contracted, off-shell 
lines. 
This picture seems unaltered by the presence of parametrically small fermion masses, because the limit of small $m$ yields the same singular pinch surfaces as the massless theory. In support of this claim, we analysed the QED massive form factor
using the method of regions~\cite{Beneke:1997zp,Smirnov:2002pj,Jantzen:2011nz}, 
finding that soft and collinear modes are sufficient to 
correctly reproduce the singularity structure in this limit. 
This analysis is presented in appendix~\ref{appx:regions}.

To carry out the power counting we use light-cone coordinates. For each external momentum $p_i$, we introduce two light-like vectors $n_i$ and $\bar n_i$, defined by
\be\label{def:ni}
n_i^\mu = \frac{1}{\sqrt{2}}\left(1,+\frac{\vec{p}_i}{|\vec{p}_i|}\right)\,,\qquad 
\bar{n}_i^\mu = \frac{1}{\sqrt{2}}\left(1,-\frac{\vec{p}_i}{|\vec{p}_i|}\right)\,.
\ee
These vectors are normalised such that 
$n_i \cdot \bar{n}_i = 1$, and by definition 
$n_i^2 = \bar n_i^2=0$. In any of these coordinates, a generic vector $v$ 
decomposes as 
\be\label{LCdecomposition}
v^{\mu} = v^+ \,n^{\mu}_i 
+ v^- \,\bar n^{\mu}_i +v^\mu_{\perp i},
\ee
where $v^+ = v\!\cdot\! \bar{n}_i$, $v^- = v\!\cdot\! n_i$\,. 
A scalar product of two vectors then reads
\be\label{LCscalarproduct}
v\!\cdot\! w = v^+w^- + v^- w^+ + v^\mu_\perp\,w_{\perp\h\mu}\,.
\ee
Of course, this needs further specification in which of the $N$ collinear directions the decomposition is carried out.

Adopting the notation $k^\mu = (k^+,\vec{k}_\perp,k^-)$, 
we associate the following scaling to lines that are soft 
or collinear to the $i$-th external leg
\begin{align} 
\text{Soft:} &\hspace{20pt} k^\mu\sim Q\left(\lambda^2,\lambda^2,\lambda^2 \right)\, ,\nn\\ \text{Collinear:} &\hspace{20pt} k^\mu\sim Q\left(1,\lambda, \lambda^2\right) \,,   
\label{softcol} 
\end{align} 
in terms of the light-cone coordinates corresponding to $p_i$. The scaling of the normal coordinates parametrises the contribution of soft and collinear lines around the singular surface, which is reached for $\lambda\rightarrow 0$. Away from this limit, power counting in $\lambda$ thus amounts to the ordering of finite contributions of different size and proves to be a valuable technique.
We focus on virtual corrections to a hard scattering configuration,
for which all invariants $s_{ij} = (p_i+p_j)^2 
\sim Q^2$ involving external momenta are large compared to the energy of the 
radiated soft photon in ${\cal M}_{n+1}$. Requiring the soft momentum 
to be of order $\lambda^2$ rather than $\lambda$ guarantees that the photon is soft with respect to all particles in the elastic amplitude. 

Whenever we refer to a NLP 
quantity in this paper, we mean that it is suppressed by \textit{up to two} powers in $\lambda$ with respect to the leading power contribution. This
nomenclature originates from strictly massless ($m=0$) QED 
 where power corrections arise only through scales associated to soft emissions $p_i\cdot k \!\sim\! \lambda^2 Q^2$. In case of parametrically small masses ($m\!\sim\!\lambda Q$) power suppressed terms at $\ord (\lambda)$ do occur, but we apply the same definition nonetheless.

Using the momentum scaling in \eq{softcol}, we start by
deriving the \textit{superficial degree of divergence} of 
a particular reduced diagram $\mathcal{G}$ contained 
in \fig{reduced_diagram}, which is simply the 
$\lambda$-scaling of this diagram, 
$\mathcal{G}\sim \lambda^{\gamma_\mathcal{G}}$. 
Specifically, we wish to determine how 
$\gamma_\mathcal{G}$ depends on the structure of 
$\mathcal{G}$. We will see that, in practice, 
$\gamma_\mathcal{G}$ can be expressed as function 
of the number of fermion and photon connections 
between the hard, soft and collinear subgraphs and, 
in presence of fermion mass, on the internal structure 
of the soft subgraph. Such a formula tells us, at any 
perturbative order, which pinch surfaces contribute 
up to NLP and guides us in setting up a consistent 
and complete NLP factorization framework for 
QED. This approach is analogous to the one taken for Yukawa theory in \cite{Gervais:2017yxv}\footnote{We summarise additional results for Yukawa theory with respect to that reference in appendix \ref{appx:Yukawa}.}, while the power counting itself is a direct application of the well-known method first developed in Ref.~\cite{Sterman:1978bi}. To support the factorization analysis in \sect{factorization}, we show in some detail the derivation leading to the final power-counting formula in \eq{eq:QEDPC_final}.

\subsection{Power counting rules for individual components} \label{sec:detailedPC}

In order to derive an expression for $\gamma_\mathcal{G}$ it is convenient to set up a catalogue of the degree of divergence of the individual components first. For massive ($m \sim \lambda Q$) and massless ($m=0$) QED, these rules vary slightly and we derive them explicitly here. 

Given \eq{softcol}, the propagator for a collinear, massive fermion scales as 
\begin{equation}\label{dropSubleading}
\frac{i\left(\slsh{p}+m\right)}{p^2-m^2}\sim\frac{\gamma^-+\lambda^2\gamma^+ + \lambda \gamma^\perp+ \lambda}{\lambda^2}\sim \frac{1}{\lambda^2}\,.
\end{equation} A massless collinear fermion obeys the same rule as the mass term is subleading in the numerator ($\ord (\lambda)$ versus $\ord (1)$) and of equal size in the denominator (both $\ord\left(\lambda^2\right)$). For soft fermion lines a difference does arise; for non-zero mass
\begin{equation}\label{anomScaling}
\frac{i\left(\slsh{p}+m\right)}{p^2-m^2}\sim\frac{\lambda^2\gamma^- +\lambda^2\gamma^+ + \lambda^2 \gamma^\perp+ \lambda}{\lambda^4+\lambda^2}\sim \frac{1}{\lambda}\,,\end{equation}
while for a massless fermion one finds instead
\begin{equation}
\frac{i\slsh{p}}{p^2}\sim\frac{\lambda^2\gamma^- +\lambda^2\gamma^+ + \lambda^2 \gamma^\perp}{\lambda^4}\sim \frac{1}{\lambda^2}\,.
\end{equation}

Since we aim at determining the order at which each configuration start contributing, we will only keep track of the most singular contribution to $\gamma_\mathcal{G}$, and discard the subleading terms in \eq{dropSubleading} and~\eqref{anomScaling}. The singular structure of \eq{anomScaling} is uncommon because the denominator is not strictly on shell, since $p^2 \sim \lambda^4$ while $m^2 \sim \lambda^2$. In fact, this singularity is entirely determined by the fermion mass. Intuitively, because of their mass, soft fermions are integrated out, an aspect that would be worth investigating from an effective theory perspective. This momentum configuration, which contributes to the singular structure of scattering amplitudes despite being off shell, bears similarity to Glauber gluons, scaling as $(\lambda^2,\lambda,\lambda^2)$. Our power counting shows that these momentum configurations could affect scattering amplitudes only beyond NLP.

In gauge theories the rules for vector boson vertices depend on the choice of gauge. For power counting purposes, the axial gauge is particularly convenient since non-physical degrees of freedom do not propagate.
The latter is a direct consequence of the form of the photon propagator, which reads \begin{equation} \Delta^{\mu\nu}(k,r)=\frac{i}{k^2+i\eta} \bigg[-\eta^{\mu\nu}+\frac{r^\mu k^\nu+r^\nu k^\mu}{r\cdot k}-\frac{r^2 k^\mu k^\nu}{\left(r \cdot k\right)^2}\bigg]\,, \label{Axprop}  \end{equation} with the choice of the reference vector $r^\mu$ fixing the gauge. \Eq{Axprop} satisfies \beq r_\mu\Delta^{\mu\nu} \label{pol1}(k)=0\,,\eeq while contracting with the propagating momentum results in \beq k_\mu \Delta^{\mu\nu}(k,r) = i\left(\frac{r^\nu}{r\cdot k}-\frac{r^2 k^\nu}{\left(r\cdot k\right)^2}\right)\,,\label{pol2} \eeq which no longer has a pole in $k^2$. Together, eqs.~\eqref{pol1} and \eqref{pol2} show that scalar and longitudinal polarisations do not propagate in the chosen gauge. 
This choice of gauge makes it particularly convenient to derive the suppression effects associated to vertices involving gauge bosons~\cite{Sterman:1978bi}. These are effective rules, in the sense that they are not evident from the vertex factor as obtained from the QED Lagrangian, but follow from an interplay with the adjacent lines. 
 To make this concrete, consider the expression for the emission of a  photon from a collinear fermion line with momentum $p^\mu$, which is proportional to 
 \beq 
 \left(\slashed{p}-\slashed{k}\right) \gamma^\mu \slashed{p}= -p^2\gamma^\mu+\gamma^\mu\slashed{k} \slashed{p} + 2 \left(p^\mu-k^\mu\right)\slashed{p}\,. \label{QEDem}
 \eeq
First, we point out that the first two terms are always power suppressed: the first one is per definition of order $\lambda^2$, while the second term is of order $\lambda$ even if the photon emission is collinear, as the dominant component vanishes due to $\left(\gamma^-\right)^2=0$ (for a soft photon the second term is manifestly of order $\lambda^2$). If the photon is soft, $p^\mu\slashed{p}$ in the third term of \eq{QEDem} dominates, being of order $\lambda^0$. In that case, no suppression is caused by the vertex. However, if the photon is collinear to the fermion lines extending from the vertex, we can write $p^\mu = \frac{p^{+}}{k^+} k^\mu + \mathcal{O}(\lambda)$. From \eq{pol2} we then conclude that there is no dominant contribution to on-shell scattering amplitudes from the third term in \eq{QEDem}.  Hence, in axial gauge, a suppression of $\lambda$ is associated to each emission of a collinear photon from a collinear fermion line. With a different choice of gauge, the presence of longitudinal polarisations would erase this suppression effect of vertices, and individual diagrams would exhibit a harder scaling. In physical observables such polarisations cancel due to Ward identities, and the extra $\lambda$ suppression would become evident when summing over a gauge-invariant set of diagrams. We summarise the rules for QED vertices in table \ref{tab:QEDrules}. The scaling of a photon line is determined by the common factor $\frac{1}{k^2}$ in \eq{Axprop}, and is therefore $\sim \frac{1}{\lambda^2}$ and $\sim \frac{1}{\lambda^4}$ for respectively collinear and soft particles.  

A further suppression of the degree of divergence results from integration over loop momenta, where the measure $\int d\ell^+ d\ell^- d^2\vec{\ell}_\perp$ provides a suppression of, respectively, $\lambda^4$ and $\lambda^8$ for collinear and soft loops. These results are, together with the rules for propagators, presented in table \ref{tab:pcrules}.
\begin{table}[t]
\begin{center}
\begin{tabular}{lll}
\hline
                             QED Vertex                                                    & \hspace{-30pt}Suppression  \\ \hline
\begin{tabular}[c]{@{}l@{}} $\bar{\psi}^{(c)}\gamma^\mu\psi^{(c)}A_{\mu}^{(c)}$ \end{tabular} & $\lambda$  \\
\begin{tabular}[c]{@{}l@{}} $\bar{\psi}^{(c)}\gamma^\mu\psi^{(c)}A_{\mu}^{(s)}$\end{tabular} & $1$  \\
\begin{tabular}[c]{@{}l@{}} $\bar{\psi}^{(s)}\gamma^\mu\psi^{(c)}A_{\mu}^{(c)}$ or  $\bar{\psi}^{(c)}\gamma^\mu\psi^{(s)}A_{\mu}^{(c)}$\end{tabular} & $1$   \\
\begin{tabular}[c]{@{}l@{}} $\bar{\psi}^{(s)}\gamma^\mu\psi^{(s)}A_{\mu}^{(s)}$\end{tabular} & $1$  \\ \hline
\end{tabular}
\end{center}
\caption{Power counting rules for QED vertices, depending on the soft or collinear nature of the field. These rules apply to massive and massless fermions alike.}
\label{tab:QEDrules} 
\end{table}
\begin{table}[t]
\begin{center}
\begin{tabular}{lll}
\hline
                          & $m=0$          & $m \sim \lambda Q$  \\
\hline 
Collinear fermion         & $\lambda^{-2}$ &  \\
Soft fermion              & $\lambda^{-2}$ & $\lambda^{-1}$ \\
Collinear photon          & $\lambda^{-2}$ &  \\
Soft photon               & $\lambda^{-4}$ &  \\
Collinear loop            & $\lambda^{4}$  &  \\
Soft loop                 & $\lambda^{8}$  &  \\
\hline
\end{tabular}
\end{center}
\caption{Power counting rules for loop integrals and propagators for photons and fermions. If no rule is specified for $m \sim \lambda Q$, the scaling is identical to $m=0$.}
\label{tab:pcrules}
\end{table}

\subsection{Constructing the overall degree of divergence}

We started the derivation of a formula for $\gamma_{\mathcal{G}}$ by obtaining the power counting rules for the basic constituents of any diagram. Here we use these results to obtain power counting formulae for the soft ($\gamma_{S}$) and collinear ($\gamma_{J_i}$) sub-diagrams independently. Subsequently, we consider the effect of connections between all sub-diagrams ($\gamma_{S\leftrightarrow H}$,$\gamma_{J_i\leftrightarrow H}$ and $\gamma_{J_i\leftrightarrow S}$) as well as the connections to the external particles ($\gamma_{J_i}^{\rm ext}$). The degree of divergence of a reduced diagram $\mathcal{G}$ with $n$-jets will thus be given by 
\begin{equation} \label{eq:QEDPCsplit}
  \gamma_\mathcal{G} = \gamma_S + \gamma_{S\leftrightarrow H} + \sum_{i=1}^n (\gamma_{J_i} + \gamma_{J_i\leftrightarrow H} + \gamma_{J_i \leftrightarrow S} + \gamma_{J_i}^{\rm ext})\, .
\end{equation}   

We begin with $\gamma_{J_i}$ and consider a blob of collinear lines, without any external attachments. According to the rules of table~\ref{tab:pcrules} the associated degree of divergence is 
\begin{equation} \gamma_{J_i} = -2I+4L+V,\label{Jint}\end{equation}
where $I = \tilde{I}_f+\tilde{I}_\gamma$ denotes the total number of fermion and photon lines internal to the isolated blob, $L$ the number of loops and $V$ the number of vertices.  We use \emph{Euler's identity} 
\begin{equation}L=1+\tilde{I}_f+\tilde{I}_\gamma-V,\label{euler}\end{equation} and note that diagrams without external legs (i.e.~vacuum bubbles) have three internal lines per pair of vertices: $I=\frac{3}{2}V$. As a result, the degree of divergence of a collinear sub-diagram is independent of its internal structure: 
\begin{equation}\gamma_{J_i} = -3V+4(1+\tfrac{1}{2}V)+V=4.\label{Jint2}\end{equation} 

For the soft sub-diagram one needs to distinguish between the different mass cases. We start with \begin{equation} \gamma_{S} = \begin{cases} -2\tilde{I}_f-4\tilde{I}_\gamma+8L \qquad \qquad & (m=0) \\
-\tilde{I}_f-4\tilde{I}_\gamma+8L \qquad \qquad & (m \sim \lambda Q) \end{cases}. \end{equation} Applying Euler's identity in \eq{euler} and exploiting the fixed ratio of the number of fermion and photon lines to the number of vertices in a QED vacuum bubble $\tilde{I}_f = 2 \tilde{I}_\gamma = V$, we obtain \begin{equation} \gamma_{S} = \begin{cases} = 8 \qquad \qquad &(m=0) \\ =8+\tilde{I}_f \qquad\qquad &(m \sim \lambda Q)\end{cases}. \label{Sint} \end{equation}
\begin{figure*}[t]
    \centering
    \subfloat[Photon insertion on a fermion line.]{
    \makebox[.21\textwidth][c]
    {\includegraphics[width=.14\textwidth]{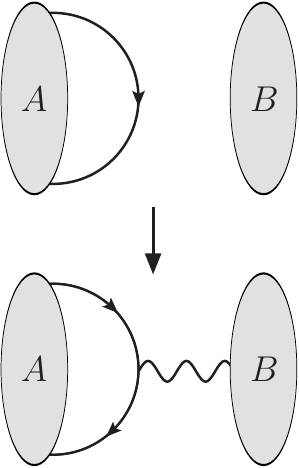}
    }
    \label{fig:insertion_a}}\hspace{30pt}
    \subfloat[Pairwise fermion insertion on a fermion line.]{
    \makebox[.21\textwidth][c]
    {\includegraphics[width=.14\textwidth]{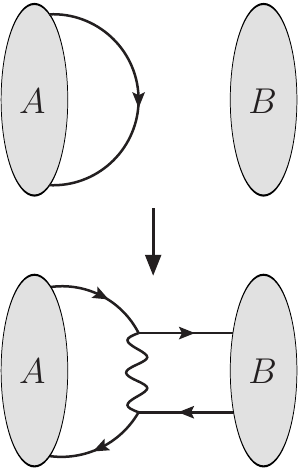}
    } 
    \label{fig:insertion_b}}\hspace{30pt}
    \subfloat[Pairwise fermion insertion on a photon line.]{
    \makebox[.21\textwidth][c]
    {\includegraphics[width=.14\textwidth]{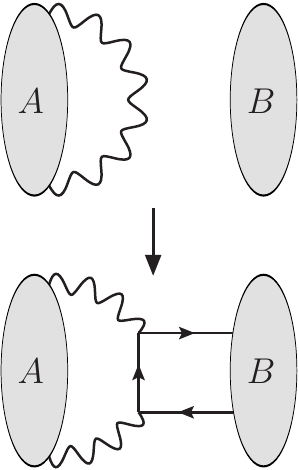}
    }
    \label{fig:insertion_c}}
    \caption{The effect of connecting lines between isolated sub-diagrams $A$ and $B$ on the former.}
    \label{fig:insertions}
\end{figure*}
Next, we must account for the contribution to the overall degree of divergence arising from the connecting lines between hard, soft and jet sub-diagrams of the general reduced diagram in \fig{reduced_diagram}. Besides the explicit powers of $\lambda$ associated to lines themselves, they affect the power counting of the disconnected sub-diagrams by splitting internal propagators and adding vertices to both sub-diagrams. In \fig{insertions} we show these effects on a generic sub-diagram A resulting from either a photon or fermion connection to a sub-diagram B, depending on the internal line that is probed. For fermion connections, a fermion anti-fermion pair is inserted to conserve charge in both sub-diagrams.\footnote{In principle the charge flow can be more involved and form, for example, a closed loop through the hard, soft and a collinear sub-diagram, or connect to external fermions of opposite charge. These configurations can nevertheless be obtained by applying the basic steps in \fig{insertions}.} The effect per fermion is simply half that of the combined fermion anti-fermion insertion. An additional suppression effect arises from the loops that are formed in this process.
Consider the connection between a jet and the hard sub-diagram first. A connecting (collinear) photon line adds also a collinear fermion line and an all-collinear QED vertex to the collinear blob, as shown in \fig{insertion_a}. According to the rules listed in table \ref{tab:pcrules} and \ref{tab:QEDrules}, such a connection enhances the degree of divergence by $-2-2+1=-3$. Each connecting fermion line gives the same effect, as found by using the aforementioned procedure: a fermion anti-fermion pair adds in total four collinear lines and two all-collinear vertices, such that the enhancement of the degree of divergence due to a single fermion line is $\frac{4\times(-2)\,+\,2\times (+1)}{2}\!=\!-3$. In addition, the $N^{(i)}\!=\!N_\gamma^{(i)}\!+\!N_f^{(i)}$ connecting lines give rise to $N^{(i)}\!-1$ collinear loops. Summing up, we find 
\begin{align}
  \gamma_{J_i\leftrightarrow H} &= -3N_\gamma^{(i)} -3 N_f^{(i)} + 4(N_\gamma^{(i)}+N_f^{(i)}-1) \nonumber\\
  &= N_\gamma^{(i)}+N_f^{(i)}-4\,.
\end{align}

The reduced diagrams considered here are amputated, meaning that there is no propagator, and thus no power counting, associated to the external leg itself. Therefore, connecting the jet to an external fermion leg gives a further enhancement of the degree of divergence of
\begin{equation}
    \gamma^{\rm ext}_{J_i} = \frac{2\!\times\!(-2)+2\!\times \!(+1)}{2}= -1\,,
\end{equation} where only the vertices and additional collinear propagators due to the (pairwise) fermion insertion are counted. Similarly, connecting the jet to an external photon gives one additional collinear fermion line an an all-collinear vertex, such that also in this case
\begin{equation}
    \gamma_{J_i}^{\rm ext} = -2 +1 = -1\, .
\end{equation}
In contrast to $\gamma_{J_i\leftrightarrow H}$ and $\gamma^{\rm ext}_{J_i}$, the degree of divergence associated to the connection between the soft and hard sub-diagram is not suppressed by vertices since all lines are soft. Therefore, we only need to count the $m_\gamma+m_f$ soft connections themselves, as well as the additional lines created in the soft blob by these insertions (one soft fermion per photon insertion; one soft photon and an additional soft fermion for a pairwise fermion insertion).\footnote{The number of soft photon and fermion lines connecting to the hard sub-diagram, denoted by $m_\gamma$ and $m_f$, should not be confused with the fermion mass $m$.} Including the loop suppression, we find
\begin{widetext}
\begin{subequations}
\begin{align}
  \gamma_{S\leftrightarrow H} & \;=\; (-2-4)m_\gamma + \Big(\frac{3\!\times\!(-2)-4}{2}\Big)m_f + 8(m_\gamma + m_f -1) 
  \;=\; 2m_\gamma + 3m_f - 8 \hspace{40pt}(m=0)\, ,\\
  \gamma_{S\leftrightarrow H} & \;=\; (-1-4)m_\gamma + \Big(\frac{3\!\times\!(-1)-4}{2}\Big)m_f + 8(m_\gamma + m_f -1)
   \;=\; 3m_\gamma + \tfrac{9}{2}m_f - 8 \hspace{39pt}(m \sim \lambda Q)\, .
\end{align}
\end{subequations}
Finally, we consider the $n_\gamma^{(i)}+n_f^{(i)}$ connections between the soft sub-diagram and the jets, which affect both the sub-diagrams involved.\footnote{The main difference in power counting compared to Yukawa theory arises from this interaction. In Yukawa theory, each scalar emission from a collinear fermion line is suppressed by a  factor of $\lambda$, such that power counting rules for all-collinear and all-soft vertices are identical in QED and Yukawa theory. However, vertices for soft-collinear interactions are suppressed by $\lambda$ in Yukawa theory, but are not suppressed in QED.} Also, these connections will form an additional loop by closing a path through $H$, $J_i$ and $S$, giving a total of $n_\gamma^{(i)}+n_f^{(i)}$ soft loops. The result is  
\vspace{-5pt} 
\begin{subequations}
\begin{align}
 \gamma_{J_i \leftrightarrow S} & \;=\; -2\overbrace{\left(n_\gamma^{(i)}+n_f^{(i)}\right)}^{\rm collinear\ effects} - \overbrace{\left(6n_\gamma^{(i)} + 5 n^{(i)}_f\right)}^{\rm soft\ effects} + 8\,(n^{(i)}_\gamma+n^{(i)}_f) \;=\; n_f^{(i)} &\hspace{20pt} &(m=0)\, , \\
\gamma_{J_i \leftrightarrow S} & \;=\; -2\left(n_\gamma^{(i)}+n_f^{(i)}\right) - \left(5n_\gamma^{(i)} + \tfrac{7}{2} n^{(i)}_f\right) + 8\,(n^{(i)}_\gamma+n^{(i)}_f)
 \;=\; n_\gamma^{(i)} + \tfrac{5}{2}n_f^{(i)} &&(m \sim \lambda Q)\, .
\end{align}
\end{subequations}
Combining ingredients according to \eq{eq:QEDPCsplit} gives
\begin{subequations} \label{eq:QEDPC}
\begin{align}
  \gamma_{\mathcal{G}} &= 2m_\gamma + 3m_f + \sum_{i=1}^{n} (N_\gamma^{(i)} + N_f^{(i)} + n_f^{(i)} -1) &\hspace{40pt}& (m=0)\,, \\
  \gamma_{\mathcal{G}} &= \tilde{I}_f + 3m_\gamma + \tfrac{9}{2}m_f + \sum_{i=1}^{n}(N_\gamma^{(i)}+N_f^{(i)}+n_\gamma^{(i)} + \tfrac{5}{2}n_f^{(i)}-1) && (m \sim \lambda Q)\,.
\end{align}
\end{subequations}
\end{widetext}
We emphasise that the number of internal fermion lines $\tilde{I}_f$ in the soft sub-diagram denotes the number of lines in the \textit{isolated} blob, before connections to the hard and jet functions have been accounted for. It is more intuitive to express this in terms of the total number of internal fermion lines in the amputated soft function, $I_f$, for which we disregard the actual fermion connections to other blobs, but retain the effect that the connections have on the soft blob itself. Either a single photon attachment or a pairwise (anti-)fermion insertion adds a fermion line to the soft sub-diagram, as indicated in \fig{insertions}, giving the relation
\begin{equation}I_f = \tilde{I}_f+m_\gamma +\tfrac{1}{2}m_f+\sum_i\left(n_\gamma^{(i)}+\tfrac{1}{2}n_f^{(i)}\right).\label{If}\end{equation} 
Inserting \eq{If} in \eq{eq:QEDPC} gives
\begin{subequations} \label{eq:QEDPC_final}
\begin{align}
\label{eq:QEDPC_final_massless}
  \gamma_{\mathcal{G}} =&\, 2m_\gamma + 3m_f && (m=0) \nonumber \\&+ \sum_{i=1}^{n} (N_\gamma^{(i)} + N_f^{(i)} + n_f^{(i)} -1) \,, \\ \label{eq:QEDPC_final_massive}
  \gamma_{\mathcal{G}} =&\, I_f + 2 m_\gamma + 4 m_f &\hspace{-20pt}& (m \sim \lambda Q)\nonumber \\&+ \sum_{i=1}^{n}(N_\gamma^{(i)}+N_f^{(i)}+2n_f^{(i)}-1)\,,
\end{align}
\end{subequations}
 which are the final expressions for the overall degree of divergence for a reduced diagram $\mathcal{G}$ with $n$-jets. The massless result in \eq{eq:QEDPC_final_massless} is the analogue for $N$-jet production in QED of the power-counting formulae first derived in \cite{Sterman:1978bi,Akhoury:1978vq} for cut vacuum polarisation diagrams and wide-angle scattering amplitudes in a broader class of theories. The massive result in \eq{eq:QEDPC_final_massive} is the equivalent of the equation obtained in \cite{Gervais:2017yxv} for Yukawa theory, which we also re-derived. We present this and other results for Yukawa theory in \appx{appx:Yukawa}.

\subsection{NLP factorization of QED amplitudes}\label{sec:factorization}

Equipped with \eq{eq:QEDPC_final}, we can determine which reduced diagrams $\mathcal{G}$ contribute up to NLP in $\lambda$. For the class of diagrams considered in the previous section, which have an arbitrary number of purely virtual corrections, we see that $\gamma_\mathcal{G} \geq 0$, independent of the number of hard particles in the final state. The $\gamma_\mathcal{G} = 0$ diagrams contain at most logarithmic singularities, while the $\gamma_\mathcal{G} > 0$ are finite and give a vanishing contribution in the $\lambda\rightarrow 0$ limit. For small but non-zero values of $\lambda$, the $\gamma_\mathcal{G} = 0$ diagrams form LP contributions, with the $\gamma_\mathcal{G} > 0$ diagrams acting as power corrections.

Eventually, we wish to develop a factorization formalism that allows one to resum NLP threshold logarithms associated to soft final-state radiation, which requires us to study the factorization of \emph{radiative} amplitudes. Dressing the non-radiated graphs with a single, soft emission will enhance the degree of divergence by $-2$, by the splitting of a soft/collinear fermion line.\footnote{For $m \!\sim\! \lambda Q$, an emission from a soft fermion would enhance the degree of divergence by $-1$ instead. However, any non-radiative diagram that allowed for such an emission would contribute beyond NLP, so we may neglect this subtlety here.} So for these radiative amplitudes, a LP contribution will be $\ord(\lambda^{-2})$ instead, with the NLP corrections of $\ord(\lambda^{0})$.\footnote{At the cross section level, this still constitutes a logarithmic divergence as the phase space integral over the soft gluon cancels the enhancement of the degree of divergence due to additional propagators in the squared amplitude.} Therefore, we will list all purely virtual reduced diagrams $\mathcal{G}$ characterised by $\gamma_\mathcal{G} \leq 2$. Since we study the abelian theory, we restrict our analysis to (anti-)fermions in the final state, although the power counting formulae of \eq{eq:QEDPC_final} describe processes involving hard final-state photons as well. As a minimal example we study the amplitude for $\gamma \rightarrow f\! \bar{f}$, but stress that the jet functions that appear there cover the general case of $n$ (anti-)fermions.

\begin{figure}[t]
    \centering
    \subfloat[ $\gamma=0$]{
    \makebox[.24\textwidth][c]{\includegraphics[width=.15\textwidth]{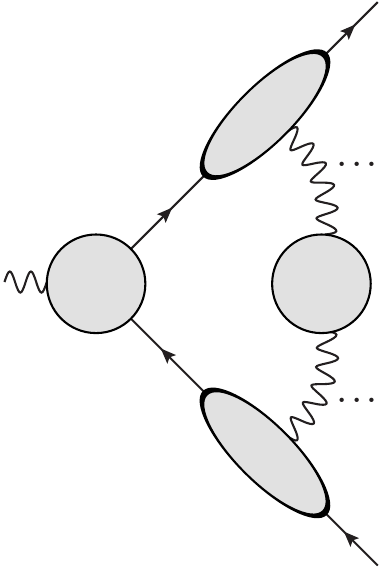}
    }
    \label{fig:red_diag_g0}}
    \subfloat[ $\gamma=1$]{
    \makebox[.24\textwidth][c]{\includegraphics[width=.15\textwidth]{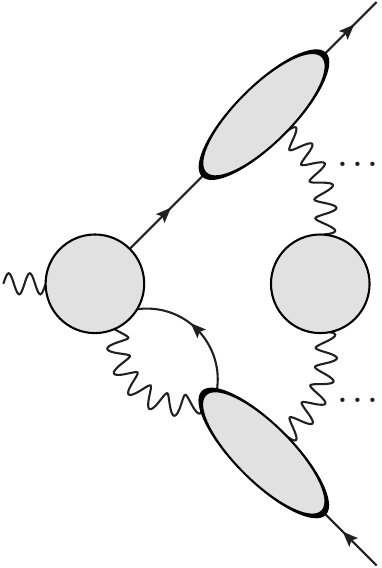}
    }
    \label{fig:red_diag_g1}}  
    \caption{Reduced diagrams for the process $\gamma \to f \bar{f}$. (a) $\gamma = 0$ (b) $\gamma = 1$, where a similar configuration exists with the double jet connection on the upper leg instead.}
  \label{fig:red_diag01}
\end{figure}

\begin{figure}[t]
    \centering
    \subfloat[]{
    \makebox[.24\textwidth][c]{\includegraphics[width=.16\textwidth]{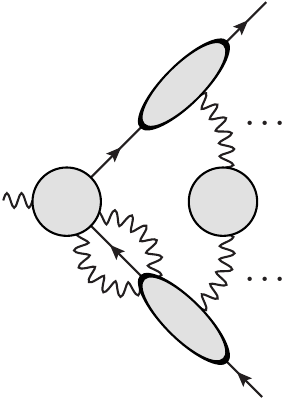}}
    \label{fig:red_diag_HJ_fgammagamma}
    }
    \subfloat[]{
    \makebox[.24\textwidth][c]{\includegraphics[width=.16\textwidth]{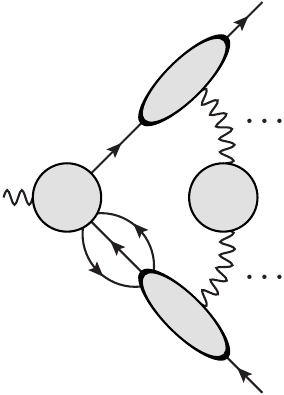}}
    \label{fig:red_diag_HJ_fff}
    }
    \vspace{-5pt}
    \subfloat[]{
    \makebox[.24\textwidth][c]{\includegraphics[width=.16\textwidth]{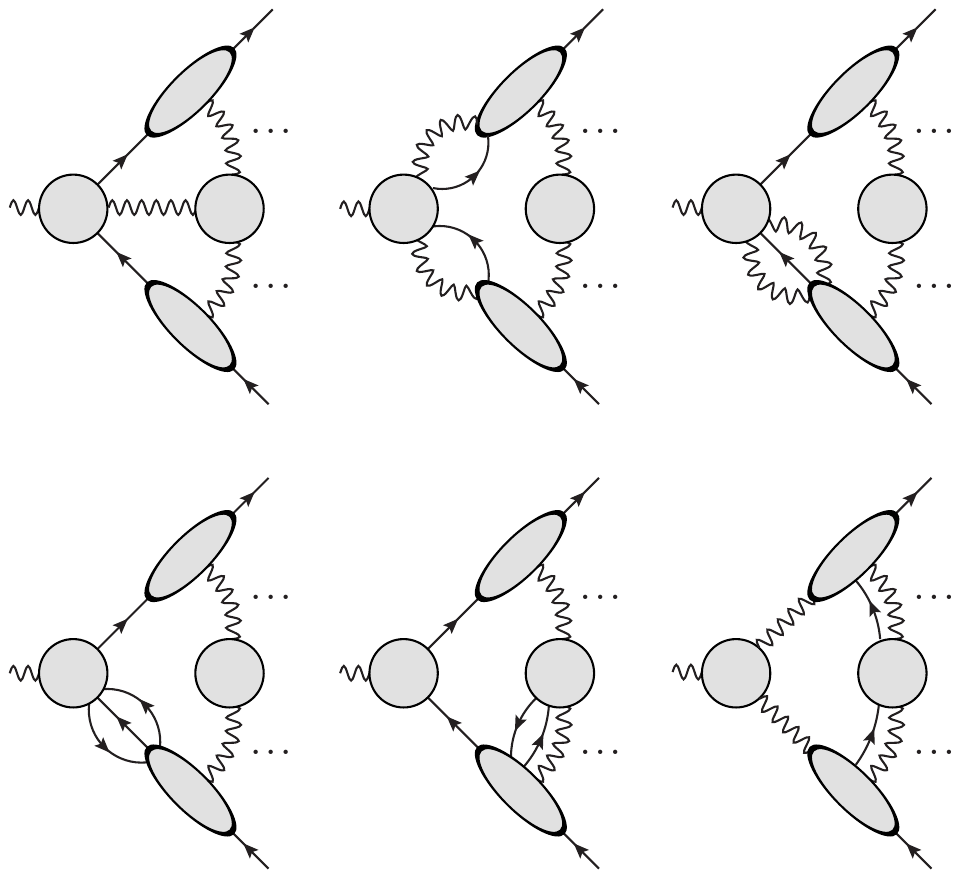}}
    \label{fig:red_diag_HJ_fgamma_x2}
    }
    \subfloat[]{
    \makebox[.24\textwidth][c]{\includegraphics[width=.16\textwidth]{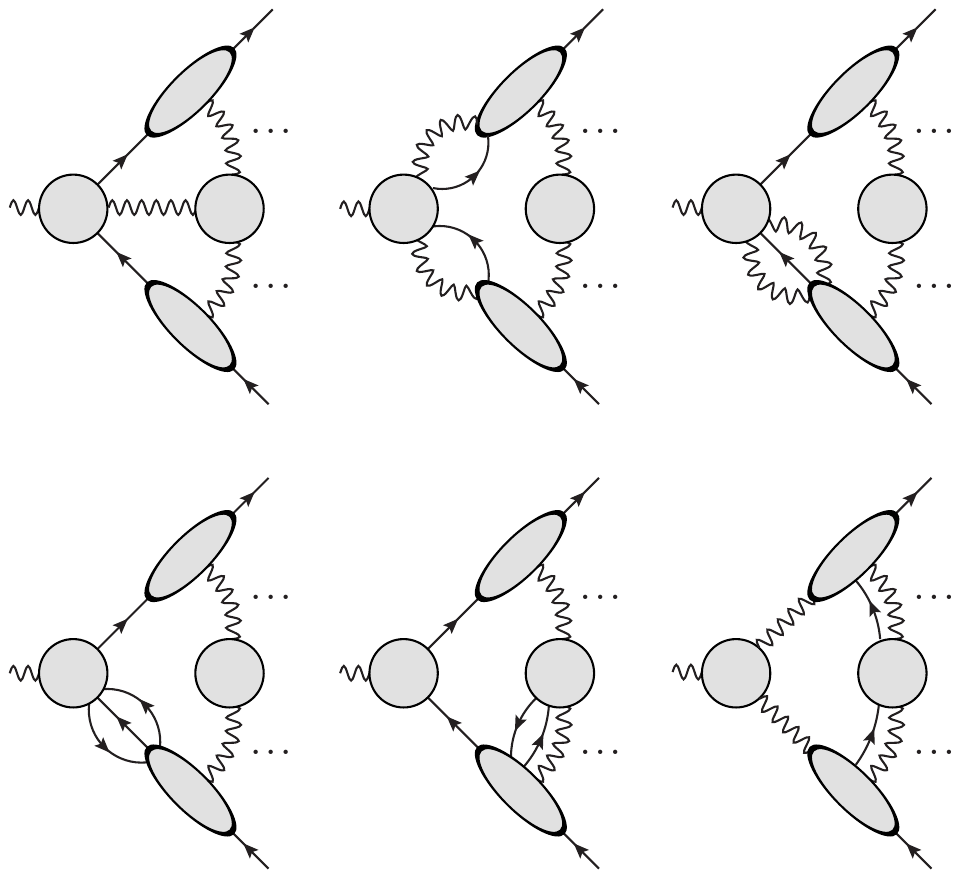}}
    \label{fig:red_diag_HS_gamma}
    }
    \vspace{-5pt}
    \subfloat[]{
    \makebox[.24\textwidth][c]{\includegraphics[width=.16\textwidth]{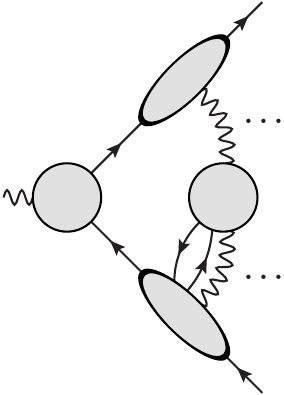}}
    \label{fig:red_diag_JS_ffgamma}
    }  
    \subfloat[]{
    \makebox[.24\textwidth][c]{\includegraphics[width=.16\textwidth]{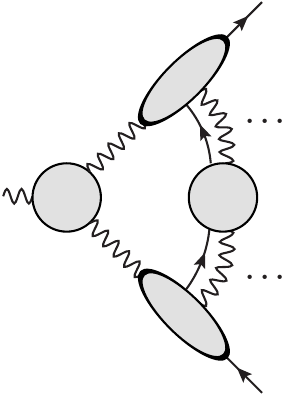}}
    \label{fig:red_diag_JS_fgamma_x2}
    }  
    \caption{Reduced diagrams contributing at $\gamma = 2$ for $m = 0$. For $m \sim \lambda Q$, diagrams (e) and (f) contribute only beyond NLP. (a) Two fermions and a photon connecting a collinear blob to the hard blob. (b) Triple fermion connection instead. (c) Two collinear blobs connected to the hard blob by a fermion and photon each. (d) A single photon connecting the hard and soft blobs. (e) Two fermions connecting a collinear blob to the soft blob. (f) Two collinear blobs connected to the soft blob by a fermion each.}
  \label{fig:red_diag2}
\end{figure}

The leading power configuration at $\gamma_\mathcal{G}=0$ is obtained for $N_f^{(i)}\!=\!1$ and $\{N_\gamma^{(i)}, n_f^{(i)},$ $m_f, m_\gamma, I_f\}\!=\!0$ for all $i$ in \eq{eq:QEDPC_final} and is depicted in \fig{red_diag_g0}. At $\gamma_\mathcal{G}=1$, the only reduced diagrams allowed by charge conservation are those with one additional photon connection between a jet and the hard sub-diagram, $N_\gamma^{(j)}\!=\!1$ and $N_\gamma^{(i)}\!=\!0$ for all $i\neq j$, as shown in \fig{red_diag_g1}. 
Finally, $\gamma_\mathcal{G}=2$ can follow from a variety of configurations, as indicated in \fig{red_diag2}. We can have a double photon connection from the hard sub-diagram to a jet in addition to the fermion line ($N_\gamma^{(j)}\!=\!2$, \fig{red_diag_HJ_fgammagamma}) or a triple collinear (anti-)fermion connection  ($N_f^{(j)}\! =\! 3$, \fig{red_diag_HJ_fff}). Naturally we can have two jets with one extra photon connection as well ($N^{(j)}_\gamma \!=\! N^{(k\neq j)}_\gamma \!=\! 1$, \fig{red_diag_HJ_fgamma_x2}). In addition, there are configurations in which the soft sub-diagram provides the suppression of the degree of divergence. This can be either through a single photon connection to the hard scattering ($m_\gamma \!=\! 1$, \fig{red_diag_HS_gamma}), a double fermion connection to a particular jet ($n_\gamma^{(j)}\! =\! 1$, \fig{red_diag_JS_ffgamma}) or fermion connections to two different jets ($n_\gamma^{(j)} \!=\! n_\gamma^{(k\neq j)} \!=\! 1$, \fig{red_diag_JS_fgamma_x2}). The latter two configurations contribute at NLP only in case $m=0$, while for $m\sim\lambda Q$ \eq{eq:QEDPC_final_massive} yields $\gamma_\mathcal{G}=5$.\footnote{The exception being a single fermion exchanged between the two jets, with no extra soft interactions, which is in fact $\gamma_\mathcal{G}=3$. In \eq{eq:QEDPC_final_massive} one should set $I_f= -1$ in order not to overcount the legs.} In either mass scenario, the soft blob may be connected to the jets by an arbitrary number of photons.
Since the reduced diagrams of \fig{red_diag01} and \fig{red_diag2} encode all relevant soft and collinear configurations up to NLP, we may immediately cast them into entries in the factorization formula. Starting at leading power,  \fig{red_diag_g0} yields the well-known factorization formula
\beq \label{LPfactorization}
    \mathcal{M}^\LP = \bigg(\prod_{i=1}^n J_{(f)}(\hat{p}_i)\bigg) \otimes H(\hat{p}_1,\dots,\hat{p}_n)\, S(n_i\! \cdot\! n_j)\,,
\eeq
where the tensor product $\otimes$ denotes a contraction of spinor indices. The hatted vectors contain the dominant momentum component only
\beq \label{eq:hatNotation}
    \hat{p}_i^{\h\mu} = p_i^+ \, n_i^\mu\,,
\eeq
where the light-cone vector $n_i^\mu$ is defined 
in \eq{def:ni}. The jet function has the operator definition
\beq
    J_{(f)}(p_i) = \langle p_i |\overline{\psi}(0)\Phi_{\bar{n}_i}(0,\infty)|0\rangle\,,
\eeq
involving a semi-infinite Wilson line in the direction $\bar{n}_i$
\beq
    \Phi_{\bar{n}_i}(0,\infty) = \mathcal{P} \exp \bigg[
    -i \,q_i\, e\int_0^\infty \!ds\, \bar{n}_i\!\cdot\! A(s\,\bar{n}_i)\bigg]\,,
\eeq
while the soft function $S$ is given by a product of Wilson lines,
\beq \label{softLP}
    S(n_i \cdot n_j) = \langle 0| \prod_{i=1}^n \Phi_{n_i}(0,\infty) | 0\rangle\,.
\eeq
For simplicity, we assume that the potential overlap between the soft and collinear regions has already been accounted for in a redefinition of the jet functions. 

Following the reasoning of~\cite{Gervais:2017yxv} for Yukawa theory, we assume that a similar factorization picture holds at next-to-leading power, with each class of reduced diagrams described by a different jet function. As far as the hard-collinear sector is concerned, this means that the leading power formula in \eq{LPfactorization} is supplemented with four types of contributions,
\begin{widetext}
\begin{align} \label{NLPfactorization}
    \mathcal{M}_{\mathrm{coll}}^\NLP &= \sum_{i=1}^n \bigg(\prod_{j\neq i}J_{(f)}^j\bigg)
    \Big[J^i_{(f\gamma)} \otimes H^i_{(f\gamma)}+J^i_{(f\partial \gamma)} \otimes H^i_{(f\partial \gamma)}\Big]\,S \nonumber
    \;+\; \sum_{i=1}^n \bigg(\prod_{j\neq i}J_{(f)}^j\bigg)
    J^i_{(f\gamma\gamma)} \otimes H^i_{(f\gamma\gamma)}\, S\\
    &\quad + \sum_{i=1}^n \bigg(\prod_{j\neq i}J_{(f)}^j\bigg)
    J^i_{(f\!f\!f)} \otimes H^i_{(f\!f\!f)}\, S \;+\!\! \sum_{1\leq i \leq j \leq n} \bigg(\prod_{k\neq i,j}J_{(f)}^k \bigg)
    J^i_{(f\gamma)}J^j_{(f\gamma)} \otimes H^{ij}_{(f\gamma)(f\gamma)}\, S\,.
\end{align}
\end{widetext}
To improve readability, we suppress the arguments of the factorization ingredients and introduce the indices $i,j$, labeling the collinear sectors. We will clarify this notation further momentarily. The first term describes the effect of \fig{red_diag_g1} and starts contributing at order $\lambda$. This implies that at order $\lambda^2$ we may expect a dependence of the hard function on the perpendicular momentum component of the collinear photon emerging from it, which can be re-expressed in terms of the $H_{(f\partial \gamma)}^i$ function, as will be shown shortly. The second and third terms describe the classes of diagrams $(\rm a)$ and $(\rm b)$ in \fig{red_diag2}, while diagram $(\rm c)$ corresponds to the last term. These contributions, as well as the $f\partial\gamma$-term, are strictly $\ord(\lambda^2)$, which implies that the soft function appearing in those terms is given by the leading-power definition of \eq{softLP}. While for massless fermions the same reasoning applies to the $f\gamma$-term, in the massive case the soft function could in principle receive $\ord(\lambda)$ corrections. Since we focus on hard-collinear factorization, we do not explore this possibility in detail. For the same reason, we will not supplement our factorization formula with terms corresponding to reduced diagrams $(\rm d)-(\rm f)$ with additional connections to the soft function. We leave the identification and investigation of the corresponding terms for future work.
\Eq{NLPfactorization} is formally identical to the counterpart for massive Yukawa theory \cite{Gervais:2017yxv}, as the collinear sectors of the two theories exhibit the same scaling modulo the replacement of scalars with photons. 

We now clarify the shorthand index notation. In the simplest non-trivial example of the $f\gamma$-jet and hard functions, we define
\begin{align} \label{eq:explicitNotation}
    J^i_{(f\gamma)}&=J_{(f\gamma)}(p_i-\hat{\ell_i},\hat{\ell_i};\eps)\,,\nonumber\\
    H^i_{(f\gamma)}&=H_{(f\gamma)}(p_1 \dots;p_i-\hat{\ell_i},\hat{\ell_i};\dots p_n;\eps)\,.
\end{align}
The last argument indicates that the factorization in \eq{NLPfactorization} is formulated for unrenormalized amplitudes, which depend on a regulator: the factorization ingredients in four dimensions are affected by UV divergences; working in $D=4-2\eps$ dimensions, these divergences take the form of poles in $\eps$. The first two arguments of the jet function denote the momentum flowing through the fermion and photon leg, respectively, while in the hard function the index $i$ also specifies which of the $n$ hard momenta has been shifted in presence of the additional collinear emission. In analogy with \eq{eq:hatNotation}, $\hat{\ell}_i^{\h\mu}=\ell^+_i n_i^\mu$ denotes the large component of the momentum flowing in the photon leg. In principle, in the spirit of the LP factorization, one would like to replace $p_i$ with $\hat{p}_i$ in the argument of the hard function, thus neglecting the small components in the external momenta too. This can be done in the massless theory, where the jet functions start contributing at $\ord(\lambda^2)$. However, the massive theory allows for odd powers in the $\lambda$ expansion, so that an overall $\ord(\lambda^2)$ term can also originate from an order $\lambda$ correction from both the hard and a jet function. This effect forces us to retain some subleading components in the argument of \eq{eq:explicitNotation}, as will be made clear in the explicit calculation in \sect{sec:oneLoopTest}.
In contrast to eq.~\eqref{LPfactorization}, the $\otimes$-product in \eq{NLPfactorization} involves, besides spinor index contractions, convolutions over the leading momentum components and additional Lorentz contractions over spacetime indices carried by the photon leg. Explicitly, for the first term in \eq{NLPfactorization}
\begin{widetext}
\begin{align} \label{eq:explicitFirstLine}
    &\bigg(\prod_{j\neq i}J_{(f)}^j\bigg)
    \Big[J^i_{(f\gamma)} \otimes H^i_{(f\gamma)}+J^i_{(f\partial \gamma)} \otimes H^i_{(f\partial \gamma)}\Big]\,S 
    \;\equiv\; 
    S(\hat{p}_i\cdot\hat{p}_j;\eps)\,
    \bigg(\prod_{j\neq i}J_{(f)}(p_{j};\eps)\bigg) \int_0^{p_i^+}d \ell^+_i \,\\ \nn
    &
    \times\bigg[J^\nu_{(f\gamma)}(p_i-\hat{\ell_i},\hat{\ell_i};\eps)\, H_{(f\gamma)\nu}(p_1 \dots;p_i-\hat{\ell_i},\hat{\ell_i};\dots p_n;\eps)\,
    +\,J^{\nu\rho}_{(f\partial\gamma)}(p_i-\hat{\ell_i},\hat{\ell_i};\eps)\, H_{(f\partial\gamma)\nu\rho}(p_1 \dots;p_i-\hat{\ell_i},\hat{\ell_i};\dots p_n;\eps)\, \bigg]\,.
\end{align}
\end{widetext}
The other terms in \eq{NLPfactorization} involve a straightforward generalisation of the notation in \eq{eq:explicitNotation}. In presence of more than two legs (as for $f\gamma\gamma$), the corresponding hard function acquires an additional argument, and the $p_i$ are shifted accordingly.

As is clear from \eq{eq:explicitFirstLine}, the hard functions depend only on the large momentum component $\hat{\ell}$ (and not on the full $\ell$). However, since we want the NLP formula to be accurate at $\ord(\lambda^2)$, we cannot set $\ell^\mu = \hat{\ell}^\mu$ at the level of amplitudes, but we need to keep also its transverse component $\ell_\perp^\mu$. This can be rephrased as a Taylor expansion in the transverse momentum around zero,
\begin{align} \label{HardExpansion}
  &\widetilde{H}^i_{(f\gamma)\h\nu} \big(p_1\dots p_n;\ell_i;\eps\big) =
  \widetilde{H}^i_{(f\gamma)\nu}\big(p_1\dots p_n;\hat{\ell}_i;\eps\big) \nonumber \\&\hspace{37pt}+
  \ell_\perp^\rho\Big[ \frac{\partial}{\partial\ell_\perp^\rho} \widetilde{H}^i_{(f\gamma)\nu}\big(p_1\dots p_n;\ell_i;\eps\big)\Big]_{\ell_\perp=0}+\ord{(\lambda^2)}\nonumber\\
  & \equiv\, H_{(f\gamma)\nu}(p_1 \dots;p_i-\hat{\ell_i},\hat{\ell_i};\dots p_n;\eps)\nonumber \\ &\hspace{37pt}+
  \ell_\perp^\rho H_{(f\partial\gamma)\nu\rho}(p_1 \dots;p_i-\hat{\ell_i},\hat{\ell_i};\dots p_n;\eps)
  \,,
\end{align}
thus identifying the two terms with respectively the $f\gamma$- and $f\partial\gamma$-contributions in \eq{NLPfactorization}, where by definition the $\ell_\perp^\rho$ in the second term is absorbed in $J_{(f\partial\gamma)}^i$. In \eq{HardExpansion}, we generically denoted with $\widetilde{H}$ the part of the amplitude that is not explicitly described by the soft and collinear functions. In the traditional factorization approach, this would be obtained via a subtraction algorithm, while in the effective field theory it corresponds to a Wilson coefficient obtained from matching to full QED. Both approaches would require matrix element definitions of the jet functions in \eq{NLPfactorization}, as well as of the NLP soft function. Gauge invariance of each separate ingredient would then be manifest. This systematic analysis requires further investigation of the interplay between jet functions and (generalised) Wilson lines, which we leave for future work. In absence of an operator definition, we will extract in section~\ref{massiveFgamma} the $f\gamma$- and $f\partial\gamma$-jet functions from a generic matrix element, assuming the validity of the picture above. This necessitates in turn a diagrammatic definition of the hard function, of which we will give explicit examples in section~\ref{sec:MassiveCase} and \ref{sec:rad_massless}. As a consistency check on this setup, we will show that \eq{NLPfactorization} with these functions reproduces the (hard-)collinear region of one-loop (two-loop) diagrams. 

In passing we note that approaches based on SCET \cite{Larkoski:2014bxa,Beneke:2019oqx} contain similar functions to describe amplitudes at the next-to-leading power. Consisting of power-suppressed operators, these functions also account for dynamical configurations in which multiple particles emerging from the hard interaction belong to the same collinear sector. The $J_{f \gamma}$, $J_{f\partial\gamma}$, and $J_{f\!f\!f}$ jets introduced in \eq{NLPfactorization} are thus related to matrix elements of the operators $J^{B1}$, $J^{B2}$, $J^{C1}$ in the position-space formulation of SCET in \cite{Beneke:2019oqx}, and to matrix elements of the part of $N$-jet operators ${\cal O}_{N}^{(1,X)}$, ${\cal O}_{N}^{(2,X\delta)}$, ${\cal O}_{N}^{(2,X^2)}$ corresponding to a specific collinear direction in the label formulation of SCET \cite{Larkoski:2014bxa}. In this work we are primarily interested to test the consistency of the factorisation formula \eq{NLPfactorization} within the current approach, and we leave a detailed comparison of the jets introduced here with the matrix elements of SCET operators to future work.

The present diagrammatic approach provides important insight into the NLP behaviour of gauge theories. Firstly, we can explicitly test the categorisation of factorization ingredients given by the power-counting in \eq{eq:QEDPC}. Secondly, we shed light on some dynamical subtleties that are not fully accounted for by simpler factorization theorems as \cite{DelDuca:1990gz,Bonocore:2015esa}. Finally, the explicit calculations presented here show how to deal with the endpoint contributions that result from a factorization structure consisting of convolutions rather than direct products.

\section{Hard-collinear factorization for massive fermions} \label{sec:MassiveCase}

We now turn to the study of some of the ingredients entering the factorization picture, in the regime where the fermion mass is parametrically small.
For definiteness we focus on the $f\gamma$- and $f\partial \gamma$-contributions to our NLP factorization formula, corresponding to $N_\gamma = 1$ and $N_f = 1$ in \eq{eq:QEDPC}. We calculate these jet functions at one-loop order in section \ref{massiveFgamma} and validate them as well as the factorization structure through one- and two-loop calculations in section \ref{sec:massive_MOR_tests}. The $f\gamma$-term is particularly relevant, according to our power counting formula, it already contributes at $\ord(\lambda)$, for a parametrically small fermion mass. However, the $f\partial\gamma$-term is further suppressed by a power of the transverse momentum component. Thus, to appreciate the interplay between the two functions, we will need to carry out the calculation to $\ord(\lambda^2)$. This level of accuracy, and the fact that we consider QED, constitutes an important generalisation of the analogous functions presented in~\cite{Gervais:2017yxv} where Yukawa theory was studied. We stress that at $\ord(\lambda^2)$ accuracy one also needs the $f\!f\!f$- and $f\gamma\gamma$-jet functions, which contribute from two-loop onwards. We leave the calculation of these ingredients for future work. At this order other interesting aspects such as endpoint contributions come into play.

Although in this section we do not consider external soft radiation, our analysis of the non-radiative factorization ingredients is an important step towards generalising soft theorems to gauge theories at NLP, in the case of parametrically small masses.
In addition, this scenario carries intrinsic interest to collider phenomenology, since precise measurements of cross sections may benefit from classifying and possibly resumming logarithms of small fermion masses at NLP. Examples are charm mass effects in B decays~\cite{Bagan:1994zd}, initial-state mass effects in heavy-quark induced processes~\cite{Aivazis:1993pi,Thorne:1997ga,Forte:2010ta}, bottom mass effects in Higgs production and decay \cite{Liu:2017vkm,Liu:2019oav}, and $t\bar{t}$ production at a future linear collider, where the top mass could serve as a soft scale. Understanding the NLP factorization structure is a necessary intermediate step towards resumming such mass effects at this level of accuracy.

\subsection{The massive
\texorpdfstring{$f\gamma$}{f gamma}-jet}
\label{massiveFgamma}

In the following we carry out an explicit derivation of the one-loop expressions for two of the jet functions that enter the NLP factorization formula for massive QED, as presented in \eq{NLPfactorization}. The detailed calculation of these quantities sheds light on the subtleties involved beyond leading power. Moreover the functions we extract are process-independent, and could therefore be used in other QED calculations.
\begin{figure}[t]
\begin{center}\vspace*{10pt}
\includegraphics[width = .18\textwidth]{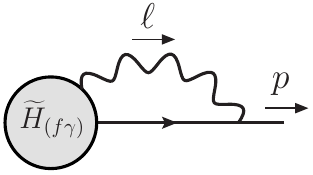}
\caption{Diagram from which the $f\gamma$- and $f\partial\gamma$-jet functions are extracted.}
\label{fig:QEDjet}
\end{center}
\end{figure}
To keep expressions compact, we choose a reference frame such that the momentum $p^\mu$ of the external particle that defines the jet has no perpendicular component $p^\mu = (p^+,\,0,\, p^-)$, with $p^-\ll p^+$. Unit vectors in the collinear and anti-collinear direction are then denoted by $n^\mu = (1,0,0)$ and $\bar{n}^\mu = (0,0,1)$. From now on, we also set the electric charge $q=-1$.
The power counting in \eq{eq:QEDPC} was derived in axial gauge, which we also use for the jet functions.
We will verify that, in a gauge which only allows for physical polarisations, the predicted power counting works on a diagram-by-diagram basis. For simplicity we select light-cone gauge, setting $r^2 = 0$ in \eq{Axprop}. Furthermore, we choose the reference vector in the anti-collinear direction, $ r^\mu = r^- \bar{n}^\mu.$ Note that $r^-$then cancels in the photon propagator in \eq{Axprop}, leaving
\begin{align} \label{eq:reducedPhotonPropagator}
    \Delta_{\nu\sigma}(\ell) = \frac{i}{\ell^2+i\eta}\bigg(-\eta_{\nu\sigma}+\frac{\ell_\nu\bar{n}_\sigma+\ell_\sigma \bar{n}_\nu}{\ell\!\cdot\!\bar{n}}\bigg)\,.
\end{align}
We extract the jet functions from the diagram in \fig{QEDjet}. Working in $D=4-2\eps$ dimensions, the corresponding amplitude is given by
\begin{align} \label{fgammaAmplitude}
  \mathcal{M}_{(f\gamma)}^{(n+1)}(p) &=
  \PSell\frac{i e \mu^{2\eps}\,\bar{u}(p)\,\mathcal{N}^\nu(p,\ell)}{\big[ (p-\ell)^2-m^2+i\eta\big]\,\big[\ell^2+i\eta\big]}\, \nn\\
  &\quad\times\widetilde{H}_{(f\gamma)\nu}^{(n)}\bigg(\frac{\ell^+}{p^+},\ell_\perp\bigg)\, ,
\end{align}
with the numerator factor
\beq
  \mathcal{N}^\nu(p,\ell)=
  \bigg(-\gamma^\nu+\frac{\slashed{\ell}r^\nu+\slashed{r}\ell^\nu}{\ell\!\cdot\! r}\bigg)(\slashed{p}-\slashed{\ell}+m)
\eeq
and a generic $n$-loop hard function $\widetilde{H}^{(n)}_{(f\gamma)\nu}$. We first rearrange its transverse-momentum dependence by Taylor expanding in $\ell_\perp^\rho$, as in \eq{HardExpansion}.\footnote{We recall that throughout this paper $\ell_\perp^\rho$ is the $D$-dimensional perpendicular component of $\ell^\rho$, as defined by means of a Sudakov decomposition: $\ell^\rho = \ell\cdot \bar{n}\, n^\rho + \ell\cdot n \,\bar{n}^\rho + \ell_\perp^\rho$. Similarly we define $\eta_{\perp}^{\nu\rho} = \eta^{\nu\rho}-n^\nu\bar{n}^\rho-\bar{n}^\nu n^\rho$.} Retaining terms up to $\ord(\lambda)$
\begin{align}
  &\widetilde{H}^{(n)}_{(f\gamma)\nu} \bigg(\frac{\ell^+}{p^+},\ell_\perp\bigg) \;=\;
  \widetilde{H}^{(n)}_{(f\gamma)\nu}\bigg(\frac{\ell^+}{p^+},0\bigg) \nonumber \\ &\hspace{31pt}+
  \ell^\rho_\perp \left[\frac{\partial}{\partial\ell_\perp^\rho} \widetilde{H}^{(n)}_{(f\gamma)\nu}\bigg(\frac{\ell^+}{p^+},\ell_\perp\bigg)\right]_{\ell_\perp=0}+\ord(\lambda^2)\nonumber \\ &\equiv\;
  H^{(n)}_{(f\gamma)\nu}(x) + \ell^\rho_\perp  H^{(n)}_{(f\partial\gamma)\nu\rho}(x)\,, \label{Hfgammadef}
\end{align}
we trade the initial hard function for two objects that depend only on the fraction of the large component of the loop momentum $x = \ell^+/p^+$. Here and in the following we shall suppress the $\eps$ dependence of the hard and jet functions for brevity. We remind the reader that we deal with unrenormalized quantities throughout this paper. Comparing \eq{fgammaAmplitude} with the first line of \eq{NLPfactorization},
\begin{align} \label{factorization}
  \mathcal{M}^{(n+1)}_{(f\gamma)}(p) =&\, \int_0^1\!\! d x\, \Big[ J_{(f\gamma)}^{(1)\h\nu}(x)\, H^{(n)}_{(f\gamma)\nu}(x) \nonumber \\ &\hspace{32pt}+ J_{(f\partial\gamma)}^{(1)\h\nu\rho} (x)  H^{(n)}_{(f \partial \gamma)\nu\rho}(x) \Big]\,,
\end{align}
allows us to extract the jet functions,
\begin{subequations}\label{fgammaJetDefinitions}\begin{align}
J_{(f\gamma)}^{(1)\h\nu}(x,p) &= \PSellRed
\frac{ i e p^+ \mu^{2\eps}
\bar{u}(p)\,\mathcal{N}^\nu(p,\ell)}{\big[ \ell^2-2\ell\!\cdot\!p+i\eta\big]\big[\ell^2+i\eta\big]}\,,
 \\
J_{(f\partial\gamma)}^{(1)\h\nu\rho}(x,p) &=  \PSellRed
\frac{i e p^+ \mu^{2\eps}\,\bar{u}(p)\,\mathcal{N}^\nu(p,\ell)\,\ell_\perp^\rho}{\big[ \ell^2-2\ell\!\cdot\!p+i\eta\big]\big[\ell^2+i\eta\big]}\,.
\end{align}\end{subequations}
In \eq{factorization}, we switched from the dominant loop momentum component $\ell^+$ to the momentum fraction $x$, which determines the convolution between hard and jet functions. The $x$-integration range is a priori $(-\infty, +\infty)$, but is in fact restricted to $(0,1)$ by noting that the integral over $\ell^-$ vanishes if the two poles lie on the same side of the integration contour. 

In \eq{fgammaJetDefinitions} the denominators have homogeneous $\lambda$-scaling, but the numerator still needs expanding. As expected from the power counting rule for an all-collinear vertex, we find, in axial gauge, that $\mathcal{N}$ is $\ord(\lambda)$; therefore, the $f\gamma$-jet starts at the same order, while the additional term $\ell_\perp$ causes the $f\partial\gamma$-jet to begin at $\ord(\lambda^2)$. Performing the expansion leaves us with three independent numerator structures,
\beq \label{eq:numStructures}
    1\,,\qquad \ell_\perp^\alpha \ell_\perp^\beta\,,\qquad \ell^-\,.
\eeq
The first two lead to straightforward integrals, and follow from closing the integration contour at infinity in the $\ell^-$ complex plane, evaluating the residue of the integrand at the pole $\ell^- = -\ell_\perp^2/(2\, x\, p^+)-i\eta$, and solving in turn the resulting integral over transverse momentum. The third one is more subtle, since the integrand does not vanish fast enough at the boundary to apply Jordan's lemma. Instead, we can isolate the troublesome term, introduce a Schwinger parameter, and integrate the minus component to a Dirac delta,
\begin{align} \label{eq:endpoint}
  &\int\! d \ell^-\, \frac{1}{-2p^+\ell^-(1-x)+\ell_\perp^2-xm^2+i\eta}
  \nonumber \\
  &\quad= -i\int_0^\infty\!\!d t \int\! d \ell^-\, e^{i t [-2p^+\ell^-(1-x) +\ell_\perp^2-x m^2]}
  \nonumber \\
  &\quad= -\frac{\pi i}{p^+}\,\delta(1-x)\int_0^\infty\frac{d t}{t} e^{it(\ell_\perp^2-x m^2)}\,.
\end{align}
This endpoint contribution at $x=1$ corresponds to the limit where the photon leg carries all the momentum along the $+$-direction and the fermion line becomes soft. 

Results for the integrals relevant to computing \eq{fgammaJetDefinitions} are collected in \eq{masterIntegrals} in appendix \ref{appx:intermediate}. Having carried them out, we conclude
\begin{widetext}
\begin{align} \label{massiveJetFunctionsResult}
    J_{(f\gamma)}^{(1)\h\nu}(x,p)  =& -\frac{e}{16\pi^2}\left(\!\frac{m^2}{4\pi\mu^2}\!\right)^{-\eps} \Gamma(\eps)\,
  \bar{u}(p)\bigg\{
  m\,x^{1-2\eps}(\slashed{\bar{n}}\, n^\nu-\gamma^\nu)
  \nonumber \\
  &+\frac{m^2}{p_+}\left[\frac{1}{2\,(1\!-\!\eps)}\left(\delta(1\!-\!x)-(1\!-\!2\eps)\,x^{1-2\eps}\right)\gamma^\nu\slashed{\bar{n}}-2\,x^{-2\eps}(1-x)\,\bar{n}^\nu\right]\!\bigg\}\,,
  \nonumber \\
  J_{(f\partial\gamma)}^{(1)\h\nu\rho}(x,p) =& - \frac{e}{16\pi^2}\left(\!\frac{m^2}{4\pi\mu^2}\!\right)^{-\eps} \Gamma(\eps)\,
  \bar{u}(p)\frac{m^2x^{2-2\eps}}{2\,(1\!-\!\eps)}\bigg\{
  \gamma_\perp^\rho(\slashed{\bar{n}}\, n^\nu-\gamma^\nu)+
  \frac{2}{x}\eta_\perp^{\nu\rho}\bigg\}\,.
\end{align}
\end{widetext}
These are the one-loop expressions for the $f\gamma$- and $f\partial\gamma$-jet functions in QED, as derived in light-cone gauge. As expected, the $f\gamma$-jet function starts at order $\lambda\!\sim\! m/Q$, while the $f\partial\gamma$-jet has pure $\lambda^2$ scaling. This is due to the additional factor $\ell_\perp^\rho$ in the expansion of $\widetilde{H}_{(f\gamma)}$ in \eq{Hfgammadef}, which it absorbs in the definition of \eq{factorization}. As a result only the structure $\ell_\perp^\alpha \ell_\perp^\beta$ in \eq{eq:numStructures} survives in the numerator. Since $m$ is the only small scale in these functions, the mass expansion coincides with the power expansion. Once expanded in $\eps$, the first line of \eq{massiveJetFunctionsResult} is the QED equivalent of the $\ord(m)$ result derived for Yukawa theory in \cite{Gervais:2017yxv}, and coincides with the amplitude computed in \cite{Beneke:2019slt} within SCET in the context of heavy quark leptonic decays.

\subsection{Testing NLP factorization with the method of regions} \label{sec:massive_MOR_tests}

Equipped with the result of \eq{massiveJetFunctionsResult}, we will now test the factorization formula \eq{NLPfactorization} in a process with two final-state jet directions, at both one- and two-loop order. Specifically, we wish to see whether this formula reproduces the (hard-)collinear limit of full, unfactorized amplitudes which at face value should be described by the $f\gamma$- and $f\partial\gamma$-jet functions. We will isolate the part of the amplitude that we want to compare with, using the method of regions \cite{Beneke:1997zp,Smirnov:2002pj,Pak:2010pt,Jantzen:2011nz}. This is a well-tested tool for expanding (loop) amplitudes in kinematic limits where the various scales entering the amplitude are largely separated in magnitude. It is particularly useful for the dissection of loop integrals, by defining regions where the virtual modes have momenta of a certain size as compared to a particular scale in the problem. In this case we use the small ratio of scales $\lambda = \frac{m}{Q}$ to select momentum regions where a virtual photon is hard, soft or collinear to either of the highly energetic particles in the final state. Once the regions have been defined, one may expand the integrand for each region in $\lambda$ (up to an arbitrary order), which simplifies its structure. The integration is still carried out over the full momentum space, which allows for easy evaluation, but one must be careful not to overcount contributions that appear in multiple regions. Most of the time this causes no issue, as each region has a specific associated energy scale (in our case, $(m^2/\mu^2)^{-\eps}$ for a collinear region and $(2p_1^+p_2^-/\mu^2)^{-\eps}$ for the hard region), which inhibits any cross-talk between such regions. Finally, by summing over all relevant regions, one obtains the result of the full integral up to the chosen order in $\lambda$.

In presence of just two jets in the final state, we choose a frame in which the jets are back to back. Given the light-cone decomposition of $p_1^\mu$, we identify the direction collinear to $p_2^\mu$ as the \emph{anti-}collinear direction. This yields the following regions
\begin{align}
   {\rm Hard:} \qquad &k^\mu\sim Q \left( 1, 1, 1 \right) \, \nn\\
  {\rm Soft:} \qquad &k^\mu\sim Q \left( \lambda^2, \lambda^2, \lambda^2 \right) \, ,
  \nn  \\
  {\rm Collinear:} \qquad &k^\mu\sim Q \left( 1, \lambda, \lambda^2 \right) \, , \quad \nn \\
\;  \text{Anti-collinear}: \qquad &k^\mu\sim Q \left( \lambda^2, \lambda, 1 \right) \, .
\label{regions}
\end{align} 
In the following sections we refrain from considering all regions, but use this tool to extract only the contribution from the collinear (hard-collinear) region of the full amplitude, which is relevant for our one-loop (two-loop) test.

\subsubsection{One-loop test} \label{sec:oneLoopTest}
\begin{figure}[b]
  \centering
  \includegraphics[width=0.14\textwidth]{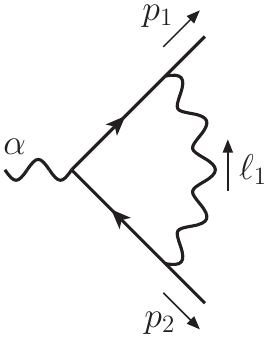}%
  \caption{One-loop diagram used for verification of the jet function results.}
  \label{fig:1v}
\end{figure}
For one-loop accuracy we calculate the collinear region of \fig{1v}, which should be described by contracting the one-loop $f\gamma$- and $f\partial\gamma$-jet functions with corresponding tree-level hard functions. We will carry out the regions calculation in axial gauge, as we did for the $f\gamma$- and $f\partial\gamma$-jet functions.
 
The full amplitude for the diagram in \fig{1v} reads
\begin{widetext}
\begin{align} \label{eq:oneLoopStartingPoint}
   \hspace{-4pt} \mathcal{M}^{(1)\h\alpha}(p_1,p_2) &= \int \!\frac{d^{4-2\eps}\ell_1}{(2\pi)^{4-2\eps}}\, \frac{e^3\, \mu^{2\eps}\,\bar{u}(p_1)\, \mathcal{N}^{(1)\h\alpha}(p_i,\ell_1)\, v(p_2) }{\big[\ell_1^2+i\eta\big]
    \big[(p_1-\ell_1)^2-m^2+i\eta\big]
    \big[(p_2^2+\ell_1)^2-m^2+i\eta\big]} \,,\\
   \hspace{-4pt} \mathcal{N}^{(1)\h\alpha}(p_i,\ell_1) &=
    \gamma^\sigma(\slashed{p}_1-\slashed{\ell}_1+m)\gamma^\alpha(-\slashed{p}_2-\slashed{\ell}_1+m)
    \left(-\gamma_\sigma +\frac{\ell_\sigma\, \slashed{\bar{n}} + \slashed{\ell}\, \bar{n}_\sigma}{\ell\!\cdot\!\bar{n}} \right)\nonumber\,.
\end{align}
\end{widetext}
The expansion of the integrand in the collinear region is obtained by rescaling the momentum components of both the (collinear) loop momentum and the (anti-)collinear external momenta, according to \eq{regions}. We further exploit our freedom of frame choice to set the perpendicular momentum components of the external momenta to zero. In practice, it is convenient to project onto a single set of light-like vectors in the plus- and minus-direction, using
\begin{align}
p_1^\mu &\,=\, \underbrace{\strut p_1\!\cdot\!\bar{n}}_{\sim\lambda^0}\,n^\mu + \underbrace{\strut p_1\!\cdot\!n\,}_{\sim\lambda^2}\bar{n}^\mu \nn\\
p_2^\mu &\,=\,  \underbrace{\strut p_2\!\cdot\!\bar{n}}_{\sim\lambda^2}\,n^\mu +\underbrace{\strut p_2\!\cdot\!n}_{\sim\lambda^0}\,\bar{n}^\mu.    \label{pdecomp}
\end{align}
The denominator in \eq{eq:oneLoopStartingPoint} is expanded as
\begin{align} \label{eq:oneLoopDenomExpansion}
    &\frac{1}{\ell_1^2+i\eta}\frac{1}{(\ell_1-p_1)^2-m^2+i\eta}\frac{1}{(\ell_1+p_2)^2-m^2+i\eta} \nn\\ &= \frac{1}{\ell_1^2+i\eta}\frac{1}{(\ell_1^2-2\,\ell_{1}\!\cdot\! \bar{n}\, p_{1}\!\cdot\! n-2\,\ell_{1}\!\cdot\!n\,p_{1}\!\cdot\!\bar{n})+i\eta}
    \\
    &\quad\times\frac{1}{2\,\ell_{1}\!\cdot\!\bar{n}\,p_{2}\!\cdot\!n+i\eta}\left[1\!-\!\frac{\,\ell^2}{2\,\ell_1\!\cdot\!\bar{n}\,p_{2}\!\cdot\!n+i\eta}+\mathcal{O}(\lambda^4)\right]\!,\nonumber
\end{align}
where all propagator denominators have now a homogenous $\lambda$-scaling. The numerator in \eq{eq:oneLoopStartingPoint} is suppressed by one power of $\lambda$, allowing us to drop every term but the leading one from the denominator expansion in \eq{eq:oneLoopDenomExpansion}, including the explicitly shown $\mathcal{O}(\lambda^2)$ term.
By discarding higher power corrections in the numerator too, we readily calculate the collinear region up to NLP from this expression using standard techniques: we perform the Dirac algebra using the \texttt{Mathematica} package \texttt{FeynCalc} \cite{Mertig:1990an,Shtabovenko:2016sxi}, Feynman parametrise the homogeneous denominators, shift the loop momentum and remove odd integrands, evaluate the momentum integrals through standard tensor integrals and finally integrate the Feynman parameters in a convenient order. We find
\begin{multline} \label{eq:oneLoopMassiveFinal}
  \mathcal{M}^{(1)\h\alpha}_{\rm C}(p_1,p_2) = 
  \frac{i\, e^3}{16\,\pi^2}\frac{m}{p_2 \!\cdot\! n}\left(\!\frac{m^2}{4\pi\mu^2}\!\right)^{-\eps}\frac{\Gamma(1+\eps)}{\eps}\\
  \quad\times\bar{u}(p_1)
  \bigg[n^\alpha -\frac{m}{2\,p_1 \!\cdot\! \bar{n}}\frac{1-2\eps+4\eps^2}{(1-\eps)(1-2\eps)}\gamma^\alpha \bigg]v(p_2)\,.
\end{multline}
The subscript $\rm C$ indicates that this result is expanded in the collinear region. As expected, in axial gauge the diagram obeys the power counting, strictly contributing only at NLP. This expression has a single pole in $\eps$, which receives both UV and IR contributions. The UV term regulates divergences that would be subtracted by one-loop renormalisation; the remainder has a collinear (rather than soft) origin. 

The vector nature of the electromagnetic current and the Sudakov decomposition we employ in \eq{pdecomp} limit the possible Dirac structures that can appear in the result to $\gamma^\alpha, n^\alpha,$ and $\bar{n}^\alpha$. In particular, we observe that $\gamma^\alpha$ occurs only in even powers of the mass expansion, while $n^\alpha$ and $\bar{n}^\alpha$ multiply odd powers of the mass. In this specific case, the structure $\bar{n}^\alpha$ is absent due to a cancellation which, as our two-loop check will make clear, is accidental.
\begin{figure}[t]
  \vspace{10pt}
  \centering
  \includegraphics[width=0.14\textwidth]{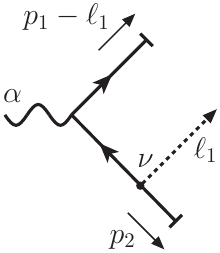}%
  \caption{Diagrammatic interpretation of the leading order hard function $\widetilde{H}_{(f\gamma)}^{(0)}$, from which $H_{(f\gamma)}^{(0)}$ and $H_{(f\partial\gamma)}^{(0)}$ are derived. The dashed line indicates where the collinear momentum $\ell$ is extracted. External lines are amputated.}
  \label{fig:Hfgamma_LO}
\end{figure}
Turning to the factorization approach, we note that the collinear photon in \fig{1v} is emitted from an anti-collinear external line, such that the propagator before the emission carries a hard momentum. Consequently, this diagram should be described by the convolution of the one-loop $f\gamma$- and $f\partial\gamma$-jets with the respective (tree-level) hard functions, as claimed above. We expect the following factorization structure up to $\mathcal{O}(\lambda^2)$
\begin{multline}
    \mathcal{M}^{(1)\h\alpha}_{\rm fact.}(p_1,p_2) =\  \int_0^1\!dx\, \Big[J^{(1)\h\nu}_{(f\gamma)}(x,p_1)\,H^{(0)\h\alpha}_{(f\gamma)\h\nu}(x,p_1,p_2) \\ +J^{(1)\h\nu\rho}_{(f\partial\gamma)}(x,p_1)\,H^{(0)\h\alpha}_{(f\partial\gamma)\h\nu\rho}(x,p_1,p_2)\Big]J^{(0)}_{(f)}(p_2)\,. \label{1v_fact}
\end{multline}
The hard functions are extracted from \fig{Hfgamma_LO}, and their $f\gamma$- and $f\partial\gamma$- parts separated according to the prescription of \eq{Hfgammadef}, yielding
\begin{align}
    H^{(0)\h\alpha}_{(f\gamma)\h\nu}(x,p_1,p_2) &= \label{eq:oneLoopFgamma} \frac{-i\,e^2\,}{2x\,p_1\!\cdot\!\bar{n}\,p_2\!\cdot\!n}\gamma^\alpha
    \\\nn &\quad\times\left(x\,p_1\!\cdot\!\bar{n}\,\slashed{n}+p_2\!\cdot\!n\,\slashed{\bar{n}}-m\right)\gamma_\nu\,, \\
    H^{(0)\h\alpha}_{(f\partial\gamma)\h\nu\rho}(x,p_1,p_2) &= \label{eq:oneLoopFdgamma} \frac{+i\,e^2\,}{2x\,p_1\!\cdot\!\bar{n}\,p_2\!\cdot\!n}\gamma^\alpha \gamma_{\perp\rho}\gamma_\nu\,.
\end{align}
We emphasise that as we are interested in the first two orders in $\lambda$, we cannot ignore $\ord(\lambda)$ terms in the numerator of \eq{eq:oneLoopFgamma}, as they will combine with the leading term in the $f\gamma$-jet (\eq{massiveJetFunctionsResult}). In particular, we cannot drop the mass term. However, we can do so in \eq{eq:oneLoopFdgamma}, since the $f\partial\gamma$-jet is proportional to two powers of the mass (thus $\mathcal{O}(\lambda^2)$). After some Dirac algebra, we find
\begin{widetext}
\begin{align} \label{eq:fGamma1loop}
    J^{(1)\h\nu}_{(f\gamma)}(x,p_1)\,
    H^{(0)\h\alpha}_{(f\gamma)\h\nu}(x,p_1,p_2)\,
    J^{(0)}_{(f)}(p_2) =&\, \frac{i\, e^3}{8\,\pi^2}\frac{m}{p_2 \!\cdot\! n}\left(\frac{m^2}{4\pi\mu^2}\right)^{-\eps}\Gamma(1+\eps)\,
  \bar{u}(p_1) \nn \\ &\times
  \left[\left(\frac{1}{\eps}-1\right)x^{1-2\eps}\,n^\alpha +\frac{m}{2\,p_1 \!\cdot\! \bar{n}}\gamma^\alpha\frac{1}{1-\eps}\left(\delta(1-x)+ x^{1-2\eps}\eps\right) \right]v(p_2)\,,\\
    J^{(1)\h\nu\rho}_{(f\partial\gamma)}(x,p_1)\,
    H^{(0)\h\alpha}_{(f\partial\gamma)\h\nu\rho}(x,p_1,p_2)\,
    J^{(0)}_{(f)}(p_2) =&\, -\frac{i\, e^3}{32\,\pi^2}\frac{m^2}{p_1 \!\cdot\! \bar{n}\,p_2 \!\cdot\! n}\left(\frac{m^2}{4\pi\mu^2}\right)^{-\eps}\Gamma(1+\eps)\,\bar{u}(p_1)\gamma^\alpha v(p_2) 
  x^{-2\eps}\nn\\&\times
  \bigg[\frac{1}{\eps}-\frac{1}{1-\eps}\left(\frac{1}{\eps}-1-\eps\right)x\bigg]\,.
\end{align}
\end{widetext}
The integral over the energy fraction $x$ is easily performed, yielding indeed the result in \eq{eq:oneLoopMassiveFinal}. This provides a first check of our jet functions. We observe that the singular structure of the collinear region is entirely reproduced by  the jet functions of \eq{eq:fGamma1loop}, while the convolution with the respective hard functions does not generate any additional pole in $\eps$. We stress that the endpoint contribution, described by the Dirac delta function in \eq{eq:fGamma1loop}, is essential to obtain the correct result. 

\subsubsection{Two-loop test} \label{sec:twoLoopMassiveTest}

\begin{figure}[t]
  \centering
  \includegraphics[width=0.13\textwidth]{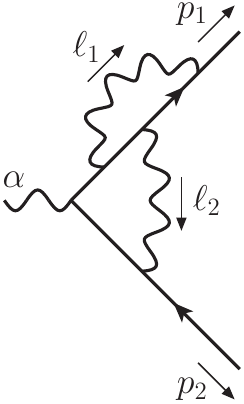}%
\caption{A typical two-loop diagram that receives contributions from the $f\gamma$- and $f\partial\gamma$-jet functions.}
  \label{fig:twoLoopMassive}
\end{figure}

We now proceed with a more strenuous test, based on the same method. The goal is to validate the factorization of a fermion-anti-fermion-production amplitude into $f\gamma$- and $f\partial\gamma$-jet functions if the hard function is loop-induced. A minimal diagram suited for this task is given in \fig{twoLoopMassive}. It consists of an off-shell, one-loop vertex correction, described by the hard loop momentum $\ell_2$, which gets probed by a collinear fermion-photon pair forming the $\ell_1$-loop on the upper leg. We recall that focusing on one particular diagram is justified in axial gauge, where the power counting holds on a diagram-by-diagram basis and the factorization picture is derived.\footnote{For covariant gauge choices, one is forced to sum over a gauge invariant set of diagrams, as we will see in section \ref{sec:rad_one_loop} and \ref{sec:rad_two_loop}. An extensive analysis of relevant momentum configurations is given in section \ref{sec:rad_two_loop}.} Naturally, a complete evaluation of such a process would require us to determine the full hard function, necessitating the calculation of additional diagrams.
We now proceed with the region expansion. The full two-loop amplitude reads
\begin{align} \label{eq:twoLoopMassiveStart}
    &\mathcal{M}^{(2)\alpha}(p_1,p_2) =  i\, e^5 \mu^{4\eps}
    \int\!\frac{d^{4-2\eps}\ell_1}{(2\pi)^{4-2\eps}}
    \int\!\frac{d^{4-2\eps}\ell_2}{(2\pi)^{4-2\eps}}\,
    \nonumber \\ &\hspace{10pt}\times\frac{\bar{u}(p_1)\,\mathcal{N}^{(2)\h\alpha}(p_i,\ell_i)\,v(p_2)}{
    \big[(\ell_1-p_1)^2-m^2\big]
    \big[(\ell_1-\ell_2-p_1)^2-m^2\big]}\nonumber \\
    &\hspace{10pt}\times 
    \frac{1}{\big[(\ell_2+p_1)^2-m^2\big]
    \big[(\ell_2-p_2)^2-m^2\big]
    \big[\ell_1^2\big]\big[\ell_2^2\big]}\,,
\end{align}
where for brevity we omitted the Feynman prescription $i\eta$ in each of the square brackets in the denominators. The numerator structure reads
\begin{align} \label{eq:twoLoopMassiveNumerator}
    &\mathcal{N}^{(2)\h\alpha}(p_i,\ell_i)  = 
    \gamma^\mu(\slashed{p}_1\!-\!\slashed{\ell}_1+m)
    \gamma^\rho  \\&\hspace{4pt}\times(\slashed{p}_1\!-\!\slashed{\ell}_1\!+\!\slashed{\ell}_2+m)
    \gamma^\nu(\slashed{p}_1\!+\!\slashed{\ell}_2+m)
    \gamma^\alpha(-\slashed{p}_2\!+\!\slashed{\ell}_2+m)\gamma^\sigma\nonumber\\
    &\hspace{4pt}\times \left( \eta_{\mu\nu} - \frac{\ell_{1\h\mu}\, \bar{n}_\nu + \ell_{1\h\nu}\, \bar{n}_\mu}{\ell_1\!\cdot\!\bar{n}}\right)\,
    \left( \eta_{\rho\sigma} - \frac{\ell_{2\h\rho}\, \bar{n}_\sigma + \ell_{2\h\sigma}\, \bar{n}_\rho}{\ell_2\!\cdot\!\bar{n}}\right)\,\nonumber.
\end{align}
To carry out the integrals in \eq{eq:twoLoopMassiveStart} we use the same techniques as the one-loop example. The main difference is the presence of two-loop integrals, but due to the regions expansion the added complexity is limited. In the presence of masses and axial-gauge propagators, numerator structures proliferate, which makes the calculation computationally more intensive. However, as in the one-loop case, the numerator~\eqref{eq:twoLoopMassiveNumerator} scales as $\lambda$, which allows us to neglect $\ord(\lambda^2)$ terms from the denominator expansion. In fact, only the denominator in \eq{eq:twoLoopMassiveStart} mixing the two loop momenta generates $\ord(\lambda)$ terms, through the expansion
\begin{multline}
    \frac{1}{(\ell_1-\ell_2-p_1)^2-m^2}\,=\,\frac{1}{\ell_2^2+2\,\ell_2\!\cdot\!(p_1-\ell_1)}\\
    +\frac{2\,\ell_{1\h\perp}\!\cdot\!\ell_{2\h\perp}}{\big[\ell_2^2+2\,\ell_2\!\cdot\!(p_1-\ell_1)\big]^2} + \ord(\lambda^2)\,,
\end{multline}
which is a consequence of our frame choice, $p_{1\perp}\!=\!p_{2\perp}\!=\!0$. For the 1-loop-hard 1-loop-collinear (HC) region we thus obtain
\begin{align} \label{eq:twoLoopMassiveFinal}
    \mathcal{M}^{(2)\h\alpha}_{\rm{HC}}(p_1,p_2) &= 
    \frac{i\,e^5}{128\,\pi^4}\left(\!\frac{-2\,p_1\!\cdot\! \bar{n} \,p_2\!\cdot\!n\,}{4\pi\mu^2}\!\right)^{-\eps}\!\left(\!\frac{m^2}{4\pi\mu^2}\!\right)^{-\eps}
    \nn\\
    &\hspace{-52pt}\bar{u}(p_1)\frac{1}{1-2\eps}\bigg\{
    m\, \Gamma_1
    \left(\frac{1}{\eps^3}+\frac{2}{\eps^2}-\frac{3}{\eps}\right)
    \left(\frac{\bar{n}^\alpha}{p_1\!\cdot\!\bar{n}} - \frac{n^\alpha}{p_2\!\cdot\!n}\right)
     \nonumber \\
    &\hspace{-55pt}+m\,\Gamma_2\bigg[\bigg(\frac{2}{\eps ^3}-\frac{1}{\eps ^2}-\frac{8}{\eps } +11 -4 \eps\bigg)\frac{n^\alpha}{p_2\!\cdot\!n}
    \nonumber \\ &
    \hspace{-55pt}\qquad\quad\ \  -\bigg(\frac{4}{\eps ^3}-\frac{8 }{\eps ^2}+\frac{1}{\eps }+3 \bigg)\frac{\bar{n}^\alpha}{p_1\!\cdot\!\bar{n}}\bigg] \nonumber \\
    &\hspace{-55pt}+\frac{m^2}{2p_1\!\cdot\! \bar{n}\,p_2\!\cdot n}\gamma^\alpha\bigg[
    \frac{\Gamma_1}{(1-\eps^2)}
    \left(\frac{3}{\eps^2}-\frac{8}{\eps}-11+14\eps+8\eps^2\right)
    \nonumber\\ &\hspace{-55pt}+\frac{\Gamma_2}{(1+\eps)}
    \left(\frac{2}{\eps^3}+\frac{1}{\eps^2}-\frac{13}{\eps}+44-28\eps-24\eps^2\right)\bigg]\bigg\}v(p_2)\,,
\end{align}
where $\Gamma_{1,2}$ denote the following combinations of Euler gamma functions
\begin{align} \label{eq:massiveGammaStructures}
    \Gamma_1 &= \frac{\Gamma^2(1-\eps)\Gamma^2(1+\eps)}{\Gamma^2(2-2\eps)}\,, \nonumber \\ \Gamma_2 &= \frac{\Gamma^3(1-\eps)\Gamma^2(1+\eps)}{\Gamma(3-3\eps)}\,.
\end{align}
It is instructive to compare this result with its one-loop equivalent in \eq{eq:oneLoopMassiveFinal}. Despite the more involved expressions for the coefficients, the basic structure is similar, but now all three different spin structures $\gamma^\alpha$, $n^\alpha$ and $\bar{n}^\alpha$ contribute, with the latter arising only at order $m^2$. There are other important differences, though. First, due to the more involved dynamical structure, the result features the two independent $\Gamma$-combinations in \eq{eq:massiveGammaStructures}. Second, since now a hard and a collinear loop are present at the same time, both scale ratios $(m^2/\mu^2)^{-\eps}$ and $(-2\, p_1^+p_2^-/\mu^2)^{-\eps}$ show up in the prefactor.  However, setting for convenience the renormalisation scale equal to the hard scale will remove the second factor. Upon expansion in $\eps$ this will result in logarithms of $m^2/(2 p_1^+p_2^-)$ at NLP, as for the one-loop case. These are small-mass logarithms that ideally would be resummed by a complete factorization framework. 
\begin{figure}[t]
  \centering
  \includegraphics[width=0.16\textwidth]{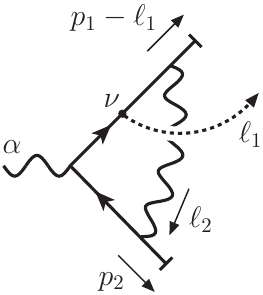}%
  \caption{Diagrammatic representation of the one-loop matrix element $\widetilde{H}_{(f\gamma)}^{(1)}$, from which we extract $H_{(f\gamma)}^{(1)}$ and $H_{(f\partial\gamma)}^{(1)}$\,.}
  \label{fig:twoLoopMassiveHardFunction}
\end{figure}
We now continue with the calculation of the corresponding hard functions, and check that the convolution with the jet functions in \eq{massiveJetFunctionsResult} reproduces our region calculation. Similar to the one-loop example, we can extract the hard functions by Taylor expanding the hard matrix element represented in \fig{twoLoopMassiveHardFunction}, according to the prescription of \eq{Hfgammadef}. The unexpanded amplitude reads
\begin{widetext}
\begin{align} \label{eq:unexpanded2loop}
    \widetilde{H}_{(f\gamma)}^{(1)\h\alpha\nu}(p_1,p_2,\ell_1) &= 
    \int\!\frac{d^{4-2\eps}\ell_2}{(2\pi)^{4-2\eps}}\,
    \frac{e^4 \mu^{2\eps}\,\bar{u}(p_1)\,\gamma^\sigma(\slashed{p}_1\!+\!\slashed{\ell}_2\!+\!m)
    \gamma^\nu(\slashed{p}_1\!+\!\slashed{\ell_2}\!+\!m)
    \gamma^\alpha(-\slashed{p}_2\!+\!\slashed{\ell}_2\!+\!m)}{
    \big[(\ell_2+p_1-\ell_1)^2-m^2\big]
    \big[(\ell_2+p_1)^2-m^2\big]
    \big[(\ell-p_2)^2-m^2\big]
    \ell_2^2}\left( -\gamma_\sigma+\frac{\slashed{\ell_2}\,\bar{n}_\sigma+\slashed{\bar{n}}\,\ell_{2\h\sigma}}{\ell_2\!\cdot\!\bar{n}}\right)\,,
\end{align}
from which we separate the $f\gamma$- and $f\partial\gamma$-term,
\begin{align} \label{eq:fGammaHard2loop}
    H_{(f\gamma)}^{(1)\h\alpha\nu}(x,p_1,p_2) =
    e^4\mu^{2\eps}\!\!\int\!\frac{d^{4-2\eps}\ell_2}{(2\pi)^{4-2\eps}}\,
    \frac{\mathcal{N}_{(f\gamma)}^{\alpha\nu}(x,p_i,\ell_2)}{\mathcal{D}_0(x,p_i,\ell_2)}\,,\quad
    H_{(f\partial\gamma)}^{(1)\h\alpha\nu\rho}(x,p_1,p_2) =
    e^4\mu^{2\eps}\!\!\int\!\frac{d^{4-2\eps}\ell_2}{(2\pi)^{4-2\eps}}\,
    \frac{\mathcal{N}_{(f\partial\gamma)}^{\alpha\nu\rho}(x,p_i,\ell_2)}{\mathcal{D}_0(x,p_i,\ell_2)}\,.
\end{align}
Here the common denominator and the numerator structures are
\begin{align}
\mathcal{D}_0(x,p_i,\ell_2) &= \big[\ell_2^2\big]
    \big[\ell_2^2+2\ell_2\!\cdot\!n\,p_1\!\cdot\!\bar{n}(1-x)\big]
    \big[\ell_2^2+2\ell_2\!\cdot\!n\,p_1\!\cdot\!\bar{n}\big]
    \big[\ell_2^2-2\ell_2\!\cdot\bar{n}\,p_2\!\cdot\!n\big]\,,
    \nonumber \\
    \mathcal{N}_{(f\gamma)}^{\alpha\nu}(x,p_i,\ell_2) &=
    \gamma^\mu\big[(1-x)p_1\!\cdot\bar{n}\,\slashed{n}+\slashed{\ell}_2+m\big]
    \gamma^\nu\big[p_1\!\cdot\bar{n}\,\slashed{n}+\slashed{\ell}_2+m\big]
    \gamma^\alpha\big[-\slashed{p}_2+\slashed{\ell}_2+m\big] \left(-\gamma_\mu + \frac{\slashed{\ell_2}\,\bar{n}_\mu + \slashed{\bar{n}}\, \ell_{2\h\mu}}{\ell_2\!\cdot\!\bar{n}}\right)\,, \nonumber\\
    \mathcal{N}_{(f\partial\gamma)}^{\alpha\nu\rho}(x,p_i,\ell_2) &= \left\{
    -\gamma^\mu\gamma^\rho_\perp\gamma^\nu
    (p_1\!\cdot\! \bar{n}+\slashed{\ell}_2) +
    \frac{2\,\ell_2^\rho\, \gamma^\mu\big[p_1\!\cdot\! \bar{n}(1-x)+\slashed{\ell}_2\big]\gamma^\nu
    (p_1\!\cdot\! \bar{n}+\slashed{\ell}_2)}{\big[\ell_2^2+2\ell_2\!\cdot\!n\,p_1\!\cdot\!\bar{n}(1-x)\big]} \right\} 
    \nonumber\\
    & \quad \times \gamma^\alpha(-p_2\!\cdot \!n\,\slashed{\bar{n}}+\slashed{\ell}_2\big)\big[-\gamma_\mu + \big(\slashed{\ell_2}\,\bar{n}_\mu + \slashed{\bar{n}}\, \ell_{2\h\mu}\big)/\ell_2\!\cdot\!\bar{n}\big]\,.
\end{align}
\end{widetext}
Note that the derivative in the transverse component defining the $f\partial\gamma$-term as in \eq{Hfgammadef} can act either on the spin structure in the numerator, or on the $\ell_{1\perp}$-dependent denominator of \eq{eq:unexpanded2loop}, generating the two structures displayed in curly brackets. In the $f\partial\gamma$-numerator structure we already dropped $\ord(\lambda)$ terms, since we know that this structure enters the factorization formula in a convolution with a jet function that is already $\ord(\lambda^2)$. We can now solve the integral with standard techniques. The presence of many different spin structures at this stage renders the intermediate expressions for the hard functions rather cumbersome, therefore we will not show them here.
Taking the convolutions with the jet functions in \eq{massiveJetFunctionsResult} yields the partial results shown in \eq{eq:convolutionsFinal} in the appendix. As one can readily verify, their sum correctly reproduces the hard-collinear region result obtained in \eq{eq:twoLoopMassiveFinal}. 

We will now examine the pole structure of \eq{eq:twoLoopMassiveFinal} in light of the equivalent factorization result. Interestingly, we note the presence of triple poles. One overall inverse power of $\eps$ is due to the single pole in the jet functions in \eq{massiveJetFunctionsResult}, while the remaining factor of $1/\eps^2$ has two distinct origins. First, the hard functions contain explicit double poles since they describe both hard and soft physics. In \sect{sec:rad_two_loop} we will extensively comment on this effect when examining the hard function in \eq{eq:twoLoopHardFunction}. Second, the hard functions contain $\frac{1}{\eps}\frac{1}{x}$ terms that produce an additional pole upon convolution with the respective jet functions. These endpoint singularities arise in the limit where the dominant momentum component of the collinear photon vanishes. Their origin is thus different from that of the (finite) endpoint contributions captured by $\delta(1-x)$ in \eq{massiveJetFunctionsResult}, which describe the soft quark limit. Endpoint singularities appear in factorization studies using SCET, too \cite{Beneke:2019kgv,Moult:2019uhz,Beneke:2019oqx,Liu:2019oav}, and seem inevitable at NLP. Since the expressions involved are unrenormalised quantities expressed in $D=4-2\eps$ dimensions, the endpoint singularities are easily regulated. 

\section{Hard-collinear factorization for massless fermions}
\label{sec:rad_massless}

In this section, we focus on the scenario of negligible fermion masses, $m=0$, which is the standard approximation in high-energy collisions for light quarks. In the previous section we have seen that the fermion mass entered the non-radiative $f\gamma$-jet function through the overall scale factor $\left(m^2/\mu^2\right)^{-\eps}$ and a second-order polynomial in $m$. Removing this scale from the problem will thus have a serious impact on the ingredients in the factorization framework. Virtual loop corrections to the $f$-jet, as well as all loop-induced, genuine NLP jet functions (like the $f\gamma$-jet), are rendered scaleless and do not contribute. However as we are ultimately interested in threshold effects associated to soft final-state radiation, we are required to compute \emph{radiative} jet functions. For such functions a new scale arises, set by the dot product of the (external) momenta of the emitting fermion and the soft photon. In massless radiative jet functions this small scale takes the place of the mass as the collinear scale. 

As in the previous section, we will focus on the $f\gamma$- and $f\partial\gamma$-jet functions. The radiative functions will be obtained from the non-radiative counterparts by inserting a soft photon on any of the collinear fermion lines. Checking the factorization properties of gauge-invariant sets of diagrams will allow us to make a convenient choice of gauge. While axial gauge proved to be practical for power counting the pinch surfaces that underlie the factorization ingredients, Feynman gauge is more suited for complex calculations. Therefore, we will extract the radiative jet functions here using the latter, and apply this gauge choice consistently in the calculation of the hard functions. The presence of longitudinally polarised photons in Feynman gauge will modify the power counting for individual diagrams. As a consequence, each diagram calculated in this section may contain spurious LP terms, which must cancel upon summing over a gauge-invariant set of diagrams. 

The radiative jet functions we extract are process-independent quantities, which describe collinear physics regardless of the underlying hard scattering event. Similar to the massive case, we validate the expressions obtained by convolving these jets with appropriate hard functions by means of a one- and two-loop method of regions calculation. These non-trivial checks show that the all-order factorization formula for elastic amplitudes in \eq{NLPfactorization} provides a good starting point for the factorization (and potentially resummation) of threshold effects due to soft final state radiation. 

In particular, our calculations show that the NLP factorization formula presented in~\cite{DelDuca:1990gz,Bonocore:2015esa} does not suffice for one-loop accuracy at the matrix element level, which has also recently been noted in a SCET context \cite{Beneke:2019oqx}, although they do work at the cross section level for the cases studied there. Beyond one-loop such factorization formulae do not capture the intricate hard-collinear interplay for matrix elements that our current approach does account for.

\subsection{The radiative, massless 
\texorpdfstring{$f\gamma$}{f gamma}-jet}
\label{sec:rad_jet}

\begin{figure}[t]
  \centering\vspace*{15pt}
  \includegraphics[width=0.37\textwidth]{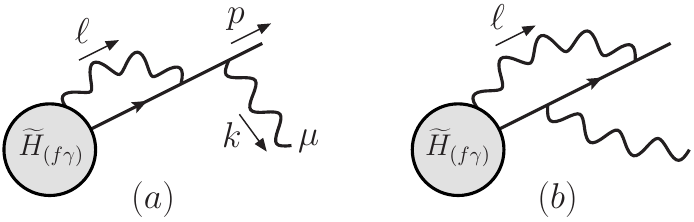}%
  \caption{Contributions to the radiative $f\gamma$- and $f\partial\gamma$-jet. (a) Outer photon attachment. (b) Inner photon attachment.}
  \label{fig:rad_fgamma_jet}
\end{figure}

At the lowest order in perturbation theory the radiative $f\gamma$- and $f\partial\gamma$-jet receive contributions from the two diagrams in \fig{rad_fgamma_jet}. Both functions are defined in the same manner as in the massive fermion case and their evaluation relies on similar techniques. As before, we drop terms beyond NLP. In this case, we apply this constraint to the more involved denominator structure too, expanding denominators whose scaling is inhomogeneous in $\lambda$, as we did in the method of regions calculation of the amplitude. For example, after rescaling the momentum components by the appropriate powers of $\lambda$, the denominator of the innermost propagator is expanded as 
\vspace{-10pt}
\begin{multline}
    \frac{1}{(\ell-p-k)^2+i\eta} \\
    =\frac{1}{D}\bigg[1+ \overbrace{\frac{\strut 2\,\ell_\perp\!\cdot\! k_\perp}{D}}^{\sim\lambda}+\bigg(\overbrace{\frac{\strut 2\,\ell^-k^+}{D}+\frac{(2\,\ell_\perp\!\cdot\! k_\perp)^2}{D^2}}^{\sim\lambda^2}\bigg)\bigg]\,,
    \label{massless_exp}\end{multline}
assuming collinear and soft scaling for $\ell$ and $k$ respectively and abbreviating the homogeneous denominator by 
    \begin{align}
    D &= 2\ell^+\ell^- \!+\ell_\perp^2 - 2p^+\ell^- \!- 2\ell^+k^- \!+ 2p^+k^- \!+i\eta\,.
\end{align}
Note that, having set $m=0$, the external momentum can be chosen to be strictly in the $+$-direction $p = (p^+,0,0)$.
Using this approach, we require six one-loop integrals to evaluate the contributions to the $f\gamma$- and $f\partial\gamma$-jet, which are listed in appendix \ref{appx:radjet_integrals}. We find the following result for these jet functions
\begin{widetext}
\begin{align}
J^{(1)\h\mu\nu}_{(f\gamma)}(x,p,k)=&-\frac{e^2}{16\pi^2} \left(\!\frac{-2\,p^+k^-}{4\pi\mu^2}\!\right)^{-\eps}\Gamma(\eps)\left[x\,\omx\right]^{-\eps}\bar{u}(p)\Big\{2\,\omx\,\eta^{\mu\nu}-\frac{\eps}{\ome}\,x\,\gamma^\nu\gamma^\mu 
 +2\,(1\!-\!2\,x)\frac{k^+}{k^-}n^\mu n^\nu
 \nonumber \\ 
 &  - 2\,(1\!-\!2\,x)\, \bar{n}^\mu n^\nu +\frac{1}{k^-}\Big[x\,  \gamma^\mu\slashed{k}\,n^\nu+2\,\frac{\eps}{\ome}\,x\,k^\mu n^\nu
+\frac{\eps}{\ome}\,x\,\gamma^\nu\slashed{k}\,n^\mu-2\,\omx\, n^\mu k^\nu  \Big]\Big\}\,, \label{rjetfgamma}
 \\
    J^{(1)\h\mu\nu \rho}_{(f\partial \gamma)}(x,p,k)=&-\frac{e^2\,p^+}{8\pi^2} \left(\!\frac{-2\,p^+k^-}{4\pi\mu^2}\!\right)^{-\eps}\frac{\Gamma(\eps)}{\ome}\left[x\,\omx\right]^{1-\eps}\bar{u}(p)\,n^\nu\left(\eta_\perp^{\mu\rho}-\frac{n^\mu k_\perp^\rho}{k^-}\right)\,. \label{rjetfdgamma}
\end{align}
\end{widetext}
We point out that \eq{rjetfgamma} is strictly $\ord(\lambda^0)$, while the individual contributions from \fig{rad_fgamma_jet}\textcolor{prd_blue}a and \fig{rad_fgamma_jet}\textcolor{prd_blue}{b} have indeed a LP component $\ord(\lambda^{-2})$.
Note that \eq{rjetfgamma} does not contain the $\delta \omx$ term which appeared in the non-radiative jet function for the massive fermion case (\eq{massiveJetFunctionsResult}) which was associated to the soft quark limit. In principle, one might expect a similar contribution here, but the numerator supplements the standard integrals of \Eqns{mIntaminus}{mIntbminus} with sufficient powers of $\omx$ to suppress such a term.

The radiative $f$-jet is known to have a Ward identity \cite{DelDuca:1990gz} relating it to its non-radiative counterpart (order by order in perturbation theory) via 
\begin{equation} \label{eq:WardId}
    k_\mu J^{(n)\h\mu}_{(f)}(p,k) = -q\,e\,J^{(n)}_{(f)}(p)\,, 
\end{equation}
where $q=-1$ for the jets considered here. Similarly, we expect
\begin{subequations}
\begin{align}
    k_\mu J^{(n)\h\mu\nu}_{(f\gamma)}(x,p,k) &= -q\,e\, J^{(n)\h\nu}_{(f\gamma)}(x,p)\,, \\ k_\mu J^{(n)\h\mu\nu\rho}_{(f\partial\gamma)}(x,p,k) &= -q\,e\, J^{(n)\h\nu\rho}_{(f\partial\gamma)}(x,p)\,.
\end{align}
\end{subequations}
In particular, since $J^{(f\gamma)\nu}$ and $J^{(f\partial\gamma)\nu\rho}$ consist solely of scaleless integrals (for $m=0$) and thus vanish in dimensional regularisation, we should find
\begin{equation}
    k_\mu J^{(n)\h\mu\nu}_{(f\gamma)}(x,k) = 0\,,\qquad k_\mu J^{(n)\h\mu\nu\rho}_{(f\partial\gamma)}(x,k) = 0\,. \label{radfgammajetWI}
\end{equation} By contracting \eq{rjetfgamma} and \eq{rjetfdgamma} with $k_\mu$, one finds that \eq{radfgammajetWI} is indeed satisfied, which serves as a first check on these jet functions.
 
\subsection{NLP factorization of the collinear sector at the one-loop level}
\label{sec:rad_one_loop}
 
Following the same approach as in \ref{sec:massive_MOR_tests}, we wish to test the factorization structure of \emph{radiative} amplitudes in the collinear sector, by means of a comparison to a method-of-regions computation of the single-real single-virtual (1R1V) correction to a dijet production process. A similar factorization/regions analysis has been carried out in refs.~\cite{Bonocore:2014wua,Bonocore:2015esa} for Drell-Yan production, at the same loop order but at the cross-section level instead. In~\cite{Bonocore:2015esa} a NLP factorization formula for radiative amplitudes was derived from the standard LP factorization picture of purely virtual amplitudes, while here we start from the generalised NLP factorization formula of \eq{NLPfactorization}. 
The difference between these two NLP approaches at the one- and two-loop level will be highlighted in the remainder of section \ref{sec:rad_massless}.
 
\begin{figure*}[t]
     \centering
     \includegraphics[width=.70\textwidth]{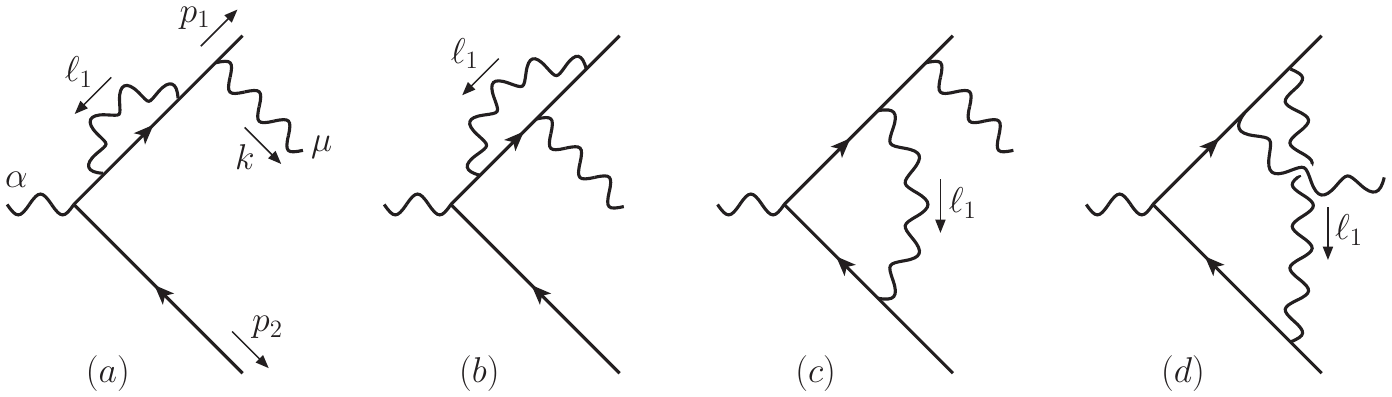}
     \caption{Diagrams contributing to the collinear region of the 1R1V dijet production amplitude. (a) Self-energy diagram, outer attachment. (b) Self-energy diagram, inner attachment. (c) Vertex diagram, outer attachment. (d) Vertex diagram, inner attachment.}
     \label{fig:1r1v}
 \end{figure*}

\begin{figure}[bh]
  \centering
  \includegraphics[width=0.37\textwidth]{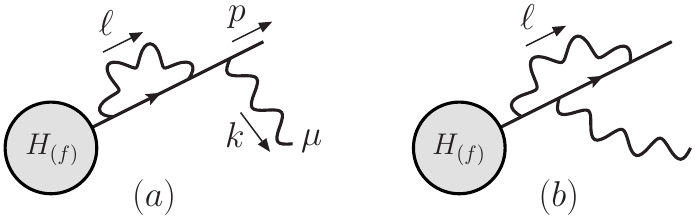}%
  \caption{One-loop contributions to the radiative $f$-jet. (a) Outer photon attachment. (b) Inner photon attachment.}
  \label{fig:rad_f_jet}
\end{figure}

The diagrams that contribute to the collinear sector at the one-loop order are shown in \fig{1r1v}.
In diagrams $(a)$ and $(b)$ the collinear loop attaches only to the upper leg, meaning that there is only one fermion connection between the hard interaction and the part of the diagram containing the collinear dynamics. Therefore, these diagrams are predicted to factorize in terms of the one-loop radiative $f$-jet (see \fig{rad_f_jet}), the Born-level hard scattering amplitude (with amputated legs) and a trivial jet function for the opposite leg:
\begin{equation}
        \mathcal{M}^{(1)\h\alpha\mu}_{\rm a+b\h |\h fact.}(p_1,p_2,k) = J^{(1)\h\mu}_{(f)}(p_1,k)H^{(0)\h\alpha}_{(f)}(p_1,p_2)J^{(0)}_{(f)}(p_2)\,, \label{1r1v_ab_fact}
\end{equation}
with
\begin{equation}
    J^{(0)}_{(f)}(p_2) = v(p_2)\,,\qquad \qquad
    H^{(0)\h \alpha}_{(f)} = -ie\gamma^\alpha.
\end{equation}
The one-loop radiative $f$-jet is readily computed by standard techniques and the result reads
\begin{align}
     \hspace{-7pt}J^{(1)\h\mu}_{(f)}(p,k) &= \frac{-\,e^3}{16\pi^2}\frac{1}{p^+k^-}\left(\!\frac{-2\,p^+k^-}{4\pi\mu^2}\!\right)^{-\eps}\frac{\Gamma^2(1\!-\!\eps)\Gamma(1\!+\!\eps)}{\Gamma(2\!-\!2\eps)}\nonumber \\&\times\bar{u}(p)\left[\left(\frac{1}{\eps}+\frac{1}{2}\right)\gamma^\mu\slashed{k}+ \left(\frac{1}{\eps}-1\right)k^\mu\right]. \label{rf_jet}
\end{align}
Note that this is strictly a NLP quantity, while from the non-radiative power counting formula (\eq{eq:QEDPC_final}) one may have expected a contribution at LP. This power suppression is a radiative effect that only starts at one-loop order and is therefore not captured by the general power counting formula (the tree-level result does have a LP contribution). It is however fully consistent with \cite{DelDuca:1990gz}, in which the radiative $f$-jet is defined to account for the NLP effects induced by soft emissions from collinear loops.
As before, we may evaluate the contributions to the collinear sector of diagrams $(a)$ and $(b)$ in \fig{1r1v}, using again the method of regions. Keeping terms up to NLP this approach yields  
 \begin{multline}
 \mathcal{M}^{(1)\h\alpha\mu}_{\rm a+b\h|\h C}(p_1,p_2,k) =  \frac{-ie^4}{8\pi^2}\frac{1}{t}\!\left(\!\frac{t}{4\pi\mu^2}\!\right)^{\!\!-\eps}\frac{\Gamma^2(1\!-\!\eps)\Gamma(1\!+\!\eps)}{\Gamma(2\!-\!2\eps)} \\ \times\bar{u}(p_1)\left[\left(\frac{1}{\eps}+\frac{1}{2}\right)\gamma^\mu\slashed{k}+ \left(\frac{1}{\eps}-1\right)k^\mu\right]\gamma^\alpha v(p_2)\,. \label{1r1v_ab_mor}
 \end{multline}
For massless fermions we may choose $p_1^\mu = (p_1^+,0,0)$ and $p_2^\mu = (0,0,p_2^-)$, such that the standard (massless) Mandelstam variable $t=(p_1-k)^2=-2p^+_1k^-$. From eqs.~\eqref{1r1v_ab_fact} and \eqref{rf_jet} on the one hand and \eq{1r1v_ab_mor} on the other, we see immediately that the regions result coincides with the factorization result which, given the trivial factorization structure of these diagrams, is perhaps not surprising.  

For diagrams $(c)$ and $(d)$ in \fig{1r1v} we expect a factorization analogous to \eq{1v_fact}
\begin{align}
    &\mathcal{M}^{(1)\h\alpha\mu}_{\rm c+d\,|\,fact.}(p_1,p_2,k)\nonumber\\
     &\quad=\int_0^1\!dx\, \Big[J^{(1)\h\mu\nu}_{(f\gamma)}(x,p_1,k)\,H^{(0)\h\alpha}_{(f\gamma)\h\nu}(x,p_1,p_2) \nonumber \\ &\quad+J^{(1)\h\mu\nu\rho}_{(f\partial\gamma)}(x,p_1,k)\,H^{(0)\h\alpha}_{(f\partial\gamma)\h\nu\rho}(x,p_1,p_2)\Big]J^{(0)}_{(f)}(p_2)\,, \label{1r1v_fact_cd1}
\end{align}
where, as in the massive case, $x=\ell_1^+/p_1^+$. The hard functions are extracted from \fig{Hfgamma_LO} (now with $m=0$) according to the definition of \eq{Hfgammadef} and read  
\begin{align}\label{eq:HfGamma0massless1}
    H^{(0)\h\alpha}_{(f\gamma)\h\nu}(x,p_1,p_2) &=  i\,e^2\,\frac{1}{x\,s}\gamma^\alpha\bigl(x\,\slashed{p}_1+\slashed{p}_2\bigr)\gamma_\nu\,, \\ \label{eq:HfGamma0massless2}
    H^{(0)\h\alpha}_{(f\partial\gamma)\h\nu\rho}(x,p_1,p_2) &= i\,e^2\,\frac{1}{x\,s}\gamma^\alpha\gamma_{\perp\rho}\gamma_\nu\,,
\end{align}
where $s=(p_1+p_2)^2=2p_1^+p_2^-$. Since the $f\gamma$- and $f\partial\gamma$-jet function are strictly NLP quantities, we have discarded NLP corrections to both hard functions, as they would affect the full amplitude only at NNLP. Substituting \Eqns{eq:HfGamma0massless1}{eq:HfGamma0massless2} together with \Eqns{rjetfgamma}{rjetfdgamma} into \eq{1r1v_fact_cd1} and simplifying the Dirac structure, we find
\begin{align}
    &\mathcal{M}^{(1)\h\alpha\mu}_{\rm c+d\,|\,fact.}(p_1,p_2,k) \;=\; 
    \frac{i\,e^4}{8\pi^2}\left(\!\frac{t}{4\pi\mu^2}\!\right)^{-\eps} \frac{\Gamma(1+\eps)}{\eps\,(\ome)}\bar{u}(p_1)\nonumber\\
    &\quad
    \times\int_0^1\!dx\left[x\, \omx\right]^{-\eps}\nonumber
    \bigg\{(\ome)\,\frac{1}{t}\,\gamma^\mu\slashed{k}\gamma^\alpha\\
    &\quad +2\bigg[\eps\, \frac{k^\mu}{t}-(1\!-\!2\eps)\,\frac{1}{s}\left(p_2^\mu-\frac{u}{t}\,p_1^\mu\right)\bigg]\gamma^\alpha \nonumber \\ &\quad-2\,\eps\,\frac{1}{s}\left[p_2^\alpha- \left(1-(\ome)\,x\right)p_1^\alpha\right]\left[2\,\frac{p_1^\mu\,\slashed{k}}{t}+\gamma^\mu\right]\!\bigg\}v(p_2)\,.
\end{align}
Upon integration over the convolution parameter $x$, we conclude that this indeed reproduces the regions result
\begin{align}
     &\mathcal{M}^{(1)\h\alpha\mu}_{\rm c+d\,|\,C}(p_1,p_2,k) \nonumber\\  &\quad=\frac{i\, e^4}{8\pi^2}\left(\!\frac{t}{4\pi\mu^2}\!\right)^{-\eps} \frac{\Gamma(1+\eps)\Gamma^2(1-\eps)}{\Gamma(2-2\eps)}
     \bar{u}(p_1)\bigg\{\frac{1}{\eps}\,\frac{\gamma^\mu\slashed{k}\gamma^\alpha}{t} \nonumber\\
     &\quad + 2\bigg[\frac{1}{1-\eps}\, \frac{k^\mu}{t}-\left(\frac{1}{\eps}-\frac{1}{\ome}\right)\frac{1}{s}\left(p_2^\mu-\frac{u}{t}\,p_1^\mu\right)\bigg]\gamma^\alpha 
     \nonumber \\ & \quad
     -\frac{1}{s}\left[\frac{2}{\ome}\,p_2^\alpha- \frac{1\!+\!\eps}{\ome}\,p_1^\alpha\right]\!\left[2\,\frac{p_1^\mu\,\slashed{k}}{t}+\gamma^\mu\right]\!\bigg\}v(p_2)\,,  \label{1r1v_cd_mor}
\end{align}
with $u=(p_2-k)^2=-2p^-_2k^+$. The collinear sector of the radiative amplitudes in \fig{1r1v} is thus, up to NLP, correctly described by dressing the jet functions appearing in \eq{NLPfactorization} with a single soft emission. This is another indication that this factorization formula indeed holds and organises NLP contributions, even in presence of soft final state radiation. Moreover, this comparison serves as an explicit verification of the process-independent jet functions in eqs.~\eqref{rjetfgamma} and \eqref{rjetfdgamma}. 

\begin{figure}[t]
  \centering
  \includegraphics[width=0.385\textwidth]{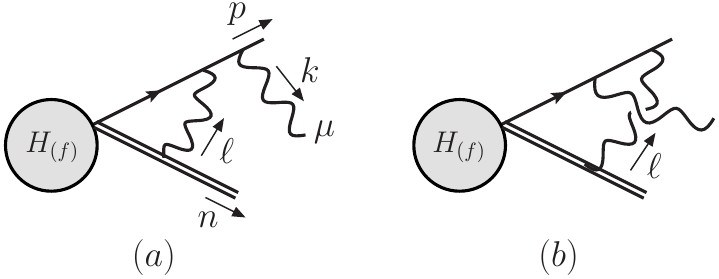}%
  \caption{Additional contributions to the radiative $f$-jet, denoted by $J^{(1)\h\mu}_{(f^{\prime})}$, in a simplified NLP factorization framework. Longitudinally polarised collinear photons that probe the hard scattering are described by the Wilson line interaction. (a) Outer photon attachment. (b) Inner photon attachment.}
  \label{fig:rad_f_jet_WL}
\end{figure}

We emphasise that the simplified radiative factorization formula of \cite{DelDuca:1990gz} does not suffice to reproduce the 1R1V amplitude, as noted recently in \cite{Beneke:2019oqx} (see in particular section 4.2.4 there). This approach relies on a direct product of the hard and (radiative) jet functions, as we do for the $f$-jet in \eq{1r1v_ab_fact}. We will illustrate this issue by supplementing our $f$-jet function with the additional contributions (denoted by $f^\prime$) shown in \fig{rad_f_jet_WL}, to recover the radiative jet that has been calculated to one-loop order in \cite{Bonocore:2015esa}. These diagrams have a Wilson line in the $\bar{n}$ direction and are the radiative equivalents of the traditional, LP jet functions.\footnote{Recall that in the derivation of factorization at LP, in a general covariant gauge, only longitudinally polarised collinear photons probe the hard function~\cite{Collins:1989gx}. By means of Ward identities these can be shown to decouple entirely and are cast into connections to a Wilson line.} This simplified factorization approach would give the following result for diagrams (c) and (d) in \fig{1r1v}
\begin{align} 
&\mathcal{M}^{(1)\h\alpha\mu}_{\rm c+d\,|\, simp.\,fact.}(p_1,p_2,k)\,
\nonumber\\
& \quad\equiv J^{(1)\h\mu}_{(f^{\prime})}(p_1,k,\bar{n}) H^{(0)\h\alpha}_{(f)}(p_1,p_2)J^{(0)}_{(f)}(p_2)  \nonumber \\ &\quad  =\frac{i\,e^4}{8\pi^2}\left(\!\frac{t}{4\pi\mu^2}\!\right)^{-\eps} \frac{\Gamma(1+\eps)\Gamma^2(1-\eps)}{\Gamma(2-2\eps)}\bar{u}(p_1)\bigg\{\frac{1}{\eps}\,\frac{\gamma^\mu\slashed{k}\gamma^\alpha}{t}  \nonumber\\      
     &\qquad+2\bigg[\frac{1}{1-\eps}\, \frac{k^\mu}{t}-\left(\frac{1}{\eps}-\frac{1}{\ome}\right)\frac{1}{s}\left(p_2^\mu-\frac{u}{t}\,p_1^\mu\right)\bigg]\gamma^\alpha \nonumber \\ 
     &\qquad-\frac{1}{s}\frac{2}{\ome}\,p_2^\alpha \left[2\,\frac{p_1^\mu\,\slashed{k}}{t}+\gamma^\mu\right]\!\bigg\}v(p_2) \label{simplifiedfactorization}\, , 
\end{align}
where we have set $\bar{n}^\mu= p_2^\mu/p_2^-$. Comparison with the collinear result of \eq{1r1v_cd_mor} shows 
\begin{align} 
&\mathcal{M}^{(1)\h\alpha\mu}_{\rm c+d\,|\,C}(p_1,p_2,k)-\mathcal{M}^{(1)\h\alpha\mu}_{\rm c+d\,|\, simp.\,fact.}(p_1,p_2,k) \,\nonumber\\
&\quad =\frac{i\, e^4}{8\pi^2}\left(\!\frac{t}{4\pi\mu^2}\!\right)^{-\eps} \frac{\Gamma(1+\eps)\Gamma^2(1-\eps)}{\Gamma(2-2\eps)}\frac{1\!+\!\eps}{\ome} \nonumber\\ &\qquad\times \bar{u}(p_1) p_1^\alpha\!\left[2\,\frac{p_1^\mu\,\slashed{k}}{t}+\gamma^\mu\right]v(p_2)\, . \end{align} Since these missing terms vanish upon contraction with the conjugate amplitude, the simplified factorization approach did suffice in the 1R1V cross-section calculation presented in \cite{Bonocore:2015esa}.
%
\subsection{Hard-collinear interplay at the two-loop level}
\label{sec:rad_two_loop}

%
\begin{figure*}[tbh]
\centering
     \includegraphics[width=.70\textwidth]{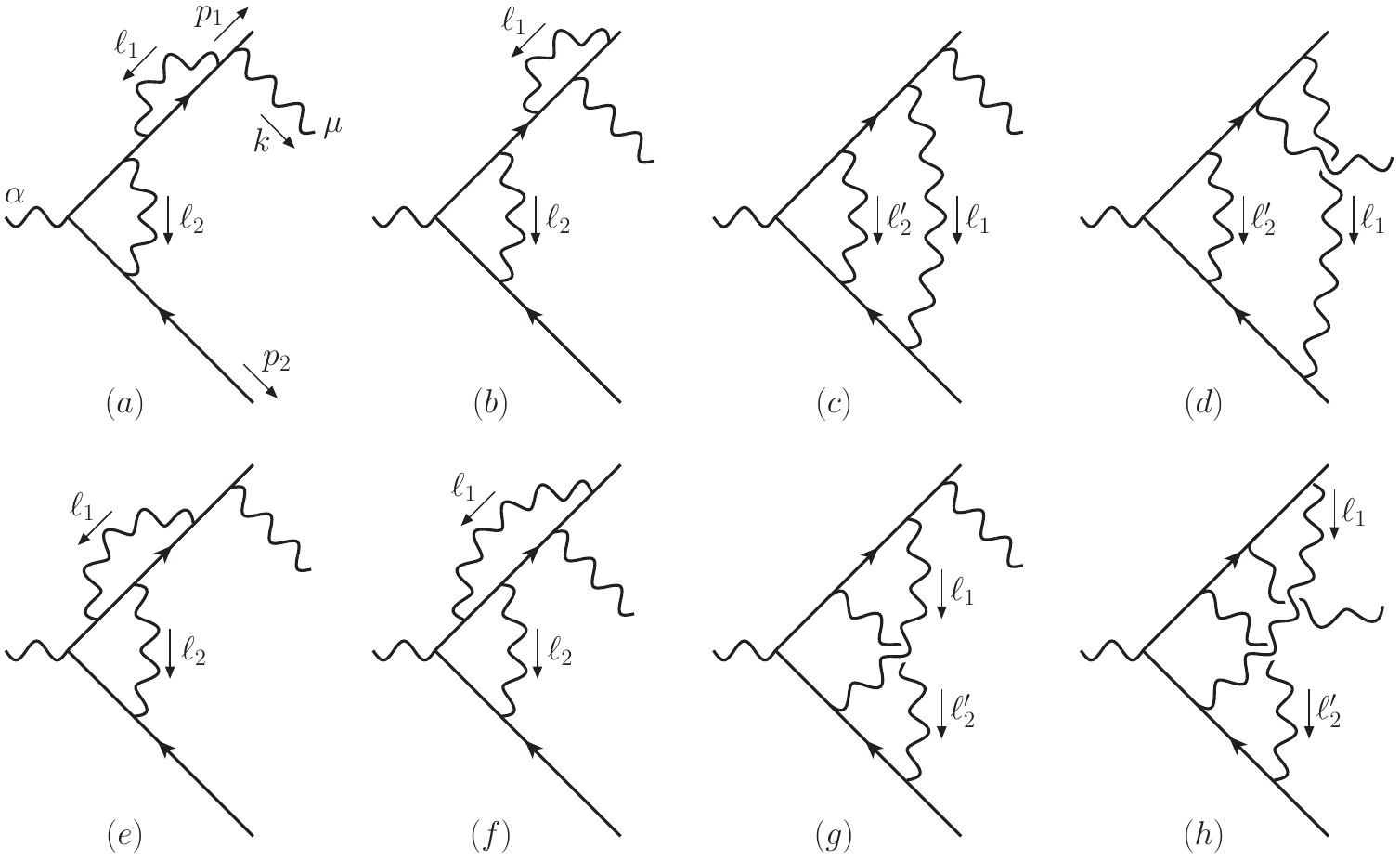}
     \caption{Diagrams contributing to the hard-collinear region of the 1R2V dijet production amplitude, with $\ell_1$ ($\ell_2$) denoting the collinear (hard) loop momentum. We use the shorthand notation $\ell^\prime_2 = \ell_2-\ell_1$. The individual diagrams (a)-(h) are labeled for reference purposes.}
     \label{fig:1r2v}
\end{figure*}
We now move to (single) radiative amplitudes at two-loop order (denoted as 1R2V) and carry out a similar test. At this loop order there is a more involved interplay between the hard and collinear sector, as the dominant component of the collinear momentum of the virtual photon may interfere with the hard loop. This hard-loop effect is not power suppressed, and has to be properly accounted for in the factorization picture in order to reproduce the exact NLP amplitude. 
 
Our main effort here will be to explore this subtle interplay and therefore we (again) compare to a hard-collinear region with a method of regions calculation. The relevant diagrams for that purpose are shown in \fig{1r2v}, where the collinear momentum is denoted by $\ell_1$ and the hard momentum by $\ell_2$. We identify these diagrams through the following considerations.

First, the soft photon must originate from the collinearly enhanced region, rather than from the hard loop. Otherwise this would be described by a different term in the factorization formula, as stated by the Low-Burnett-Kroll theorem \cite{Low:1958sn, Burnett:1967km}: a soft final-state emission from the hard scattering is described by a derivative with respect to either one of the external hard momenta, acting on the non-radiative hard scattering amplitude.\footnote{Note that even if formally needed these diagrams would not contribute, since the collinear loop integral in those configurations would be insensitive to the soft emission and therefore be scaleless.}
 
\begin{figure}[tbh]
    \centering \vspace*{5pt}
    \includegraphics[width=0.36\textwidth]{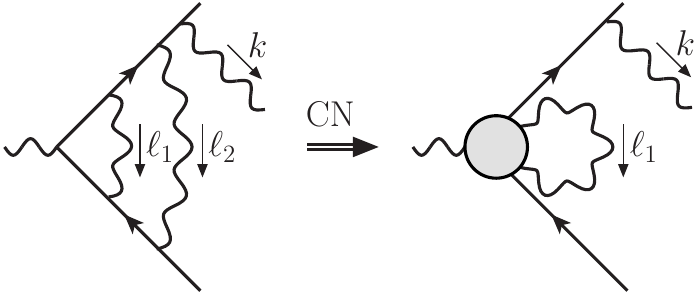}
    \caption{Coleman-Norton picture that arises from attaching the hard photon to the right of the collinear photon on the upper leg. The tadpole-like configuration of the supposedly collinear photon does not coincide with a classical trajectory. Hence this ordering of attachments does not contribute to the scattering amplitude.}
    \label{fig:inverted_loops_CN}
\end{figure}

Second, the ordering of the virtual photon attachments is crucial. This is best seen from a Coleman-Norton analysis, in which hard, off-shell lines are shrunk to a point. In fact, it is strictly the attachment on the upper leg that matters, since the fermion propagators on the lower leg are shrunk to a point irrespective of the ordering. A propagator that is not part of the hard loop but which carries both an anti-collinear external momentum as well as a collinear loop momentum, obeys a hard scaling too. This implies that we can treat the planar-topology diagrams $(c)$ and $(d)$ in \fig{1r2v} as well as the crossed-topology diagrams $(g)$ and $(h)$ on equal footing. To see what happens if one inverts the order of attachments on the upper leg, let us consider diagram $(c)$ as an example. In that case the outer loop would be hard and therefore shrunk to the tree-level hard scattering vertex, as shown in \fig{inverted_loops_CN}. The supposedly collinear photon line would now form a tadpole-like attachment to the hard scattering vertex. However, this configuration cannot describe an on-shell line since it does not coincide with any classical trajectory~\cite{Coleman:1965xm}, and does not contribute to the scattering amplitude. The collinear photon must thus attach to the upper leg outside of the hard loop, in order for the diagram to develop a hard-collinear region. 
Lastly, we note that these diagrams naturally contain a doubly collinear region too, which would be described by a higher-order radiative $f\gamma$- and corresponding $f\partial\gamma$-jet, as well as the radiative $f\gamma\gamma$-jet, all contracted with a tree-level hard function. In a complete description of the doubly collinear region at this loop order, one would even expect contributions from the radiative $f\!f\!f$-jet, which would be an interesting analysis by itself. This region does not overlap with the hard-collinear region we explore here, and thus we leave it to future work.

Analogously to the one-loop order, we foresee a pair-wise factorization of the diagrams in \fig{1r2v}, by collecting those graphs differing only by the position of the radiated photon. For diagrams $(a)$ and $(b)$, we have 
\begin{equation}
    \mathcal{M}^{(2)\h \alpha\mu}_{\rm a+b\h|\h fact.}(p_1,p_2,k) = J^{(1)\h \mu}_{(f)}(p_1,k)H^{(1)\h \alpha}_{(f)}(p_1,p_2)J^{(0)}_{(f)}(p_2)\,, \label{1r2v_ab_fact}
\end{equation}
with $H^{(1)\h \alpha}_{(f)}(p_1,p_2)$ the one-loop form factor. The hard function combined with the trivial jet function on the anti-collinear leg reads
\begin{align} \label{eq:twoLoopHardFunction}
    &H^{(1)\h \alpha}_{(f)}(p_1,p_2)J^{(0)}_{(f)}(p_2) \,
    \nonumber\\
    &\quad = \frac{i\,e^3}{8\pi^2}\left(\!\frac{-s}{4\pi\mu^2}\!\right)^{\!\!-\eps}
    \frac{\Gamma^2(\ome)\Gamma(1\!+\!\eps)}{\Gamma(2\!-\!2\eps)}
    \bigg\{\!\left(\frac{1}{\eps^2}-\frac{1}{2\eps}+1\right)\gamma^\alpha
    \nonumber\\
    &\qquad+\frac{1}{s}\left[\left(\frac{1}{\eps}-1\right)p_1^\alpha-\left(\frac{2}{\eps^2}+1\right)p_2^\alpha\right]\!\bigg\}v(p_2)\,.
\end{align}
Eq.~\eqref{eq:twoLoopHardFunction} contains explicit double poles, while the unrenormalised hard function may only contain single poles of a UV nature; this double pole is thus of IR origin. In the method of regions, the appearance of IR poles in the hard region is a common phenomenon if the soft region is scaleless. (For the diagrams defined in \fig{1r2v} this is indeed the case, as is easily verified by assigning $\ell_2$ a soft scaling according to \eq{regions} and expanding denominators in $\lambda$.) Scaleless integrals are set to zero, which typically follows from a cancellation of IR and UV poles. Isolating this UV pole in the soft region and absorbing it in the hard region would cancel the double pole there, thus moving the double pole associated to soft physics from the hard to the soft region. We do not address this mixing of the hard and soft physics, as it affects the method-of-regions calculation and the hard function in the exact same way, while the collinear sectors, which are the focus of this study, \emph{do} factorize entirely from the rest.

\begin{figure*}[tbh]
    \centering
    \includegraphics[width=.62\textwidth]{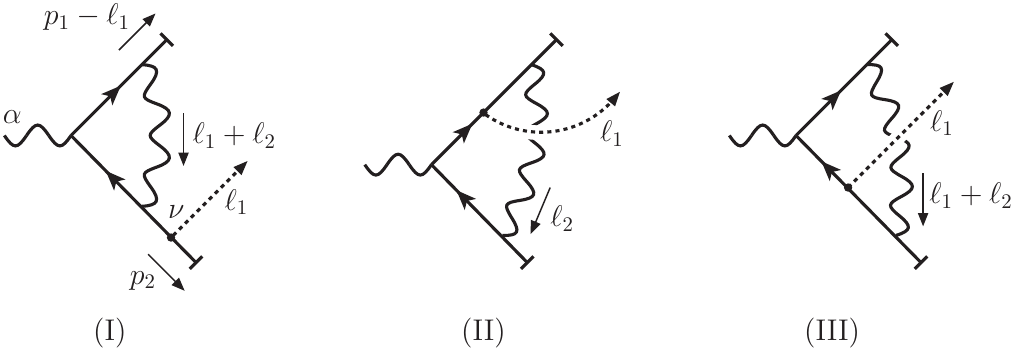}
    \caption{Diagrams contributing to the one-loop matrix element $\widetilde{H}^{(1)}_{(f\gamma)}$ (I) Momentum extraction from the anti-collinear leg external to the loop, defining $H^{(1)\alpha}_{(f\gamma|\rm I)\nu}$ and $H^{(1)\alpha}_{(f\partial\gamma|\rm I)\nu\rho}$. (II) Momentum extraction from the collinear leg, defining $H^{(1)\alpha}_{(f\gamma|\rm II)\nu}$ and $H^{(1)\alpha}_{(f\partial\gamma|\rm II)\nu\rho}$. (III) Momentum extraction from the anti-collinear leg internal to the loop, defining $H^{(1)\alpha}_{(f\gamma|\rm III)\nu}$ and $H^{(1)\alpha}_{(f\partial\gamma|\rm III)\nu\rho}$.}
    \label{fig:Hfgamma_NLO}
\end{figure*}

Turning to the remaining diagrams in \fig{1r2v}, $(c)$ to $(h)$, we expect these to factorize according to

\begin{align}
\label{1r2v_c-h_fact}
    &\mathcal{M}^{(2)\h\alpha\mu}_{\rm \{c+d,\,e+f,\,g+h\}\,|\,fact.}(p_1,p_2,k) \ 
    \\
    &=\int_0^1\!dx\, \Big[J^{(1)\h\mu\nu}_{(f\gamma)}(x,p_1,k)\,H_{(f\gamma\,|\,\rm\{I,\,II,\,III\})\,\nu}^{(1)\h \alpha}(x,p_1,p_2)\nonumber \\&+J^{(1)\h \mu\nu\rho}_{(f\partial\gamma)}(x,p_1,k)\,H_{(f\partial\gamma\,|\,\rm\{I,\,II,\,III\})\,\nu\rho}^{(1)\h \alpha}(x,p_1,p_2)\Big]J^{(0)}_{(f)}(p_2)\,\nonumber, 
\end{align}
with the one-loop $f\gamma$- and $f\partial\gamma$-hard functions extracted from \fig{Hfgamma_NLO}. The calculation of these functions is deferred to appendix \ref{appx:hard_functions} for conciseness.  
Upon evaluation of eqs.~\eqref{1r2v_ab_fact} and \eqref{1r2v_c-h_fact} we find
\begin{widetext}
\begin{align}
     &\mathcal{M}^{(2)\h\alpha\mu}_{\rm a+b\,|\,fact.}(p_1,p_2,k)=
     -\frac{i e^6}{(4\pi)^4} \left(\!\frac{-s}{4\pi\mu^2}\!\right)^{-\eps}\!\left(\!\frac{t}{4\pi\mu^2}\!\right)^{-\eps}\Gamma_3\,\bar{u}(p_1)\bigg\{2\bigg[\left(\frac{2}{\eps^2}-\frac{1}{\eps}-1\right)p_1^\alpha -\left(\frac{4}{\eps^3}+\frac{2}{\eps^2}+\frac{2}{\eps}+1\right)p_2^\alpha\bigg]\frac{\gamma^\mu}{s}\nonumber \\ &\hspace{101pt} +2\bigg[-\frac{2}{\eps^3}+\frac{3}{\eps^2}-\frac{3}{\eps}+2\bigg]\frac{k^\mu \gamma^\alpha}{t}-\bigg[\frac{4}{\eps^3}+\frac{3}{\eps}+2\bigg]\frac{\gamma^\mu\slashed{k}\gamma^\alpha}{t}
     \nonumber \\&\hspace{101pt}+4\bigg[\left(\frac{2}{\eps^2}-\frac{1}{\eps}-1\right)p_1^\alpha-\left(\frac{4}{\eps^3}+\frac{2}{\eps^2}+\frac{2}{\eps}+1\right)p_2^\alpha\bigg]\frac{p_1^\mu\slashed{k}}{s\,t}\bigg\}v(p_2)\,, \label{1r2v_ab_fact2}\\
     &\mathcal{M}^{(2)\h\alpha\mu}_{\rm c+\ldots+h\,|\,fact.}(p_1,p_2,k)=
     \,\frac{i e^6}{(4\pi)^4} \left(\!\frac{-s}{4\pi\mu^2}\!\right)^{-\eps}\!\left(\!\frac{t}{4\pi\mu^2}\!\right)^{-\eps} 
     \nonumber \\
     &\quad \times\bar{u}(p_1)\bigg\{\frac{4\,\Gamma_3}{1\!-\!\eps}\bigg[\left(\frac{2}{\epsilon ^2}-\frac{1}{\epsilon }+2\right)\frac{k^\mu}{t}+\left(\frac{2}{\epsilon ^3}-\frac{5}{\epsilon ^2}+\frac{4}{\epsilon }-4\right)\!\frac{1}{s}\left(\frac{u}{t}\,p_1^\mu-p_2^\mu\right)\bigg]\!\gamma^\alpha + 2\,\Gamma_3\left[\frac{2}{\epsilon ^3}-\frac{1}{\epsilon ^2}+\frac{2}{\epsilon }\right]\frac{\gamma^\mu\slashed{k}\gamma^\alpha}{t}\nonumber\\ &\quad+\frac{2}{(\ome)^2}\frac{1}{s}\bigg[\bigg(\!\,\Gamma_3\!\left(\!-\frac{4}{\epsilon ^3}+\frac{8}{\epsilon ^2} +\frac{2}{\epsilon }-4-8 \epsilon-2 \epsilon ^2\!\right)-\frac{2\,\Gamma_2}{1\!-\!2\eps}\bigg(\!-\frac{4}{\epsilon ^3}+\frac{18}{\epsilon ^2}-\frac{20}{\epsilon }-11+15 \epsilon +6 \epsilon ^2+4 \epsilon ^3\bigg)\!\bigg)\,p_1^\alpha
     \nonumber \\ &\quad
     +\bigg(\!\,\Gamma_3\bigg(\frac{2}{\eps^2}-\frac{14}{\eps}+14+4\eps+2\eps^2\!\bigg)
     +\frac{4\,\Gamma_2}{1\!-\!2\eps}\bigg(\!\!-\frac{2}{\epsilon ^3}+\frac{9}{\epsilon ^2}-\frac{9}{\epsilon }-6+5 \epsilon +4 \epsilon ^2+3 \epsilon ^3\!\bigg)\!\bigg)\,p_2^\alpha\bigg]\!\left[2\,\frac{p_1^\mu\,\slashed{k}}{t}+\gamma^\mu\right]\! \bigg\}v(p_2)\,. \label{1r2v_c-h_fact2}
\end{align}
\end{widetext}
We have combined the diagrams $(c)$ to $(h)$ rather than giving results per diagram pair, and have denoted combinations of gamma functions by 
\begin{equation}
    \Gamma_2 = \frac{\Gamma^3(1-\eps)\Gamma^2(1+\eps)}{\Gamma(3-3\eps)}\,,\qquad
    \Gamma_3 = \frac{\Gamma^4(1-\eps)\Gamma^2(1+\eps)}{\Gamma^2(2-2\eps)}\,, 
\end{equation}
the former coinciding with the second combination in \eq{eq:massiveGammaStructures}.

The results of \eq{1r2v_ab_fact2} and \eq{1r2v_c-h_fact2} are verified by calculating the hard-collinear region of the diagrams in \fig{1r2v}. Given that the calculation is set up in a similar way as for the massive case, we will not provide further details for the sake of brevity. In particular, we find agreement between the factorization and regions results per diagram pair $(a)\!+\!(b)$, $(c)\!+\!(d)$, $(e)\!+\!(f)$ and $(g)\!+\!(h)$. For the first three pairs we verified the exact agreement to all orders in $\epsilon$, while for the last pair we compared series expansions in $\eps$ instead. This is due to the crossed topology of diagrams $(g)$ and $(h)$, which complicates the regions calculation by entangling Feynman parameters, yielding hypergeometric functions of the form $_3F_2(a_1,a_2,a_3;b_1,b_2;1)$ upon integration. These multiply the second gamma function combination $\Gamma_2$ and are expanded up to and including finite terms $(\mathcal{O}\left(\eps^0\right))$ using \texttt{HypExp} \cite{Huber:2005yg,Huber:2007dx}. By expanding the (exact) coefficients of the $\Gamma_2$ combination in the factorization result up to the same order, we verified their consistency.

The $\Gamma_2$ combination is in fact the signature of the mixing between hard and collinear loop momenta: starting at two loops, it originates from terms in the $f\gamma$- and $f\partial\gamma$-hard functions that carry an additional factor of $x^{-\eps}$, as seen in \eq{H1_fg_explicit}. The appearance of those terms is, in turn, tied to an effective shift in the scale of the hard function; while the loop integration in $H_{(f)}$ knows only the single scale $2\,p_1^+p_2^- \!=\! s$, the $H_{(f\gamma)}$ and $H_{(f\partial\gamma)}$ functions are sensitive to the dominant component of the collinear photon through $2\,\ell_1^+p_2^-\! =\! x \,s,$ giving additional $x$ dependence.\footnote{The exact form of this $x$-dependence varies by diagram, as it is dictated by the denominators in the loop, and thus made explicit upon integration over the Feynman parameters that combine them.} Indeed, no $\Gamma_2$ combination is present for the $f$-jet factorization of diagrams $(a)\!+\!(b)$ in \eq{1r2v_ab_fact2}. As we see here, this effect is naturally captured by the NLP factorization formula of \eq{NLPfactorization}. A simplified NLP factorization as in \eq{simplifiedfactorization}, strictly in terms of $f$-jets, cannot do so: the complete factorization of the collinear and hard sector is an over-simplification of the intricate dynamics at play here. 

Lastly, we point out that \Eqns{1r2v_ab_fact2}{1r2v_c-h_fact2} contain at most $1/\eps^{3}$ poles, while at two-loop order a maximal soft-collinear overlap would generate $1/\eps^{4}$ poles. These leading singularities are captured by the soft function.\footnote{Upon a correct assignment of poles, they would appear in the hard-hard region instead, by the same mechanism discussed before.} This means that for NLP threshold resummation purposes the collinear sector is needed starting at NLL accuracy. This has been noted before in the calculation of the collinear region of the 1R1V and 2R1V correction to Drell-Yan production in \cite{Bonocore:2014wua} and \cite{Bahjat-Abbas:2018hpv} respectively. Indeed, \cite{Bahjat-Abbas:2019fqa} showed that NLP threshold logarithms in Drell-Yan and single Higgs production are resummed at LL accuracy through an exponential next-to-soft function.\footnote{Note that for the Drell-Yan process the resummation of NLP threshold logarithms at LL accuracy has been achieved before in Ref.~\cite{Beneke:2018gvs}}
 
\section{Conclusions} \label{sec:conclusions}

In this paper we have formulated a next-to-leading power factorization
formula for $n$-jet production processes in QED, based on a power
counting analysis for both zero and parametrically small fermion masses. We have thus begun the generalisation of the original Yukawa theory analysis of
\cite{Gervais:2017yxv} to gauge theories.
A factorization of degrees of freedom at NLP, following arguments
such as in \cite{Contopanagos:1996nh}, could be an important step
towards resummation of NLP (threshold) logarithms beyond
leading-logarithmic accuracy. 

We focused on the interaction that contributes at the first sub-leading
power in $\lambda$, and computed the $f\gamma$- and
$f\partial\gamma$-jets, which are universal quantities, up to order
$\lambda^2$. We first considered massive fermions, for which we calculated the non-radiative jet functions. To have a direct correspondence between the (next-to-)leading regions and their 
power counting we used axial gauge. 
We were able to test the factorization formula by
comparing the convolution of jet functions and hard parts with a 
regions calculation of the two-jet amplitude to one loop.
Subsequently we successfully tested these parts of the predicted
factorization formula at the two-loop level by comparing the combined result against the
hard-collinear region of a two-loop diagram. In particular, we pointed out the existence of two classes of endpoint contributions, one of which singular, that we deal with within dimensional regularisation.

For massless fermions we ensured the presence of a small scale, following \cite{Gervais:2017yxv}, by adding an extra soft photon emission, so that the 
invariant of this photon momentum
with a jet direction provides an analogue to the squared small fermion mass. We thus presented results for the radiative
$f\gamma$- and $f\partial\gamma$-jet instead and tested the
factorization of a radiative amplitude in a similar way as before. Here we found that the present approach reproduces all collinear contributions in the one-loop matrix element, contrary to the simplified NLP factorization of \cite{DelDuca:1990gz,Bonocore:2015esa}. In addition, we noted a subtle interplay between hard and collinear modes, which is correctly accounted for in our
factorization formula. We conjecture that our factorization
formula is sufficiently general to factorize QED
amplitudes up to NLP at arbitrary loop orders, and may thereby pave the way
for the development of a similar factorization for QCD amplitudes.

We focused in our analysis on testing a specific part of the 
factorization framework. That is, the one that accounts for non-trivial hard-collinear interplay at the two-loop level, via novel jet functions which describe double hard-collinear interactions. In addition, similar tests should be carried out for the other ingredients (such as the triple hard-collinear interactions and the soft sector). Moreover, the jet functions used in this analysis are extracted from a generic jet-like scattering amplitude, rather than being derived from an operator definition. Such definitions would make the gauge invariance of the separate factorization ingredients manifest, thereby formalizing the NLP factorization framework for QED. These further steps, together with the extension to QCD, are part of ongoing work.

\subsection*{Acknowledgments} \label{sec:acknowledgements}

EL acknowledges support from the Dutch NWO-I program 156, ``Higgs as Probe and Portal". JSD and WW are supported by the D-ITP consortium, a program of NWO funded by the Dutch Ministry of Education, Culture and Science (OCW). LV acknowledges support from the Fellini - Fellowship for Innovation at INFN, funded by the European Union's Horizon 2020 research programme under the Marie Sk\l{}odowska-Curie Cofund Action, grant agreement no. 754496. WW and LZ are supported by the ERC grant ERC-STG-2015-67732. This is also based upon work from COST Action CA16201 PARTICLEFACE supported by COST (European Cooperation in Science and Technology).
 
\appendix

\section{\boldmath Momentum regions for parametrically small fermion mass
\texorpdfstring{$m\sim\lambda Q$}{m~lambda*Q})}
\label{appx:regions}

The power-counting analysis in section 
\ref{sec:power counting} assumes
the scaling $k^\mu\sim Q\left(\lambda^2,\lambda^2,\lambda^2 \right)$,
$k^\mu\sim Q\left(1,\lambda, \lambda^2\right)$ for soft and collinear momenta respectively. 
This follows from an analysis of the infrared structure
of the scattering amplitude, which allows one to 
associate the pinch surfaces to momenta 
configurations that are soft and collinear.
Here we complement this analysis by 
performing an expansion of the amplitude in momentum 
regions. This method provides an alternative approach for 
singling out the momentum configurations which are relevant
for a given amplitude, in the presence of parametrically 
different scales, constituting a useful check for
our assumptions in section \ref{sec:power counting}.
It also gives us the opportunity to briefly discuss 
differences between the two approaches. 

To illustrate this second method, we focus on the scalar integral associated
to the 1-loop diagram in figure \ref{fig:1v}. 
For fermions with mass $m>0$ the integral reads
\begin{widetext}
\begin{align} \label{eq:oneLoopScalar} \nonumber
 T_m &= \int \!\frac{d^{4-2\eps}\ell}{(2\pi)^{4-2\eps}}\, 
 \frac{\mu^{2\eps}}{\big[\ell^2+i\eta\big]
    \big[(p_1-\ell)^2-m^2+i\eta\big]
    \big[(p_2+\ell)^2-m^2+i\eta\big]} 
    \\ \nn &
    =  \frac{i}{(4\pi)^2}  
\frac{e^{\eps \gamma_E} \Gamma(1+\eps)}{2\eps \, m^2}
\bigg( \frac{\bar \mu^2}{m^2}\bigg)^{\eps} \, 
_2F_1 \bigg(1,1+\eps,\frac{3}{2}; \frac{(\hat s+m^2)^2}{4 m^2\hat s} \bigg)\, \\ \nn
&=  - \frac{i}{(4\pi)^2} 
\frac{\hat s}{2(\hat s^2-m^4)} \Bigg\{
\left[\frac{2}{\eps} + 2 \log\bigg(\frac{\bar\mu^2}{2 m^2}\bigg)\right]
\log\left(-\frac{\hat s}{m^2}\right)
+\log^2\left(\frac{2\hat s}{\hat s-m^2}\right) \\ 
& \hspace{3.5cm}
-\,\log^2\left(\frac{2m^2}{m^2-\hat s}\right)
+2{\rm Li}_2\left(\frac{\hat s}{\hat s-m^2}\right)
-2{\rm Li}_2\left(\frac{m^2}{m^2-\hat s}\right)\Bigg\}\, ,
\end{align}
\end{widetext}
with $\bar{\mu}^2 = 4\pi e^{-\gamma_E} \mu^2$,
where in the last line we expand the result in powers
of $\eps$, showing that the integral
has a single soft pole. In the second and third line 
we write the result in terms of the variable $\hat s$, 
defined through 
\begin{align}\label{shatDef}
Q^2 &= (p_1+p_2)^2 
\nonumber\\&= 2m^2 + 2\,p_{1}^+ p_{2}^- 
+ 2\,p_{1}^- p_{2}^+   
\equiv 2m^2 + \hat s 
+ \frac{m^4}{\hat s},
\end{align}
where the momentum components $ p_i^{\pm}$ 
are given by the decomposition in 
\eq{LCdecomposition}. For small mass 
$m \ll \hat s  \sim Q^2$ we can expand 
 \eq{eq:oneLoopScalar}, 
obtaining
\begin{align}\label{massivetriangle1expandedb} \nn
&T_m = \frac{i}{(4\pi)^2}  
\frac{e^{\eps \gamma_E} \Gamma(1+\eps)}{\eps^2 \, \hat s}
\bigg\{- \bigg( \frac{\bar \mu^2}{m^2}\bigg)^{\eps} \bigg[ 
1+\frac{2}{1-\eps} \frac{m^2}{\hat s} \nonumber \\&\;+\frac{2-\eps+3\eps^2}{(1-\eps)(2-\eps)}\frac{m^4}{\hat s^2}
+ \ord \bigg(\frac{m^6}{\hat s^3} \bigg)\bigg]+ 
\bigg( \frac{\bar \mu^2}{-\hat s}\bigg)^{\eps} 
\frac{\Gamma^2(1-\eps)}{\Gamma(1-2\eps)}\nonumber \\&\;\times \bigg[ 
1+2\eps\frac{m^2}{\hat s} + (1+\eps+2\eps^2) \frac{m^4}{\hat s^2} 
+ \ord \bigg(\frac{m^6}{\hat s^3} \bigg)\bigg] \bigg\}.
\end{align}
Note that result in \eq{eq:oneLoopScalar} for finite, nonzero mass is free of collinear singularities, but exhibits mass thresholds when $\hat{s} = m^2$. When moving to the case of parametrically small masses by performing the mass expansion in \eq{massivetriangle1expandedb}, the branch cuts responsible for mass thresholds collapse to a point, which is manifest in the presence of a double pole at order $m^2/\hat{s}$. This indicates that the theory with parametrically small masses has collinear singularities analogous to the massless theory.\footnote{Alternatively, we may argue that a collinear singularity requires both $\ell^\mu \propto p_1^\mu$ and $\ell^2=0$, as dictated by the Landau equations. A large fermion mass violates these conditions as $p_1^2 = m^2$, but consistency is retrieved in the small mass limit. Again, this suggests that the singular structure for parametrically small fermion masses is comparable to that for massless fermions.}

We will now investigate which momentum
regions reproduce the result in 
\eq{massivetriangle1expandedb}, given that 
the external momenta are $p_{1}^+ \sim p_{2}^{-} \sim Q$, 
$p_{1}^{-} \sim p_{2}^{+} \sim \lambda^2 Q$, with the parameter
$\lambda$ fixed by the condition $\lambda \sim m/Q$.   
In principle several regions can be considered:
\begin{align} \label{eq:regions} \nn
   \mbox{hard:}                       \qquad & \ell^\mu  \,\sim\,  Q \left( 1, 1, 1 \right)\, , \\ \nn
   \mbox{semi-hard:}                 \qquad & \ell^\mu  \,\sim\,  Q \left( \lambda, \lambda, \lambda \right) \, , \\ \nn
   \mbox{collinear:}                 \qquad & \ell^\mu  \,\sim\,  Q \left( 1, \lambda, \lambda^2 \right) \, , \\ \nn
   \mbox{anti-collinear:}          \qquad & \ell^\mu  \,\sim \, Q \left( \lambda^2, \lambda, 1 \right)  \, , \\ 
   \mbox{soft:}                        \qquad & \ell^\mu  \,\sim \, Q \left( \lambda^2, \lambda^2, \lambda^2 \right) \, , \\ \nn
   \mbox{ultra-collinear:}        \qquad & \ell^\mu  \,\sim\,   Q \left( 1, \lambda^2, \lambda^4 \right) \, , \\ \nn
   \mbox{anti-ultra-collinear:}  \qquad & \ell^\mu  \,\sim \, Q \left( \lambda^4, \lambda^2, 1 \right)  \, , \\  \nn
   \mbox{ultra-soft:}                 \qquad & \ell^\mu  \,\sim \, Q \left( \lambda^4, \lambda^4, \lambda^4 \right),
\end{align} 
and in general one can have $n$-ultra-collinear regions with scaling  
$\ell^\mu  \,\sim\,   Q \left( 1, \lambda^{n}, \lambda^{2n} \right)$, with 
similarly defined $n$-ultra-anti-collinear and $n$-ultra-soft regions.
It is easy to check that only the hard, collinear and anti-collinear
region are not scaleless, each contributing as follows:
\begin{widetext}
\begin{align} \label{massivetriangle1h} 
\nonumber
T_{m}^{\rm h} &=\, \int 
\frac{d^{4-2\eps}\ell}{(2\pi)^{4-2\eps}} \, 
\frac{\mu^{2\eps}}{\ell^2 \, \big[\ell^2 + 2\ell^- p^{+}_1 \big] 
\big[\ell^2 - 2\ell^+ p^{-}_2 \big]}
\nonumber\\ &\quad\times
\bigg[ 1
-\frac{2\ell^+ p^{-}_1}{\ell^2 + 2\ell^- p^{+}_1}
+\frac{2\ell^- p^{+}_2}{\ell^2 - 2\ell^+ p_2^-}+\frac{(2\ell^+ p_1^-)^2}{\big[\ell^2 + 2\ell^- p_1^+\big]^2}
+\frac{(2\ell^+ p_1^-) (2\ell^- p_2^+)}{\big[\ell^2 + 2\ell^- p_1^+\big] \big[\ell^2 - 2\ell^+ p_2^-\big]}
+\frac{(2\ell^- p_2^+)^2}{\big[\ell^2 - 2\ell^+ p_2^-\big]^{2}}
+ \ord \bigg(\frac{m^6}{\hat s^3}\bigg) \bigg] \nonumber \\
&=\, \frac{i}{(4\pi)^2}  
\bigg( \frac{\bar \mu^2}{-\hat s}\bigg)^{\eps} 
\frac{1}{\hat s}
\frac{e^{\eps \gamma_E} \Gamma(1+\eps)\Gamma^2(1-\eps)}{\eps^2 \Gamma(1-2\eps)}
\bigg[ 1+2\eps\frac{m^2}{\hat s} + (1+\eps+2\eps^2) \frac{m^4}{\hat s^2} 
+ \ord \bigg(\frac{m^6}{\hat s^3} \bigg)\bigg] \nonumber\, \\
&=\, \frac{i}{(4\pi)^2}  \frac{1}{\hat s}
\left\{ \frac{1}{\eps^2} + \frac{1}{\eps}
\log \left(-\frac{\bar\mu^2}{\hat s} \right)
- \frac{\pi^2}{12}  + \frac{1}{2}
\log^2 \left(-\frac{\bar\mu^2}{\hat s} \right)
+\frac{m^2}{\hat s} \Bigg[
\frac{2}{\eps} 
+ 2 \log \left(-\frac{\bar\mu^2}{\hat s} \right)
\Bigg] \right. \nonumber \\ 
&+ \frac{m^4}{\hat s^2} \Bigg[\frac{1}{\eps^2}
+ \frac{1}{\eps} \bigg[1+
\log \left(-\frac{\bar\mu^2}{\hat s} \right) \bigg]
+2 - \frac{\pi^2}{12} 
+ \log \left(-\frac{\bar\mu^2}{\hat s} \right)  
+\frac{1}{2} \log^2 \left(-\frac{\bar\mu^2}{\hat s} \right) 
\Bigg]
+\ord(\eps) + \ord \bigg(\frac{m^6}{\hat s^3} \bigg) 
\Bigg\}\, , 
\end{align}
for the hard region, and 
\begin{align} \label{massivetriangle1c} 
\nonumber
T_{m}^{\rm c} &=\, \int 
\frac{d^{4-2\eps}\ell}{(2\pi)^{4-2\eps}}\, 
\frac{\mu^{2\eps}}{\ell^2 \, \big[\ell^2 + 2\ell^- p_1^{+}+ 2\ell^+ p_1^{-} \big] 
\big[ - 2\ell^+ p_2^{-} \big]} 
 \bigg[1-  \frac{\ell^2}{- 2\ell^+ p^{-}_2 }
+ \frac{ 2\ell^- p^{+}_2}{-2 \ell^+ p^{-}_2} +\frac{\ell^4}{\big[-2 \ell^+ p^{-}_2 \big]^2}
+ \ord \bigg(\frac{m^6}{\hat s^3}\bigg) \bigg]  \nonumber \\ 
&= - \frac{i}{(4\pi)^2}  
\frac{e^{\eps \gamma_E} \Gamma(1+\eps)}{2\eps^2 \, \hat s}
\bigg( \frac{\bar \mu^2}{m^2}\bigg)^{\eps} \bigg[ 
1+\frac{2}{1-\eps} \frac{m^2}{\hat s} +\frac{2-\eps+3\eps^2}{(1-\eps)(2-\eps)}\frac{m^4}{\hat s^2}
+ \ord \bigg(\frac{m^6}{\hat s^3} \bigg)\bigg] \nonumber \\
&=\ \frac{i}{(4\pi)^2}  \frac{1}{\hat s}
\Bigg\{ -\frac{1}{\eps^2} 
-\frac{1}{\eps}\log \left(\frac{\bar\mu^2}{m^2} \right)
- \frac{\pi^2}{12} -\frac{1}{2}\log^2 \left(\frac{\bar\mu^2}{m^2} \right)
+ \frac{m^2}{\hat s} \Bigg[ - \frac{2}{\eps} - 2 
- 2 \log \left(\frac{\bar\mu^2}{m^2} \right) \Bigg]
\nonumber \\ 
&
\quad+\frac{m^4}{\hat s^2}\Bigg[ -\frac{1}{\eps^2}
- \frac{1}{\eps}\bigg[1 + \log \left(\frac{\bar\mu^2}{m^2} \right) \bigg]
-\frac{5}{2} -\frac{\pi^2}{12} 
- \log \left(\frac{\bar\mu^2}{m^2} \right) 
- \frac{1}{2}\log^2 \left(\frac{\bar\mu^2}{m^2} \right)
\Bigg] +\ord(\eps) + \ord \bigg(\frac{m^6}{\hat s^3} \bigg) 
\Bigg\}\, , 
\end{align}
\end{widetext}
for the collinear region. The
anti-collinear region is identical, $T_{m}^{\rm \bar c} = T_{m}^{\rm c}$.
None of the other regions give a contribution, because they are 
of the form \vspace*{-1pt}
\begin{align}
T_{m}^{\rm  uc,\overline{uc} } &= \int 
\frac{d^{4-2\eps}\ell}{(2\pi)^{4-2\eps}}\, \frac{\mu^{2\eps}
\{ 1, \ell^{\mu}, \ell^{\mu} \ell^{\nu}, \ldots \}}{\ell^2 \, \big[2\ell^{\mp} p_1^{ \pm} \big] 
\big[-2\ell^{\mp} p_2^{ \pm} \big]}\,,
\nonumber\\
T_{m}^{\rm sh,s,us} &= \int 
\frac{d^{4-2\eps}\ell}{(2\pi)^{4-2\eps}} \, \frac{\mu^{2\eps}
\{ 1, \ell^{\mu}, \ell^{\mu} \ell^{\nu}, \ldots \}}{\ell^2 \, \big[2\ell^{-} p_{1}^+ \big]
\big[- 2\ell^{+} p_{2}^- \big]} \, ,
\end{align}
which are scaleless.
This analysis allows us to conclude that 
the relevant collinear region (third line in 
\eq{eq:regions}) has indeed the same scaling 
as the collinear momentum in \eq{softcol}, 
whose scaling has been determined by 
investigating the pinch surfaces of the 
amplitude. 

 \section{\boldmath
 Intermediate expressions for elastic amplitudes with
 \texorpdfstring{$m\sim \lambda Q$}{m ~ lambda*Q}} 
 \label{appx:intermediate}

 In this appendix we collect intermediate expressions needed for the calculations performed in \sect{sec:MassiveCase}. Specifically, in section~\ref{appx:jet_integrals} we list results for the integrals needed in the computation of the $f\gamma$- and $f\partial\gamma$-jet, in the massive theory, while in section~\ref{appx:inter2loop}, we show the partial results for the two-loop check of factorization performed there.

\subsection{Integrals for jet functions} \label{appx:jet_integrals}

For the calculation of the jet functions in the massive fermion case, we need
\begin{widetext}
\vspace*{20pt}
 \begin{subequations} \label{masterIntegrals}\begin{align}
    \PSellRed\frac{(i\mu^{2\eps}p^+)}{[2xp^+\ell^- \!+ \ell_\perp^2 \!+ i\eta]\,[2(1-x)\ell^- p^+ \!-\ell_\perp^2\!+xm^2\!-i\eta]} \!&=\! I_0\,,
     \\
    \PSellRed\frac{(i\mu^{2\eps}p^+)\,\ell^\alpha_\perp\ell^\beta_\perp}{[2xp^+\ell^- \!+ \ell_\perp^2 \!+ i\eta]\,[2(1-x)\ell^- p^+ \!-\ell_\perp^2\!+xm^2\!-i\eta]} \!&= \!\frac{x^2m^2\eta_\perp^{\alpha\beta}}{2\!-\!2\eps}I_0\,,
    \\ 
    \PSellRed\frac{(i\mu^{2\eps}p^+)\,\ell^-}{[2xp^+\ell^- \!+ \ell_\perp^2 \!+ i\eta]\,[2(1-x)\ell^- p^+ \!-\ell_\perp^2\!+xm^2\!-i\eta]} \!&=\!\frac{m^2}{2p^+}\bigg[\frac{\delta(1-x)}{1-\eps}-x\bigg]I_0\, ,
\end{align}\end{subequations}
\end{widetext}
with a common factor
\beq
    I_0 = \frac{\Gamma(\eps)}{16\pi^2}\bigg(\frac{4\pi\mu^2}{x^2m^2}\bigg)^{\eps}\,.
\eeq
Note the different signs for the $i\eta$ prescriptions in the denominators. As a consequence, the poles lie on opposite sides of the $\ell^-$ integration contour if and only if $0<x<1$, which restricts the convolution domain in $x$ to that range.

\subsection{Partial two-loop results} \label{appx:inter2loop}

In the following we show expressions for the convolution of the jet and hard functions, that serve as a two-loop check of the result obtained in \sect{sec:twoLoopMassiveTest}. Here we list the $f\gamma$- and $f\partial\gamma$-terms separately:
\begin{widetext}
\begin{align} \label{eq:convolutionsFinal}
    &\int_0^1\!\! d x\, J^{(1)}_{(f\gamma)\h\nu}(x,p_1)\,
    H_{(f\gamma)}^{(1)\h\alpha\nu}(x,p_1,p_2)\,J^{(0)}_{(f)}(p_2) 
   \;=\;
   \frac{i\,e^5}{128\,\pi^4}\left(\!\frac{-2\,p_1\!\cdot\! \bar{n} \,p_2\!\cdot\!n\,}{4\pi\mu^2}\!\right)^{-\eps}\!\left(\!\frac{m^2}{4\pi\mu^2}\!\right)^{-\eps}\,\frac{\bar{u}(p_1)}{1-2\eps} \nonumber\\ &\quad\times
    \bigg\{\left(\frac{m}{p_1\!\cdot\!\bar{n}}\bar{n}^\alpha-\frac{m}{p_2\!\cdot\!n}n^\alpha\right)\,\Gamma_1
    \left(\frac{1}{\eps^3}+\frac{2}{\eps^2}-\frac{3}{\eps}\right)
    + \frac{m}{p_2\!\cdot\!n}\,\Gamma_2
    \left(\frac{2}{\eps^3}-\frac{1}{\eps^2}-\frac{8}{\eps}+11-4\eps\right)n^\alpha
    \nonumber \\
    &\quad
    -\frac{m}{p_1\!\cdot\!\bar{n}}\,\Gamma_2
    \left(\frac{4}{\eps^3}-\frac{8}{\eps^2}+\frac{1}{\eps}+3\right)\bar{n}^\alpha
    +\frac{m^2}{2\,p_1\!\cdot\! \bar{n}\,p_2\!\cdot n}\gamma^\alpha\bigg[
    \frac{\Gamma_1}{(1-\eps^2)(1-\eps)}
    \left(\frac{6}{\eps^4}-\frac{9}{\eps^3}+\frac{2}{\eps^2}-16+33\eps-8\eps^2-4\eps^3\right)
    \nonumber\\ &\quad-\frac{\Gamma_2}{1-\eps^2}
    \left(\frac{8}{\eps^4}-\frac{12}{\eps^3}-\frac{30}{\eps^2}+\frac{112}{\eps}-158+92\eps+4\eps^2-8\eps^3\right)\bigg]\bigg\}v(p_2)\,,
    \\
    &\int_0^1\!\! d x\, J^{(1)}_{(f\partial\gamma)\h\nu\rho}(x,p_1)\,
    H_{(f\partial\gamma)}^{(1)\h\alpha\nu\rho}(x,p_1,p_2)\,J^{(0)}_{(f)}(p_2) 
    \; = \;
    -\frac{i\,e^5}{128\,\pi^4}\left(\!\frac{-2\,p_1\!\cdot\! \bar{n} \,p_2\!\cdot\!n\,}{4\pi\mu^2}\!\right)^{-\eps}\!\left(\!\frac{m^2}{4\pi\mu^2}\!\right)^{-\eps}\frac{\bar{u}(p_1)\gamma^\alpha v(p_2)}{1-2\eps}\nonumber\\ &\quad \times\frac{m^2}{2\,p_1\!\cdot\! \bar{n}\,p_2\!\cdot n}\bigg\{
    \frac{\Gamma_1}{(1-\eps^2)(1-\eps)}
    \left(\frac{6}{\eps^4}-\frac{9}{\eps^3}-\frac{1}{\eps^2}+\frac{11}{\eps}-13+8\eps-2\eps^2+4\eps^3\right)\nonumber\\
    &\quad-\frac{\Gamma_2}{(1-\eps^2)(1+\eps)}
    \left(\frac{8}{\eps^4}-\frac{2}{\eps^3}-\frac{41}{\eps^2}+\frac{67}{\eps}-3-81\eps+28\eps^2+24\eps^3+16\eps^4\right)\bigg\}\,.
\end{align}
\end{widetext}
 
 \section{\boldmath 
 Intermediate expressions for radiative amplitudes \texorpdfstring{$m=0$}{m=0}}
 \label{appx:massless_case}
 
 In this appendix we collect intermediate expressions needed for the calculations performed in \sect{sec:rad_massless}. Section \ref{appx:radjet_integrals} lists integrals that enter the calculation of the one-loop radiative $f\gamma$- and $f\partial\gamma$-jet functions, for massless fermions. In section~\ref{appx:hard_functions}, we present one-loop expressions for the corresponding hard functions.   
 
\subsection{Integrals for the radiative jet functions}\label{appx:radjet_integrals}
 
In the calculation of the radiative, massless $f\gamma$- and $f\partial\gamma$-jet we expand denominators in $\lambda$. To keep expressions compact, we define the following notation for the homogeneous propagator denominators appearing in the diagrams of \fig{rad_fgamma_jet}
\begin{align}
    D_1 &= 2 \,x \,p^+\ell^-+\ell_\perp^2+i\eta\,, \nonumber \\ 
    D_2 &= 2(1\!-\!x)p^+\ell^- -\ell_\perp^2-i\eta\,, \nonumber \\ 
    D_3 &= 2(1\!-\!x)p^+\ell^- -\ell_\perp^2-2(1\!-\!x)p^+k^--i\eta\,. \nonumber
\end{align}
For diagram $(a)$ in \fig{rad_fgamma_jet} we need
\begin{subequations}\begin{align}
 &\hspace{-5pt}(i\mu^{2\eps}p^+) \PSellRed \frac{1}{D_1\,D_3^a} = x^{-\eps}\,(1-x)^{1-a-\eps} I_1(a),  \label{mInta}
 \\
  &\hspace{-5pt}(i\mu^{2\eps}p^+)  \PSellRed \frac{\ell_\perp^\alpha\ell_\perp^\beta}{D_1\,D_3^a}
  \nonumber \\ 
  & \qquad = -\frac{p^+k^-}{2-a-\eps}\eta_\perp^{\alpha\beta}x^{1-\eps}\,\,(1-x)^{2-a-\eps}\,I_1(a)\,, \label{mIntaperp} \\ &\hspace{-5pt}(i\mu^{2\eps}p^+)\PSellRed \frac{\ell^-}{D_1\,D_3^a} =
  -\frac{k^-\,(1-x)^{2-a-\eps}}{2-a-\eps}
  \nonumber\\
  &\qquad\quad\times
  \label{mIntaminus} \left(\frac{1}{x}\,\delta(1-x)-(1-\eps)\,x^{-\eps}\right)I_1(a)\,,
\end{align}\end{subequations} 
with
\begin{equation}
    I_1(a) = \frac{1}{16\pi^2}\left(\frac{-2p^+k^-}{4\pi\mu^2}\right)^{-\eps}\frac{\Gamma(a-1+\eps)}{\Gamma(a)}(-2p^+k^-)^{1-a}\,.
\end{equation}
For diagram $(b)$ in \fig{rad_fgamma_jet} a set of slightly more involved integrals is needed:
 \begin{subequations}
 \begin{align}
  &(i\mu^{2\eps}p^+) \!\PSellRed \frac{1}{D_1\,D_2\,D_3^a}&
  \nonumber\\
  &\qquad =-\frac{1}{\eps}(1-x)^{-a-\eps}x^{-\eps}I_2(a)\,,  \label{mIntb}\\
  &(i\mu^{2\eps}p^+)  \!\PSellRed \frac{\ell_\perp^\alpha\ell_\perp^\beta}{D_1\,D_2\,D_3^a} &
  \nonumber\\
  &\qquad = -\frac{p^+k^-}{1-a-\eps}\frac{\eta_\perp^{\alpha\beta}}{1-\eps}x^{1-\eps}(1-x)^{1-a-\eps}\,I_2(a)\,, \label{mIntbperp} \\ &(i\mu^{2\eps}p^+)\!\PSellRed\! \frac{\ell^-}{D_1\,D_2\,D_3^a} = \frac{-k^-(1-x)^{1-a-\eps}}{(1-a-\eps)(1-\eps)}\label{mIntbminus}\nonumber\\ 
  &\qquad\quad\times\left(\frac{1}{x}\,\delta(1-x)-(1-\eps)\,x^{-\eps}\right)I_2(a)\,,
\end{align}
\end{subequations}
with
\begin{equation}
    \hspace{-3pt}I_2(a) =  \frac{1}{16\pi^2}\!\left(\frac{-2p^+k^-}{4\pi\mu^2}\right)^{-\eps}\!\frac{\Gamma(a+\eps)}{\Gamma(a)}(-2p^+k^-)^{-a}.
\end{equation}

 \subsection{One-loop hard functions} \label{appx:hard_functions}
 
Below we collect expressions for the one-loop $f\gamma$- and $f\partial\gamma$-hard functions used in the main text. We extract these functions from the diagrams shown in \fig{Hfgamma_NLO}, according to \eq{Hfgammadef}. For the $f\gamma$-hard function defined by diagram $(\rm I)$, the simplest topology, we will quote an explicit result to give an impression of the form of these functions. For the remaining contributions to the $f\gamma$- and $f\partial\gamma$-hard functions we give expressions  prior to any processing for brevity. The evaluation itself is a simple one-loop calculation that relies on standard techniques, but the resulting expressions are rather lengthy due to the numerous open indices. We obtain  
\begin{widetext}
 \begin{align}
\hspace{-6pt} H^{(1)\alpha}_{(f\gamma\,|\,\rm I)\,\nu}(x,p_1,p_2) =& -\frac{e^4}{x\,s}\! \int\! \frac{d^{4-2\eps}\ell_2}{(2\pi)^{4-2\eps}}\frac{N_{\rm I}^\alpha (x\,\slashed{p}_1+\slashed{p}_2)\gamma_\nu}{\left[\ell_2^2+2\,\ell_2\cdot p_1\right]\!\left[\ell_2^2+2x\,\ell_2\cdot p_1\right]\!\left[\ell_2^2-2\,\ell_2\cdot p_2\right]}, \\
 \hspace{-6pt}H^{(1)\alpha}_{(f\partial\gamma\,|\,\rm I)\,\nu\rho}(x,p_1,p_2) =& -\frac{e^4}{x\,s} \!\int\! \frac{d^{4-2\eps}\ell_2}{(2\pi)^{4-2\eps}}\frac{N_{\rm I}^\alpha\left[\gamma_{\perp\rho}-2\,\ell_{2\perp\rho}\,\frac{x\,\slashed{p}_1+\slashed{p}_2}{\ell_2^2+2x\,\ell_2\cdot p_1}\right]\gamma_\nu}{\left[\ell_2^2+2\,\ell_2\cdot p_1\right]\!\left[\ell_2^2+2x\,\ell_2\cdot p_1\right]\!\left[\ell_2^2-2\,\ell_2\cdot p_2\right]}, \\
 N_{\rm I}^\alpha=&\ (D-4)(\slashed{\ell}_2-\slashed{p}_2)\gamma^\alpha(\slashed{\ell}_2+\slashed{p}_1)+2(\slashed{\ell}_2+\slashed{p}_1)\gamma^\alpha(\slashed{\ell}_2-\slashed{p}_2), \nonumber
 \end{align}
 \end{widetext}
 after some Dirac algebra. Any perpendicular quantity $a_{\perp}^\rho$ can be rewritten as 
 \vspace{-2pt}\begin{align}
 a_{\perp}^\rho &= a^\rho - a\cdot \bar{n}\, n^\rho - a\cdot n \,\bar{n}^\rho \nonumber \\
 &= a^\rho - 2\frac{a\cdot p_2}{s}p_1^\rho - 2\frac{a\cdot p_1}{s}p_2^\rho,
 \end{align} 
 such that the loop integral can be carried out using standard integrals. Anticipating the contraction with the lowest order $f$-jet on the $p_2$ leg, we obtain a reasonably compact expression for the combination
 \begin{widetext}
 \begin{align}
    &H^{(1)\alpha}_{(f\gamma\,|\,\rm I)\,\nu}(x,p_1,p_2)J^{(0)}_{(f)}(p_2) 
    = \frac{i\, e^4}{8\pi^2}\frac{1}{s}\left(\!\frac{-s}{4\pi\mu^2}\!\right)^{-\eps}\frac{\Gamma^2(\ome)\Gamma(1\!+\!\eps)}{\Gamma(2\!-\!2\eps)}\frac{1}{1\!-\!x}
    \bigg\{\!\!\left(\frac{1}{\eps}-1\!\right)\!\bigg[\!\left(x^{-\eps}-1\right)p_2^\alpha +\left(1-x^{1-\eps}\right)p_1^\alpha\bigg]\gamma_\nu
    \nonumber\\
    &\quad+2\bigg[\left(
    \frac{1}{x}\left(-\frac{1}{\eps}+1\right)+\frac{2}{\eps}-1-\frac{x^{1-\eps}}{\eps}\right)p_1^\alpha+\bigg(\!\left(-\frac{2}{\eps^2}+\frac{2}{\eps}\right)x^{-1-\eps}+\frac{1}{x}\left(\frac{2}{\eps^2}+1\right)-\frac{3}{\eps}x^{-\eps}+\frac{1}{\eps}-1\bigg)p_2^\alpha\bigg]
    \frac{\slashed{p}_1\,p_{2\h\nu}}{s}
    \nonumber \\ 
    &\quad+\bigg[\left(\frac{2}{\eps^2}-\frac{2}{\eps}\right)x^{-1-\eps}-\left(\frac{2}{\eps^2}-\frac{1}{\eps}+2\right)\frac{1}{x} +\left(\frac{1}{\eps}+2\right)x^{-\eps}\bigg]p_{2\nu}\gamma^\alpha+\left[\left(\frac{1}{\eps}-1\right)x^{-\eps}-\frac{3}{2\eps}+1+\frac{x^{1-\eps}}{2\eps}\right]\slashed{p}_1\gamma^\alpha\gamma_\nu\bigg\}v(p_2). \label{H1_fg_explicit}
 \end{align}
\end{widetext}
We stress that the inverse powers of $1\!-\!x$ and $x$ present here, are associated to soft-collinear singularities caused by either the fermion or photon becoming soft in addition to being collinear. These endpoint singularities in the convolution variable are regulated by the $f\gamma$-jet through the overall factor $\left[x\,\omx\right]^{-\eps}$ in \eq{rjetfgamma}.  
The second diagram in \fig{Hfgamma_NLO} provides us with
\begin{widetext}
 \begin{align}
H^{(1)\alpha}_{(f\gamma\,|\,\rm II)\,\nu}(x,p_1,p_2) =&\,-e^4 \int\! \frac{d^{4-2\eps}\ell_2}{(2\pi)^{4-2\eps}}\frac{\gamma^\sigma(\slashed{\ell}_2+\omx\slashed{p}_1)\gamma_\nu(\slashed{\ell}_2+\slashed{p}_1)\gamma^\alpha(\slashed{\ell}_2-\slashed{p}_2)\gamma_\sigma}{\left[\ell_2^2+2\,\ell_2\cdot p_1\right]\left[\ell_2^2+2\omx\,\ell_2\cdot p_1 \right]
\left[\ell_2^2\right]\left[\ell_2^2-2\,\ell_2\cdot p_2\right]}\, , 
\\
H^{(1)\alpha}_{(f\partial\gamma\,|\,\rm II)\,\nu\rho}(x,p_1,p_2) =&\,-e^4 \int\! \frac{d^{4-2\eps}\ell_2}{(2\pi)^{4-2\eps}}\frac{\gamma^\sigma\!\left[2\ell_{2\perp\rho}\,\frac{\slashed{\ell}_2+\omx\slashed{p}_1}{\ell_2^2+2\omx\,\ell_2\cdot p_1}-\gamma_{\perp\rho}\right]
\gamma_\nu(\slashed{\ell}_2+\slashed{p}_1)\gamma^\alpha(\slashed{\ell}_2-\slashed{p}_2)\gamma_\sigma}{\left[\ell_2^2+2\,\ell_2\cdot p_1\right]\left[\ell_2^2+2\omx\,\ell_2\cdot p_1 \right]
\left[\ell_2^2\right]\left[\ell_2^2-2\,\ell_2\cdot p_2\right]}\, ,
\end{align}
while the third diagram gives
\begin{align}
 H^{(1)\alpha}_{(f\gamma\,|\,\rm III)\,\nu}(x,p_1,p_2) =&-e^4 \int\! \frac{d^{4-2\eps}\ell_2}{(2\pi)^{4-2\eps}}\frac{\gamma^\sigma(\slashed{\ell}_2+\slashed{p}_1)\gamma^\alpha(\slashed{\ell}_2-\slashed{p}_2)\gamma_\nu (\slashed{\ell}_2+x\,\slashed{p}_1-\slashed{p}_2)\gamma_\sigma}{\left[\ell_2^2+2\,\ell_2\cdot p_1\right]\left[\ell_2^2-2\,\ell_2\cdot p_2\right]
 \left[\ell_2^2+2x\,\ell_2\cdot p_1\right]\left[\ell_2^2+2x\,\ell_2\cdot p_1 -2\,\ell_2\cdot p_2- x\,s\right]}\, , \\
 H^{(1)\alpha}_{(f\partial\gamma\,|\,\rm III)\,\nu\rho}(x,p_1,p_2) =&-e^4 \int\! \frac{d^{4-2\eps}\ell_2}{(2\pi)^{4-2\eps}}\frac{\gamma^\sigma(\slashed{\ell}_2+\slashed{p}_1)\gamma^\alpha(\slashed{\ell}_2-\slashed{p}_2)\gamma_\nu}{\left[\ell_2^2+2\,\ell_2\cdot p_1\right]\left[\ell_2^2-2\,\ell_2\cdot p_2\right]}\nonumber \\  &\times\frac{\left[\gamma_{\perp\rho}-2\,\ell_{2\perp\rho}(\slashed{\ell}_2+x\,\slashed{p}_1-\slashed{p}_2)\,\left(\frac{1}{\ell_2^2+2x\,\ell_2\cdot p_1}+\frac{1}{\ell_2^2+2x\,\ell_2\cdot p_1 -2\,\ell_2\cdot p_2- x\,s}\right)\right]\gamma_\sigma}{\left[\ell_2^2+2x\,\ell_2\cdot p_1\right]\left[\ell_2^2+2x\,\ell_2\cdot p_1 -2\,\ell_2\cdot p_2- x\,s\right]}\,.
\end{align}\vspace{10pt}
\end{widetext}

\section{Results for Yukawa theory}
\label{appx:Yukawa}

As a byproduct of our studies, we obtained results for Yukawa theory in presence of parametrically small fermion masses, analogous to the case of massive QED considered in \sect{sec:MassiveCase}. Although this is not our main focus, the jet functions that we computed are a nontrivial generalisation of some of the results presented in~\cite{Gervais:2017yxv}, so we briefly report our findings.

The vertex content of Yukawa theory is the same as QED, with photons replaced by scalars. In fact, following~\cite{Gervais:2017yxv}, we will consider pseudoscalars (rather than scalars). The power counting procedure, extensively described in \sect{sec:detailedPC}, also applies step by step to Yukawa theory. In fact, at the level of Feynman rules, only the scaling of the fermion-scalar vertex is altered: the emission of a scalar with momentum $k$ from a collinear fermion line with momentum $p$ contributes with 
\begin{equation} \label{eq:effectiveRuleYukawa}
    \left(\slashed{p}-\slashed{k}\right)\gamma_5\slashed{p} = (-p^2+\slashed{k}\slashed{p})\gamma_5\,.
\end{equation}

The first term then scales as $\lambda^2$, while the second one is $\ord(\lambda)$ when $k$ is collinear and $\ord(\lambda^2)$ when this is soft. This causes an enhancement of at least one power of $\lambda$ with respect to the naive scaling, which is predicted to be $\ord(\lambda^0)$ when only propagators are accounted for. As in QED (\eq{QEDem}), this effective enhancement follows from $(\gamma^-)^2 = 0$. However, different from massless QED, the suppression occurs for both soft and collinear emissions. The consequent collinear power counting is unaltered, while the scaling in \eq{eq:effectiveRuleYukawa} affects the connections between soft and collinear subgraphs. Following the QED analysis, we obtain
\begin{subequations} \label{eq:YPC_final}
\begin{align}
  \gamma_{\mathcal{G}} =&\, 2m_s + 3m_f \hspace{109pt} (m=0) \nonumber \\ &+ \sum_{i=1}^{n} (N_s^{(i)} + N_f^{(i)} +n_s^{(i)}+ 3n_f^{(i)} -1) \\
  \gamma_{\mathcal{G}} =&\, I_f + 2 m_s + 4 m_f +\hspace{75pt}  (m \neq 0)\nonumber  \\ &+ \sum_{i=1}^{n}(N_s^{(i)}+N_f^{(i)}+n_s^{(i)}+3n_f^{(i)}-1)\,,
\end{align}
\end{subequations}
where the subscript $s$ identifies scalar particles. This reproduces the results derived in~\cite{Akhoury:1978vq} and~\cite{Gervais:2017yxv} for respectively the massless and massive case.

The NLP factorization formula for the collinear sector of Yukawa theory has the same structure as \eq{NLPfactorization}, and simply requires relabelling $\gamma \to s$. In particular, we focused on the fermion-scalar term
\begin{align} \label{eq:YukawaFactorization}
M_{(fs)} &= \sum_{i=1}^n \!\bigg(\!\prod_{j\neq i}J_{(f)}^j\!\bigg)\!
    \Big[J^i_{(fs)} \otimes H^i_{(fs)}+J^i_{(f\partial s)} \otimes H^i_{(f\partial s)}\Big]S
\end{align}
and extracted the jet functions $J^i_{(fs)}$ and $J^i_{(f\partial s)}$ from the convolution with a generic hard function. The calculation follows step by step the one presented in \sect{massiveFgamma}. In particular, the integrals in \eq{masterIntegrals} suffice to obtain the result, and one needs to carefully include endpoint contributions. We obtain
\begin{align} \label{eq:fs_true_result}
  \hspace{-3pt}J_{(fs)}(x) &= - \frac{g\, m}{16\pi^2}\Big(\frac{m^2}{4\pi\mu^2}\Big)^{-\eps} \Gamma(\eps)\,\bar{u}(p)
  \\&\quad\times \bigg\{x^{1-2\eps}-\frac{m}{p^+}\bar{n}\bigg[x^{1-2\eps}-\frac{\delta(1-x)}{2-2\eps}\bigg]\bigg\}\gamma_5\,, \nonumber\\
    \hspace{-3pt}J_{(f\partial s)}^\rho(x) &= \frac{g\, m^2}{16\pi^2} \Big(\frac{m^2}{4\pi\mu^2}\Big)^{-\eps}\Gamma(\eps)\,\bar{u}(p)\,\frac{x^{2-2\eps}}{2-2\eps}\,\gamma_\perp^\rho \gamma_5\,,
\end{align}
where $g$ is the coupling constant of the theory, and the notation is otherwise the same as for the QED massive jet functions in \eq{massiveJetFunctionsResult}. The $\ord(\lambda)$ result in the $f s$-function agrees with~\cite{Gervais:2017yxv}, where the $\lambda^2$ correction we computed is needed to appreciate the interplay with the $f\partial s$-function we derived. As for QED, we remark that a full treatment of the collinear sector at this order would require including $fss$- and $f\!f\!f$-jets, which however start contributing at two-loop order. Similar to \sect{sec:twoLoopMassiveTest}, we validated the factorization formula~\eqref{eq:YukawaFactorization} using the method of regions. To this end, we expanded the two-loop diagram analogous to \fig{twoLoopMassive} in the hard-collinear region, where now photons are replaced by scalars, and verified that such a region is reproduced by the convolution between the jet presented in \eq{eq:YukawaFactorization} and the hard functions. We thus provided a check of the formalism of \cite{Gervais:2017yxv} beyond one loop and beyond $\ord(\lambda)$.

\bibliographystyle{asprev4-1}
\bibliography{refs}

\begin{thebibliography}{101}%
\makeatletter
\providecommand \@ifxundefined [1]{%
 \@ifx{#1\undefined}
}%
\providecommand \@ifnum [1]{%
 \ifnum #1\expandafter \@firstoftwo
 \else \expandafter \@secondoftwo
 \fi
}%
\providecommand \@ifx [1]{%
 \ifx #1\expandafter \@firstoftwo
 \else \expandafter \@secondoftwo
 \fi
}%
\providecommand \natexlab [1]{#1}%
\providecommand \enquote  [1]{``#1''}%
\providecommand \bibnamefont  [1]{#1}%
\providecommand \bibfnamefont [1]{#1}%
\providecommand \citenamefont [1]{#1}%
\providecommand \href@noop [0]{\@secondoftwo}%
\providecommand \href [0]{\begingroup \@sanitize@url \@href}%
\providecommand \@href[1]{\@@startlink{#1}\@@href}%
\providecommand \@@href[1]{\endgroup#1\@@endlink}%
\providecommand \@sanitize@url [0]{\catcode `\\12\catcode `\$12\catcode
  `\&12\catcode `\#12\catcode `\^12\catcode `\_12\catcode `\%12\relax}%
\providecommand \@@startlink[1]{}%
\providecommand \@@endlink[0]{}%
\providecommand \url  [0]{\begingroup\@sanitize@url \@url }%
\providecommand \@url [1]{\endgroup\@href {#1}{\urlprefix }}%
\providecommand \urlprefix  [0]{URL }%
\providecommand \Eprint [0]{\href }%
\providecommand \doibase [0]{http://dx.doi.org/}%
\providecommand \selectlanguage [0]{\@gobble}%
\providecommand \bibinfo  [0]{\@secondoftwo}%
\providecommand \bibfield  [0]{\@secondoftwo}%
\providecommand \translation [1]{[#1]}%
\providecommand \BibitemOpen [0]{}%
\providecommand \bibitemStop [0]{}%
\providecommand \bibitemNoStop [0]{.\EOS\space}%
\providecommand \EOS [0]{\spacefactor3000\relax}%
\providecommand \BibitemShut  [1]{\csname bibitem#1\endcsname}%
\let\auto@bib@innerbib\@empty
\bibitem [{\citenamefont {Yennie}\ \emph {et~al.}(1961)\citenamefont {Yennie},
  \citenamefont {Frautschi},\ and\ \citenamefont {Suura}}]{Yennie:1961ad}%
  \BibitemOpen
  \bibfield  {author} {\bibinfo {author} {\bibfnamefont {D.}~\bibnamefont
  {Yennie}}, \bibinfo {author} {\bibfnamefont {S.~C.}\ \bibnamefont
  {Frautschi}}, \ and\ \bibinfo {author} {\bibfnamefont {H.}~\bibnamefont
  {Suura}},\ }\href {\doibase 10.1016/0003-4916(61)90151-8} {\bibfield
  {journal} {\bibinfo  {journal} {Annals Phys.}\ }\textbf {\bibinfo {volume}
  {13}},\ \bibinfo {pages} {379} (\bibinfo {year} {1961})}\BibitemShut
  {NoStop}%
\bibitem [{\citenamefont {Grammer}\ and\ \citenamefont
  {Yennie}(1973)}]{Grammer:1973db}%
  \BibitemOpen
  \bibfield  {author} {\bibinfo {author} {\bibfnamefont {J.}~\bibnamefont
  {Grammer}, \bibfnamefont {G.}}\ and\ \bibinfo {author} {\bibfnamefont
  {D.}~\bibnamefont {Yennie}},\ }\href {\doibase 10.1103/PhysRevD.8.4332}
  {\bibfield  {journal} {\bibinfo  {journal} {Phys. Rev. D}\ }\textbf {\bibinfo
  {volume} {8}},\ \bibinfo {pages} {4332} (\bibinfo {year} {1973})}\BibitemShut
  {NoStop}%
\bibitem [{\citenamefont {Bodwin}(1985)}]{Bodwin:1984hc}%
  \BibitemOpen
  \bibfield  {author} {\bibinfo {author} {\bibfnamefont {G.~T.}\ \bibnamefont
  {Bodwin}},\ }\href {\doibase 10.1103/PhysRevD.34.3932} {\bibfield  {journal}
  {\bibinfo  {journal} {Phys. Rev. D}\ }\textbf {\bibinfo {volume} {31}},\
  \bibinfo {pages} {2616} (\bibinfo {year} {1985})},\ \bibinfo {note}
  {[Erratum: Phys.Rev.D 34, 3932 (1986)]}\BibitemShut {NoStop}%
\bibitem [{\citenamefont {Collins}\ \emph {et~al.}(1985)\citenamefont
  {Collins}, \citenamefont {Soper},\ and\ \citenamefont
  {Sterman}}]{Collins:1985ue}%
  \BibitemOpen
  \bibfield  {author} {\bibinfo {author} {\bibfnamefont {J.~C.}\ \bibnamefont
  {Collins}}, \bibinfo {author} {\bibfnamefont {D.~E.}\ \bibnamefont {Soper}},
  \ and\ \bibinfo {author} {\bibfnamefont {G.~F.}\ \bibnamefont {Sterman}},\
  }\href {\doibase 10.1016/0550-3213(85)90565-6} {\bibfield  {journal}
  {\bibinfo  {journal} {Nucl. Phys. B}\ }\textbf {\bibinfo {volume} {261}},\
  \bibinfo {pages} {104} (\bibinfo {year} {1985})}\BibitemShut {NoStop}%
\bibitem [{\citenamefont {Collins}\ \emph {et~al.}(1988)\citenamefont
  {Collins}, \citenamefont {Soper},\ and\ \citenamefont
  {Sterman}}]{Collins:1988ig}%
  \BibitemOpen
  \bibfield  {author} {\bibinfo {author} {\bibfnamefont {J.~C.}\ \bibnamefont
  {Collins}}, \bibinfo {author} {\bibfnamefont {D.~E.}\ \bibnamefont {Soper}},
  \ and\ \bibinfo {author} {\bibfnamefont {G.~F.}\ \bibnamefont {Sterman}},\
  }\href {\doibase 10.1016/0550-3213(88)90130-7} {\bibfield  {journal}
  {\bibinfo  {journal} {Nucl. Phys. B}\ }\textbf {\bibinfo {volume} {308}},\
  \bibinfo {pages} {833} (\bibinfo {year} {1988})}\BibitemShut {NoStop}%
\bibitem [{\citenamefont {Sterman}(1987)}]{Sterman:1986aj}%
  \BibitemOpen
  \bibfield  {author} {\bibinfo {author} {\bibfnamefont {G.}~\bibnamefont
  {Sterman}},\ }\href@noop {} {\bibfield  {journal} {\bibinfo  {journal} {Nucl.
  Phys.}\ }\textbf {\bibinfo {volume} {B281}},\ \bibinfo {pages} {310}
  (\bibinfo {year} {1987})}\BibitemShut {NoStop}%
\bibitem [{\citenamefont {Catani}\ and\ \citenamefont
  {Trentadue}(1989)}]{Catani:1989ne}%
  \BibitemOpen
  \bibfield  {author} {\bibinfo {author} {\bibfnamefont {S.}~\bibnamefont
  {Catani}}\ and\ \bibinfo {author} {\bibfnamefont {L.}~\bibnamefont
  {Trentadue}},\ }\href@noop {} {\bibfield  {journal} {\bibinfo  {journal}
  {Nucl. Phys.}\ }\textbf {\bibinfo {volume} {B327}},\ \bibinfo {pages} {323}
  (\bibinfo {year} {1989})}\BibitemShut {NoStop}%
\bibitem [{\citenamefont {Korchemsky}\ and\ \citenamefont
  {Marchesini}(1993{\natexlab{a}})}]{Korchemsky:1992xv}%
  \BibitemOpen
  \bibfield  {author} {\bibinfo {author} {\bibfnamefont {G.}~\bibnamefont
  {Korchemsky}}\ and\ \bibinfo {author} {\bibfnamefont {G.}~\bibnamefont
  {Marchesini}},\ }\href {\doibase 10.1016/0550-3213(93)90167-N} {\bibfield
  {journal} {\bibinfo  {journal} {Nucl. Phys. B}\ }\textbf {\bibinfo {volume}
  {406}},\ \bibinfo {pages} {225} (\bibinfo {year} {1993}{\natexlab{a}})},\
  \Eprint {http://arxiv.org/abs/hep-ph/9210281} {arXiv:hep-ph/9210281}
  \BibitemShut {NoStop}%
\bibitem [{\citenamefont {Korchemsky}\ and\ \citenamefont
  {Marchesini}(1993{\natexlab{b}})}]{Korchemsky:1993uz}%
  \BibitemOpen
  \bibfield  {author} {\bibinfo {author} {\bibfnamefont {G.~P.}\ \bibnamefont
  {Korchemsky}}\ and\ \bibinfo {author} {\bibfnamefont {G.}~\bibnamefont
  {Marchesini}},\ }\href {\doibase 10.1016/0370-2693(93)90015-A} {\bibfield
  {journal} {\bibinfo  {journal} {Phys. Lett.}\ }\textbf {\bibinfo {volume}
  {B313}},\ \bibinfo {pages} {433} (\bibinfo {year}
  {1993}{\natexlab{b}})}\BibitemShut {NoStop}%
\bibitem [{\citenamefont {Contopanagos}\ \emph {et~al.}(1997)\citenamefont
  {Contopanagos}, \citenamefont {Laenen},\ and\ \citenamefont
  {Sterman}}]{Contopanagos:1996nh}%
  \BibitemOpen
  \bibfield  {author} {\bibinfo {author} {\bibfnamefont {H.}~\bibnamefont
  {Contopanagos}}, \bibinfo {author} {\bibfnamefont {E.}~\bibnamefont
  {Laenen}}, \ and\ \bibinfo {author} {\bibfnamefont {G.~F.}\ \bibnamefont
  {Sterman}},\ }\href {\doibase 10.1016/S0550-3213(96)00567-6} {\bibfield
  {journal} {\bibinfo  {journal} {Nucl. Phys. B}\ }\textbf {\bibinfo {volume}
  {484}},\ \bibinfo {pages} {303} (\bibinfo {year} {1997})},\ \Eprint
  {http://arxiv.org/abs/hep-ph/9604313} {arXiv:hep-ph/9604313} \BibitemShut
  {NoStop}%
\bibitem [{\citenamefont {Catani}\ \emph {et~al.}(1996)\citenamefont {Catani},
  \citenamefont {Mangano}, \citenamefont {Nason},\ and\ \citenamefont
  {Trentadue}}]{Catani:1996yz}%
  \BibitemOpen
  \bibfield  {author} {\bibinfo {author} {\bibfnamefont {S.}~\bibnamefont
  {Catani}}, \bibinfo {author} {\bibfnamefont {M.~L.}\ \bibnamefont {Mangano}},
  \bibinfo {author} {\bibfnamefont {P.}~\bibnamefont {Nason}}, \ and\ \bibinfo
  {author} {\bibfnamefont {L.}~\bibnamefont {Trentadue}},\ }\href@noop {}
  {\bibfield  {journal} {\bibinfo  {journal} {Nucl. Phys.}\ }\textbf {\bibinfo
  {volume} {B478}},\ \bibinfo {pages} {273} (\bibinfo {year} {1996})},\ \Eprint
  {http://arxiv.org/abs/hep-ph/9604351} {hep-ph/9604351} \BibitemShut {NoStop}%
\bibitem [{\citenamefont {Kidonakis}\ and\ \citenamefont
  {Sterman}(1997)}]{Kidonakis:1997gm}%
  \BibitemOpen
  \bibfield  {author} {\bibinfo {author} {\bibfnamefont {N.}~\bibnamefont
  {Kidonakis}}\ and\ \bibinfo {author} {\bibfnamefont {G.}~\bibnamefont
  {Sterman}},\ }\href@noop {} {\bibfield  {journal} {\bibinfo  {journal} {Nucl.
  Phys.}\ }\textbf {\bibinfo {volume} {B505}},\ \bibinfo {pages} {321}
  (\bibinfo {year} {1997})},\ \Eprint {http://arxiv.org/abs/hep-ph/9705234}
  {hep-ph/9705234} \BibitemShut {NoStop}%
\bibitem [{\citenamefont {Kidonakis}\ \emph {et~al.}(1998)\citenamefont
  {Kidonakis}, \citenamefont {Oderda},\ and\ \citenamefont
  {Sterman}}]{Kidonakis:1998nf}%
  \BibitemOpen
  \bibfield  {author} {\bibinfo {author} {\bibfnamefont {N.}~\bibnamefont
  {Kidonakis}}, \bibinfo {author} {\bibfnamefont {G.}~\bibnamefont {Oderda}}, \
  and\ \bibinfo {author} {\bibfnamefont {G.}~\bibnamefont {Sterman}},\
  }\href@noop {} {\bibfield  {journal} {\bibinfo  {journal} {Nucl. Phys.}\
  }\textbf {\bibinfo {volume} {B531}},\ \bibinfo {pages} {365} (\bibinfo {year}
  {1998})},\ \Eprint {http://arxiv.org/abs/hep-ph/9803241} {hep-ph/9803241}
  \BibitemShut {NoStop}%
\bibitem [{\citenamefont {Laenen}\ \emph {et~al.}(1998)\citenamefont {Laenen},
  \citenamefont {Oderda},\ and\ \citenamefont {Sterman}}]{Laenen:1998qw}%
  \BibitemOpen
  \bibfield  {author} {\bibinfo {author} {\bibfnamefont {E.}~\bibnamefont
  {Laenen}}, \bibinfo {author} {\bibfnamefont {G.}~\bibnamefont {Oderda}}, \
  and\ \bibinfo {author} {\bibfnamefont {G.}~\bibnamefont {Sterman}},\
  }\href@noop {} {\bibfield  {journal} {\bibinfo  {journal} {Phys. Lett.}\
  }\textbf {\bibinfo {volume} {B438}},\ \bibinfo {pages} {173} (\bibinfo {year}
  {1998})},\ \Eprint {http://arxiv.org/abs/hep-ph/9806467} {hep-ph/9806467}
  \BibitemShut {NoStop}%
\bibitem [{\citenamefont {Catani}\ \emph {et~al.}(2003)\citenamefont {Catani},
  \citenamefont {de~Florian}, \citenamefont {Grazzini},\ and\ \citenamefont
  {Nason}}]{Catani:2003zt}%
  \BibitemOpen
  \bibfield  {author} {\bibinfo {author} {\bibfnamefont {S.}~\bibnamefont
  {Catani}}, \bibinfo {author} {\bibfnamefont {D.}~\bibnamefont {de~Florian}},
  \bibinfo {author} {\bibfnamefont {M.}~\bibnamefont {Grazzini}}, \ and\
  \bibinfo {author} {\bibfnamefont {P.}~\bibnamefont {Nason}},\ }\href@noop {}
  {\bibfield  {journal} {\bibinfo  {journal} {JHEP}\ }\textbf {\bibinfo
  {volume} {0307}},\ \bibinfo {pages} {028} (\bibinfo {year} {2003})},\ \Eprint
  {http://arxiv.org/abs/hep-ph/0306211} {hep-ph/0306211} \BibitemShut {NoStop}%
\bibitem [{\citenamefont {Forte}\ and\ \citenamefont
  {Ridolfi}(2003)}]{Forte:2002ni}%
  \BibitemOpen
  \bibfield  {author} {\bibinfo {author} {\bibfnamefont {S.}~\bibnamefont
  {Forte}}\ and\ \bibinfo {author} {\bibfnamefont {G.}~\bibnamefont
  {Ridolfi}},\ }\href@noop {} {\bibfield  {journal} {\bibinfo  {journal} {Nucl.
  Phys.}\ }\textbf {\bibinfo {volume} {B650}},\ \bibinfo {pages} {229}
  (\bibinfo {year} {2003})},\ \Eprint {http://arxiv.org/abs/hep-ph/0209154}
  {hep-ph/0209154} \BibitemShut {NoStop}%
\bibitem [{\citenamefont {Ravindran}(2006{\natexlab{a}})}]{Ravindran:2005vv}%
  \BibitemOpen
  \bibfield  {author} {\bibinfo {author} {\bibfnamefont {V.}~\bibnamefont
  {Ravindran}},\ }\href {\doibase 10.1016/j.nuclphysb.2006.04.008} {\bibfield
  {journal} {\bibinfo  {journal} {Nucl. Phys. B}\ }\textbf {\bibinfo {volume}
  {746}},\ \bibinfo {pages} {58} (\bibinfo {year} {2006}{\natexlab{a}})},\
  \Eprint {http://arxiv.org/abs/hep-ph/0512249} {arXiv:hep-ph/0512249}
  \BibitemShut {NoStop}%
\bibitem [{\citenamefont {Ravindran}(2006{\natexlab{b}})}]{Ravindran:2006cg}%
  \BibitemOpen
  \bibfield  {author} {\bibinfo {author} {\bibfnamefont {V.}~\bibnamefont
  {Ravindran}},\ }\href {\doibase 10.1016/j.nuclphysb.2006.06.025} {\bibfield
  {journal} {\bibinfo  {journal} {Nucl. Phys. B}\ }\textbf {\bibinfo {volume}
  {752}},\ \bibinfo {pages} {173} (\bibinfo {year} {2006}{\natexlab{b}})},\
  \Eprint {http://arxiv.org/abs/hep-ph/0603041} {arXiv:hep-ph/0603041}
  \BibitemShut {NoStop}%
\bibitem [{\citenamefont {Bauer}\ \emph
  {et~al.}(2002{\natexlab{a}})\citenamefont {Bauer}, \citenamefont {Fleming},
  \citenamefont {Pirjol}, \citenamefont {Rothstein},\ and\ \citenamefont
  {Stewart}}]{Bauer:2002nz}%
  \BibitemOpen
  \bibfield  {author} {\bibinfo {author} {\bibfnamefont {C.~W.}\ \bibnamefont
  {Bauer}}, \bibinfo {author} {\bibfnamefont {S.}~\bibnamefont {Fleming}},
  \bibinfo {author} {\bibfnamefont {D.}~\bibnamefont {Pirjol}}, \bibinfo
  {author} {\bibfnamefont {I.~Z.}\ \bibnamefont {Rothstein}}, \ and\ \bibinfo
  {author} {\bibfnamefont {I.~W.}\ \bibnamefont {Stewart}},\ }\href {\doibase
  10.1103/PhysRevD.66.014017} {\bibfield  {journal} {\bibinfo  {journal} {Phys.
  Rev. D}\ }\textbf {\bibinfo {volume} {66}},\ \bibinfo {pages} {014017}
  (\bibinfo {year} {2002}{\natexlab{a}})},\ \Eprint
  {http://arxiv.org/abs/hep-ph/0202088} {arXiv:hep-ph/0202088} \BibitemShut
  {NoStop}%
\bibitem [{\citenamefont {Manohar}(2003)}]{Manohar:2003vb}%
  \BibitemOpen
  \bibfield  {author} {\bibinfo {author} {\bibfnamefont {A.~V.}\ \bibnamefont
  {Manohar}},\ }\href {\doibase 10.1103/PhysRevD.68.114019} {\bibfield
  {journal} {\bibinfo  {journal} {Phys. Rev. D}\ }\textbf {\bibinfo {volume}
  {68}},\ \bibinfo {pages} {114019} (\bibinfo {year} {2003})},\ \Eprint
  {http://arxiv.org/abs/hep-ph/0309176} {arXiv:hep-ph/0309176} \BibitemShut
  {NoStop}%
\bibitem [{\citenamefont {Idilbi}\ and\ \citenamefont
  {Ji}(2005)}]{Idilbi:2005ky}%
  \BibitemOpen
  \bibfield  {author} {\bibinfo {author} {\bibfnamefont {A.}~\bibnamefont
  {Idilbi}}\ and\ \bibinfo {author} {\bibfnamefont {X.-d.}\ \bibnamefont
  {Ji}},\ }\href {\doibase 10.1103/PhysRevD.72.054016} {\bibfield  {journal}
  {\bibinfo  {journal} {Phys. Rev. D}\ }\textbf {\bibinfo {volume} {72}},\
  \bibinfo {pages} {054016} (\bibinfo {year} {2005})},\ \Eprint
  {http://arxiv.org/abs/hep-ph/0501006} {arXiv:hep-ph/0501006} \BibitemShut
  {NoStop}%
\bibitem [{\citenamefont {Chay}\ and\ \citenamefont {Kim}(2007)}]{Chay:2005rz}%
  \BibitemOpen
  \bibfield  {author} {\bibinfo {author} {\bibfnamefont {J.}~\bibnamefont
  {Chay}}\ and\ \bibinfo {author} {\bibfnamefont {C.}~\bibnamefont {Kim}},\
  }\href {\doibase 10.1103/PhysRevD.75.016003} {\bibfield  {journal} {\bibinfo
  {journal} {Phys. Rev. D}\ }\textbf {\bibinfo {volume} {75}},\ \bibinfo
  {pages} {016003} (\bibinfo {year} {2007})},\ \Eprint
  {http://arxiv.org/abs/hep-ph/0511066} {arXiv:hep-ph/0511066} \BibitemShut
  {NoStop}%
\bibitem [{\citenamefont {Becher}\ and\ \citenamefont
  {Neubert}(2006)}]{Becher:2006nr}%
  \BibitemOpen
  \bibfield  {author} {\bibinfo {author} {\bibfnamefont {T.}~\bibnamefont
  {Becher}}\ and\ \bibinfo {author} {\bibfnamefont {M.}~\bibnamefont
  {Neubert}},\ }\href {\doibase 10.1103/PhysRevLett.97.082001} {\bibfield
  {journal} {\bibinfo  {journal} {Phys. Rev. Lett.}\ }\textbf {\bibinfo
  {volume} {97}},\ \bibinfo {pages} {082001} (\bibinfo {year} {2006})},\
  \Eprint {http://arxiv.org/abs/hep-ph/0605050} {arXiv:hep-ph/0605050}
  \BibitemShut {NoStop}%
\bibitem [{\citenamefont {Becher}\ \emph {et~al.}(2007)\citenamefont {Becher},
  \citenamefont {Neubert},\ and\ \citenamefont {Pecjak}}]{Becher:2006mr}%
  \BibitemOpen
  \bibfield  {author} {\bibinfo {author} {\bibfnamefont {T.}~\bibnamefont
  {Becher}}, \bibinfo {author} {\bibfnamefont {M.}~\bibnamefont {Neubert}}, \
  and\ \bibinfo {author} {\bibfnamefont {B.~D.}\ \bibnamefont {Pecjak}},\
  }\href {\doibase 10.1088/1126-6708/2007/01/076} {\bibfield  {journal}
  {\bibinfo  {journal} {JHEP}\ }\textbf {\bibinfo {volume} {01}},\ \bibinfo
  {pages} {076} (\bibinfo {year} {2007})},\ \Eprint
  {http://arxiv.org/abs/hep-ph/0607228} {arXiv:hep-ph/0607228} \BibitemShut
  {NoStop}%
\bibitem [{\citenamefont {Becher}\ \emph {et~al.}(2008)\citenamefont {Becher},
  \citenamefont {Neubert},\ and\ \citenamefont {Xu}}]{Becher:2007ty}%
  \BibitemOpen
  \bibfield  {author} {\bibinfo {author} {\bibfnamefont {T.}~\bibnamefont
  {Becher}}, \bibinfo {author} {\bibfnamefont {M.}~\bibnamefont {Neubert}}, \
  and\ \bibinfo {author} {\bibfnamefont {G.}~\bibnamefont {Xu}},\ }\href
  {\doibase 10.1088/1126-6708/2008/07/030} {\bibfield  {journal} {\bibinfo
  {journal} {JHEP}\ }\textbf {\bibinfo {volume} {07}},\ \bibinfo {pages} {030}
  (\bibinfo {year} {2008})},\ \Eprint {http://arxiv.org/abs/0710.0680}
  {arXiv:0710.0680 [hep-ph]} \BibitemShut {NoStop}%
\bibitem [{\citenamefont {Ahrens}\ \emph {et~al.}(2010)\citenamefont {Ahrens},
  \citenamefont {Ferroglia}, \citenamefont {Neubert}, \citenamefont {Pecjak},\
  and\ \citenamefont {Yang}}]{Ahrens:2010zv}%
  \BibitemOpen
  \bibfield  {author} {\bibinfo {author} {\bibfnamefont {V.}~\bibnamefont
  {Ahrens}}, \bibinfo {author} {\bibfnamefont {A.}~\bibnamefont {Ferroglia}},
  \bibinfo {author} {\bibfnamefont {M.}~\bibnamefont {Neubert}}, \bibinfo
  {author} {\bibfnamefont {B.~D.}\ \bibnamefont {Pecjak}}, \ and\ \bibinfo
  {author} {\bibfnamefont {L.~L.}\ \bibnamefont {Yang}},\ }\href {\doibase
  10.1007/JHEP09(2010)097} {\bibfield  {journal} {\bibinfo  {journal} {JHEP}\
  }\textbf {\bibinfo {volume} {09}},\ \bibinfo {pages} {097} (\bibinfo {year}
  {2010})},\ \Eprint {http://arxiv.org/abs/1003.5827} {arXiv:1003.5827
  [hep-ph]} \BibitemShut {NoStop}%
\bibitem [{\citenamefont {Low}(1958)}]{Low:1958sn}%
  \BibitemOpen
  \bibfield  {author} {\bibinfo {author} {\bibfnamefont {F.}~\bibnamefont
  {Low}},\ }\href {\doibase 10.1103/PhysRev.110.974} {\bibfield  {journal}
  {\bibinfo  {journal} {Phys. Rev.}\ }\textbf {\bibinfo {volume} {110}},\
  \bibinfo {pages} {974} (\bibinfo {year} {1958})}\BibitemShut {NoStop}%
\bibitem [{\citenamefont {Burnett}\ and\ \citenamefont
  {Kroll}(1968)}]{Burnett:1967km}%
  \BibitemOpen
  \bibfield  {author} {\bibinfo {author} {\bibfnamefont {T.}~\bibnamefont
  {Burnett}}\ and\ \bibinfo {author} {\bibfnamefont {N.~M.}\ \bibnamefont
  {Kroll}},\ }\href {\doibase 10.1103/PhysRevLett.20.86} {\bibfield  {journal}
  {\bibinfo  {journal} {Phys. Rev. Lett.}\ }\textbf {\bibinfo {volume} {20}},\
  \bibinfo {pages} {86} (\bibinfo {year} {1968})}\BibitemShut {NoStop}%
\bibitem [{\citenamefont {Del~Duca}(1990)}]{DelDuca:1990gz}%
  \BibitemOpen
  \bibfield  {author} {\bibinfo {author} {\bibfnamefont {V.}~\bibnamefont
  {Del~Duca}},\ }\href {\doibase 10.1016/0550-3213(90)90392-Q} {\bibfield
  {journal} {\bibinfo  {journal} {Nucl. Phys. B}\ }\textbf {\bibinfo {volume}
  {345}},\ \bibinfo {pages} {369} (\bibinfo {year} {1990})}\BibitemShut
  {NoStop}%
\bibitem [{\citenamefont {Bonocore}\ \emph
  {et~al.}(2015{\natexlab{a}})\citenamefont {Bonocore}, \citenamefont {Laenen},
  \citenamefont {Magnea}, \citenamefont {Melville}, \citenamefont {Vernazza},\
  and\ \citenamefont {White}}]{Bonocore:2015esa}%
  \BibitemOpen
  \bibfield  {author} {\bibinfo {author} {\bibfnamefont {D.}~\bibnamefont
  {Bonocore}}, \bibinfo {author} {\bibfnamefont {E.}~\bibnamefont {Laenen}},
  \bibinfo {author} {\bibfnamefont {L.}~\bibnamefont {Magnea}}, \bibinfo
  {author} {\bibfnamefont {S.}~\bibnamefont {Melville}}, \bibinfo {author}
  {\bibfnamefont {L.}~\bibnamefont {Vernazza}}, \ and\ \bibinfo {author}
  {\bibfnamefont {C.}~\bibnamefont {White}},\ }\href {\doibase
  10.1007/JHEP06(2015)008} {\bibfield  {journal} {\bibinfo  {journal} {JHEP}\
  }\textbf {\bibinfo {volume} {06}},\ \bibinfo {pages} {008} (\bibinfo {year}
  {2015}{\natexlab{a}})},\ \Eprint {http://arxiv.org/abs/1503.05156}
  {arXiv:1503.05156 [hep-ph]} \BibitemShut {NoStop}%
\bibitem [{\citenamefont {Bonocore}\ \emph {et~al.}(2016)\citenamefont
  {Bonocore}, \citenamefont {Laenen}, \citenamefont {Magnea}, \citenamefont
  {Vernazza},\ and\ \citenamefont {White}}]{Bonocore:2016awd}%
  \BibitemOpen
  \bibfield  {author} {\bibinfo {author} {\bibfnamefont {D.}~\bibnamefont
  {Bonocore}}, \bibinfo {author} {\bibfnamefont {E.}~\bibnamefont {Laenen}},
  \bibinfo {author} {\bibfnamefont {L.}~\bibnamefont {Magnea}}, \bibinfo
  {author} {\bibfnamefont {L.}~\bibnamefont {Vernazza}}, \ and\ \bibinfo
  {author} {\bibfnamefont {C.~D.}\ \bibnamefont {White}},\ }\href {\doibase
  10.1007/JHEP12(2016)121} {\bibfield  {journal} {\bibinfo  {journal} {JHEP}\
  }\textbf {\bibinfo {volume} {12}},\ \bibinfo {pages} {121} (\bibinfo {year}
  {2016})},\ \Eprint {http://arxiv.org/abs/1610.06842} {arXiv:1610.06842
  [hep-ph]} \BibitemShut {NoStop}%
\bibitem [{\citenamefont {Del~Duca}\ \emph {et~al.}(2017)\citenamefont
  {Del~Duca}, \citenamefont {Laenen}, \citenamefont {Magnea}, \citenamefont
  {Vernazza},\ and\ \citenamefont {White}}]{DelDuca:2017twk}%
  \BibitemOpen
  \bibfield  {author} {\bibinfo {author} {\bibfnamefont {V.}~\bibnamefont
  {Del~Duca}}, \bibinfo {author} {\bibfnamefont {E.}~\bibnamefont {Laenen}},
  \bibinfo {author} {\bibfnamefont {L.}~\bibnamefont {Magnea}}, \bibinfo
  {author} {\bibfnamefont {L.}~\bibnamefont {Vernazza}}, \ and\ \bibinfo
  {author} {\bibfnamefont {C.~D.}\ \bibnamefont {White}},\ }\href {\doibase
  10.1007/JHEP11(2017)057} {\bibfield  {journal} {\bibinfo  {journal} {JHEP}\
  }\textbf {\bibinfo {volume} {11}},\ \bibinfo {pages} {057} (\bibinfo {year}
  {2017})},\ \Eprint {http://arxiv.org/abs/1706.04018} {arXiv:1706.04018
  [hep-ph]} \BibitemShut {NoStop}%
\bibitem [{\citenamefont {van Beekveld}\ \emph
  {et~al.}(2019{\natexlab{a}})\citenamefont {van Beekveld}, \citenamefont
  {Beenakker}, \citenamefont {Laenen},\ and\ \citenamefont
  {White}}]{vanBeekveld:2019prq}%
  \BibitemOpen
  \bibfield  {author} {\bibinfo {author} {\bibfnamefont {M.}~\bibnamefont {van
  Beekveld}}, \bibinfo {author} {\bibfnamefont {W.}~\bibnamefont {Beenakker}},
  \bibinfo {author} {\bibfnamefont {E.}~\bibnamefont {Laenen}}, \ and\ \bibinfo
  {author} {\bibfnamefont {C.~D.}\ \bibnamefont {White}},\ }\href@noop {} {\
  (\bibinfo {year} {2019}{\natexlab{a}})},\ \Eprint
  {http://arxiv.org/abs/1905.08741} {arXiv:1905.08741 [hep-ph]} \BibitemShut
  {NoStop}%
\bibitem [{\citenamefont {Bahjat-Abbas}\ \emph {et~al.}(2019)\citenamefont
  {Bahjat-Abbas}, \citenamefont {Bonocore}, \citenamefont
  {Sinninghe~Damst\'{e}}, \citenamefont {Laenen}, \citenamefont {Magnea},
  \citenamefont {Vernazza},\ and\ \citenamefont
  {White}}]{Bahjat-Abbas:2019fqa}%
  \BibitemOpen
  \bibfield  {author} {\bibinfo {author} {\bibfnamefont {N.}~\bibnamefont
  {Bahjat-Abbas}}, \bibinfo {author} {\bibfnamefont {D.}~\bibnamefont
  {Bonocore}}, \bibinfo {author} {\bibfnamefont {J.}~\bibnamefont
  {Sinninghe~Damst\'{e}}}, \bibinfo {author} {\bibfnamefont {E.}~\bibnamefont
  {Laenen}}, \bibinfo {author} {\bibfnamefont {L.}~\bibnamefont {Magnea}},
  \bibinfo {author} {\bibfnamefont {L.}~\bibnamefont {Vernazza}}, \ and\
  \bibinfo {author} {\bibfnamefont {C.}~\bibnamefont {White}},\ }\href
  {\doibase 10.1007/JHEP11(2019)002} {\bibfield  {journal} {\bibinfo  {journal}
  {JHEP}\ }\textbf {\bibinfo {volume} {11}},\ \bibinfo {pages} {002} (\bibinfo
  {year} {2019})},\ \Eprint {http://arxiv.org/abs/1905.13710} {arXiv:1905.13710
  [hep-ph]} \BibitemShut {NoStop}%
\bibitem [{\citenamefont {Beneke}\ \emph
  {et~al.}(2019{\natexlab{a}})\citenamefont {Beneke}, \citenamefont {Broggio},
  \citenamefont {Garny}, \citenamefont {Jaskiewicz}, \citenamefont {Szafron},
  \citenamefont {Vernazza},\ and\ \citenamefont {Wang}}]{Beneke:2018gvs}%
  \BibitemOpen
  \bibfield  {author} {\bibinfo {author} {\bibfnamefont {M.}~\bibnamefont
  {Beneke}}, \bibinfo {author} {\bibfnamefont {A.}~\bibnamefont {Broggio}},
  \bibinfo {author} {\bibfnamefont {M.}~\bibnamefont {Garny}}, \bibinfo
  {author} {\bibfnamefont {S.}~\bibnamefont {Jaskiewicz}}, \bibinfo {author}
  {\bibfnamefont {R.}~\bibnamefont {Szafron}}, \bibinfo {author} {\bibfnamefont
  {L.}~\bibnamefont {Vernazza}}, \ and\ \bibinfo {author} {\bibfnamefont
  {J.}~\bibnamefont {Wang}},\ }\href {\doibase 10.1007/JHEP03(2019)043}
  {\bibfield  {journal} {\bibinfo  {journal} {JHEP}\ }\textbf {\bibinfo
  {volume} {03}},\ \bibinfo {pages} {043} (\bibinfo {year}
  {2019}{\natexlab{a}})},\ \Eprint {http://arxiv.org/abs/1809.10631}
  {arXiv:1809.10631 [hep-ph]} \BibitemShut {NoStop}%
\bibitem [{\citenamefont {Kramer}\ \emph {et~al.}(1998)\citenamefont {Kramer},
  \citenamefont {Laenen},\ and\ \citenamefont {Spira}}]{Kramer:1996iq}%
  \BibitemOpen
  \bibfield  {author} {\bibinfo {author} {\bibfnamefont {M.}~\bibnamefont
  {Kramer}}, \bibinfo {author} {\bibfnamefont {E.}~\bibnamefont {Laenen}}, \
  and\ \bibinfo {author} {\bibfnamefont {M.}~\bibnamefont {Spira}},\
  }\href@noop {} {\bibfield  {journal} {\bibinfo  {journal} {Nucl. Phys.}\
  }\textbf {\bibinfo {volume} {B511}},\ \bibinfo {pages} {523} (\bibinfo {year}
  {1998})},\ \Eprint {http://arxiv.org/abs/hep-ph/9611272} {hep-ph/9611272}
  \BibitemShut {NoStop}%
\bibitem [{\citenamefont {Catani}\ \emph {et~al.}(2001)\citenamefont {Catani},
  \citenamefont {de~Florian},\ and\ \citenamefont {Grazzini}}]{Catani:2001ic}%
  \BibitemOpen
  \bibfield  {author} {\bibinfo {author} {\bibfnamefont {S.}~\bibnamefont
  {Catani}}, \bibinfo {author} {\bibfnamefont {D.}~\bibnamefont {de~Florian}},
  \ and\ \bibinfo {author} {\bibfnamefont {M.}~\bibnamefont {Grazzini}},\
  }\href {\doibase 10.1088/1126-6708/2001/05/025} {\bibfield  {journal}
  {\bibinfo  {journal} {JHEP}\ }\textbf {\bibinfo {volume} {05}},\ \bibinfo
  {pages} {025} (\bibinfo {year} {2001})},\ \Eprint
  {http://arxiv.org/abs/hep-ph/0102227} {arXiv:hep-ph/0102227} \BibitemShut
  {NoStop}%
\bibitem [{\citenamefont {Laenen}\ \emph {et~al.}(2008)\citenamefont {Laenen},
  \citenamefont {Magnea},\ and\ \citenamefont {Stavenga}}]{Laenen:2008ux}%
  \BibitemOpen
  \bibfield  {author} {\bibinfo {author} {\bibfnamefont {E.}~\bibnamefont
  {Laenen}}, \bibinfo {author} {\bibfnamefont {L.}~\bibnamefont {Magnea}}, \
  and\ \bibinfo {author} {\bibfnamefont {G.}~\bibnamefont {Stavenga}},\ }\href
  {\doibase 10.1016/j.physletb.2008.09.037} {\bibfield  {journal} {\bibinfo
  {journal} {Phys. Lett.}\ }\textbf {\bibinfo {volume} {B669}},\ \bibinfo
  {pages} {173} (\bibinfo {year} {2008})},\ \Eprint
  {http://arxiv.org/abs/0807.4412} {arXiv:0807.4412 [hep-ph]} \BibitemShut
  {NoStop}%
\bibitem [{\citenamefont {Moch}\ and\ \citenamefont
  {Vogt}(2009)}]{Moch:2009hr}%
  \BibitemOpen
  \bibfield  {author} {\bibinfo {author} {\bibfnamefont {S.}~\bibnamefont
  {Moch}}\ and\ \bibinfo {author} {\bibfnamefont {A.}~\bibnamefont {Vogt}},\
  }\href {\doibase 10.1088/1126-6708/2009/11/099} {\bibfield  {journal}
  {\bibinfo  {journal} {JHEP}\ }\textbf {\bibinfo {volume} {11}},\ \bibinfo
  {pages} {099} (\bibinfo {year} {2009})},\ \Eprint
  {http://arxiv.org/abs/0909.2124} {arXiv:0909.2124 [hep-ph]} \BibitemShut
  {NoStop}%
\bibitem [{\citenamefont {de~Florian}\ \emph {et~al.}(2014)\citenamefont
  {de~Florian}, \citenamefont {Mazzitelli}, \citenamefont {Moch},\ and\
  \citenamefont {Vogt}}]{deFlorian:2014vta}%
  \BibitemOpen
  \bibfield  {author} {\bibinfo {author} {\bibfnamefont {D.}~\bibnamefont
  {de~Florian}}, \bibinfo {author} {\bibfnamefont {J.}~\bibnamefont
  {Mazzitelli}}, \bibinfo {author} {\bibfnamefont {S.}~\bibnamefont {Moch}}, \
  and\ \bibinfo {author} {\bibfnamefont {A.}~\bibnamefont {Vogt}},\ }\href
  {\doibase 10.1007/JHEP10(2014)176} {\bibfield  {journal} {\bibinfo  {journal}
  {JHEP}\ }\textbf {\bibinfo {volume} {10}},\ \bibinfo {pages} {176} (\bibinfo
  {year} {2014})},\ \Eprint {http://arxiv.org/abs/1408.6277} {arXiv:1408.6277
  [hep-ph]} \BibitemShut {NoStop}%
\bibitem [{\citenamefont {Lo~Presti}\ \emph {et~al.}(2014)\citenamefont
  {Lo~Presti}, \citenamefont {Almasy},\ and\ \citenamefont
  {Vogt}}]{Presti:2014lqa}%
  \BibitemOpen
  \bibfield  {author} {\bibinfo {author} {\bibfnamefont {N.}~\bibnamefont
  {Lo~Presti}}, \bibinfo {author} {\bibfnamefont {A.}~\bibnamefont {Almasy}}, \
  and\ \bibinfo {author} {\bibfnamefont {A.}~\bibnamefont {Vogt}},\ }\href
  {\doibase 10.1016/j.physletb.2014.08.044} {\bibfield  {journal} {\bibinfo
  {journal} {Phys. Lett. B}\ }\textbf {\bibinfo {volume} {737}},\ \bibinfo
  {pages} {120} (\bibinfo {year} {2014})},\ \Eprint
  {http://arxiv.org/abs/1407.1553} {arXiv:1407.1553 [hep-ph]} \BibitemShut
  {NoStop}%
\bibitem [{\citenamefont {Ajjath}\ \emph
  {et~al.}(2020{\natexlab{a}})\citenamefont {Ajjath}, \citenamefont
  {Mukherjee},\ and\ \citenamefont {Ravindran}}]{Ajjath:2020ulr}%
  \BibitemOpen
  \bibfield  {author} {\bibinfo {author} {\bibfnamefont {A.}~\bibnamefont
  {Ajjath}}, \bibinfo {author} {\bibfnamefont {P.}~\bibnamefont {Mukherjee}}, \
  and\ \bibinfo {author} {\bibfnamefont {V.}~\bibnamefont {Ravindran}},\
  }\href@noop {} {\  (\bibinfo {year} {2020}{\natexlab{a}})},\ \Eprint
  {http://arxiv.org/abs/2006.06726} {arXiv:2006.06726 [hep-ph]} \BibitemShut
  {NoStop}%
\bibitem [{\citenamefont {Ajjath}\ \emph
  {et~al.}(2020{\natexlab{b}})\citenamefont {Ajjath}, \citenamefont
  {Mukherjee}, \citenamefont {Ravindran}, \citenamefont {Sankar},\ and\
  \citenamefont {Tiwari}}]{Ajjath:2020sjk}%
  \BibitemOpen
  \bibfield  {author} {\bibinfo {author} {\bibfnamefont {A.}~\bibnamefont
  {Ajjath}}, \bibinfo {author} {\bibfnamefont {P.}~\bibnamefont {Mukherjee}},
  \bibinfo {author} {\bibfnamefont {V.}~\bibnamefont {Ravindran}}, \bibinfo
  {author} {\bibfnamefont {A.}~\bibnamefont {Sankar}}, \ and\ \bibinfo {author}
  {\bibfnamefont {S.}~\bibnamefont {Tiwari}},\ }\href@noop {} {\  (\bibinfo
  {year} {2020}{\natexlab{b}})},\ \Eprint {http://arxiv.org/abs/2007.12214}
  {arXiv:2007.12214 [hep-ph]} \BibitemShut {NoStop}%
\bibitem [{\citenamefont {Grunberg}\ and\ \citenamefont
  {Ravindran}(2009)}]{Grunberg:2009yi}%
  \BibitemOpen
  \bibfield  {author} {\bibinfo {author} {\bibfnamefont {G.}~\bibnamefont
  {Grunberg}}\ and\ \bibinfo {author} {\bibfnamefont {V.}~\bibnamefont
  {Ravindran}},\ }\href {\doibase 10.1088/1126-6708/2009/10/055} {\bibfield
  {journal} {\bibinfo  {journal} {JHEP}\ }\textbf {\bibinfo {volume} {10}},\
  \bibinfo {pages} {055} (\bibinfo {year} {2009})},\ \Eprint
  {http://arxiv.org/abs/0902.2702} {arXiv:0902.2702 [hep-ph]} \BibitemShut
  {NoStop}%
\bibitem [{\citenamefont {van Beekveld}\ \emph
  {et~al.}(2019{\natexlab{b}})\citenamefont {van Beekveld}, \citenamefont
  {Beenakker}, \citenamefont {Basu}, \citenamefont {Laenen}, \citenamefont
  {Misra},\ and\ \citenamefont {Motylinski}}]{vanBeekveld:2019cks}%
  \BibitemOpen
  \bibfield  {author} {\bibinfo {author} {\bibfnamefont {M.}~\bibnamefont {van
  Beekveld}}, \bibinfo {author} {\bibfnamefont {W.}~\bibnamefont {Beenakker}},
  \bibinfo {author} {\bibfnamefont {R.}~\bibnamefont {Basu}}, \bibinfo {author}
  {\bibfnamefont {E.}~\bibnamefont {Laenen}}, \bibinfo {author} {\bibfnamefont
  {A.}~\bibnamefont {Misra}}, \ and\ \bibinfo {author} {\bibfnamefont
  {P.}~\bibnamefont {Motylinski}},\ }\href@noop {} {\  (\bibinfo {year}
  {2019}{\natexlab{b}})},\ \Eprint {http://arxiv.org/abs/1905.11771}
  {arXiv:1905.11771 [hep-ph]} \BibitemShut {NoStop}%
\bibitem [{\citenamefont {Gervais}(2017)}]{Gervais:2017yxv}%
  \BibitemOpen
  \bibfield  {author} {\bibinfo {author} {\bibfnamefont {H.}~\bibnamefont
  {Gervais}},\ }\href {\doibase 10.1103/PhysRevD.95.125009} {\bibfield
  {journal} {\bibinfo  {journal} {Phys. Rev.}\ }\textbf {\bibinfo {volume}
  {D95}},\ \bibinfo {pages} {125009} (\bibinfo {year} {2017})},\ \Eprint
  {http://arxiv.org/abs/1704.00806} {arXiv:1704.00806 [hep-th]} \BibitemShut
  {NoStop}%
\bibitem [{\citenamefont {Bauer}\ \emph {et~al.}(2000)\citenamefont {Bauer},
  \citenamefont {Fleming},\ and\ \citenamefont {Luke}}]{Bauer:2000ew}%
  \BibitemOpen
  \bibfield  {author} {\bibinfo {author} {\bibfnamefont {C.~W.}\ \bibnamefont
  {Bauer}}, \bibinfo {author} {\bibfnamefont {S.}~\bibnamefont {Fleming}}, \
  and\ \bibinfo {author} {\bibfnamefont {M.~E.}\ \bibnamefont {Luke}},\ }\href
  {\doibase 10.1103/PhysRevD.63.014006} {\bibfield  {journal} {\bibinfo
  {journal} {Phys. Rev. D}\ }\textbf {\bibinfo {volume} {63}},\ \bibinfo
  {pages} {014006} (\bibinfo {year} {2000})},\ \Eprint
  {http://arxiv.org/abs/hep-ph/0005275} {arXiv:hep-ph/0005275} \BibitemShut
  {NoStop}%
\bibitem [{\citenamefont {Bauer}\ \emph {et~al.}(2001)\citenamefont {Bauer},
  \citenamefont {Fleming}, \citenamefont {Pirjol},\ and\ \citenamefont
  {Stewart}}]{Bauer:2000yr}%
  \BibitemOpen
  \bibfield  {author} {\bibinfo {author} {\bibfnamefont {C.~W.}\ \bibnamefont
  {Bauer}}, \bibinfo {author} {\bibfnamefont {S.}~\bibnamefont {Fleming}},
  \bibinfo {author} {\bibfnamefont {D.}~\bibnamefont {Pirjol}}, \ and\ \bibinfo
  {author} {\bibfnamefont {I.~W.}\ \bibnamefont {Stewart}},\ }\href@noop {}
  {\bibfield  {journal} {\bibinfo  {journal} {Phys. Rev.}\ }\textbf {\bibinfo
  {volume} {D63}},\ \bibinfo {pages} {114020} (\bibinfo {year} {2001})},\
  \Eprint {http://arxiv.org/abs/hep-ph/0011336} {hep-ph/0011336} \BibitemShut
  {NoStop}%
\bibitem [{\citenamefont {Bauer}\ and\ \citenamefont
  {Stewart}(2001)}]{Bauer:2001ct}%
  \BibitemOpen
  \bibfield  {author} {\bibinfo {author} {\bibfnamefont {C.~W.}\ \bibnamefont
  {Bauer}}\ and\ \bibinfo {author} {\bibfnamefont {I.~W.}\ \bibnamefont
  {Stewart}},\ }\href {\doibase 10.1016/S0370-2693(01)00902-9} {\bibfield
  {journal} {\bibinfo  {journal} {Phys. Lett. B}\ }\textbf {\bibinfo {volume}
  {516}},\ \bibinfo {pages} {134} (\bibinfo {year} {2001})},\ \Eprint
  {http://arxiv.org/abs/hep-ph/0107001} {arXiv:hep-ph/0107001} \BibitemShut
  {NoStop}%
\bibitem [{\citenamefont {Bauer}\ \emph
  {et~al.}(2002{\natexlab{b}})\citenamefont {Bauer}, \citenamefont {Pirjol},\
  and\ \citenamefont {Stewart}}]{Bauer:2001yt}%
  \BibitemOpen
  \bibfield  {author} {\bibinfo {author} {\bibfnamefont {C.~W.}\ \bibnamefont
  {Bauer}}, \bibinfo {author} {\bibfnamefont {D.}~\bibnamefont {Pirjol}}, \
  and\ \bibinfo {author} {\bibfnamefont {I.~W.}\ \bibnamefont {Stewart}},\
  }\href {\doibase 10.1103/PhysRevD.65.054022} {\bibfield  {journal} {\bibinfo
  {journal} {Phys. Rev. D}\ }\textbf {\bibinfo {volume} {65}},\ \bibinfo
  {pages} {054022} (\bibinfo {year} {2002}{\natexlab{b}})},\ \Eprint
  {http://arxiv.org/abs/hep-ph/0109045} {arXiv:hep-ph/0109045} \BibitemShut
  {NoStop}%
\bibitem [{\citenamefont {Beneke}\ \emph {et~al.}(2002)\citenamefont {Beneke},
  \citenamefont {Chapovsky}, \citenamefont {Diehl},\ and\ \citenamefont
  {Feldmann}}]{Beneke:2002ph}%
  \BibitemOpen
  \bibfield  {author} {\bibinfo {author} {\bibfnamefont {M.}~\bibnamefont
  {Beneke}}, \bibinfo {author} {\bibfnamefont {A.}~\bibnamefont {Chapovsky}},
  \bibinfo {author} {\bibfnamefont {M.}~\bibnamefont {Diehl}}, \ and\ \bibinfo
  {author} {\bibfnamefont {T.}~\bibnamefont {Feldmann}},\ }\href {\doibase
  10.1016/S0550-3213(02)00687-9} {\bibfield  {journal} {\bibinfo  {journal}
  {Nucl. Phys. B}\ }\textbf {\bibinfo {volume} {643}},\ \bibinfo {pages} {431}
  (\bibinfo {year} {2002})},\ \Eprint {http://arxiv.org/abs/hep-ph/0206152}
  {arXiv:hep-ph/0206152} \BibitemShut {NoStop}%
\bibitem [{\citenamefont {Larkoski}\ \emph {et~al.}(2015)\citenamefont
  {Larkoski}, \citenamefont {Neill},\ and\ \citenamefont
  {Stewart}}]{Larkoski:2014bxa}%
  \BibitemOpen
  \bibfield  {author} {\bibinfo {author} {\bibfnamefont {A.~J.}\ \bibnamefont
  {Larkoski}}, \bibinfo {author} {\bibfnamefont {D.}~\bibnamefont {Neill}}, \
  and\ \bibinfo {author} {\bibfnamefont {I.~W.}\ \bibnamefont {Stewart}},\
  }\href {\doibase 10.1007/JHEP06(2015)077} {\bibfield  {journal} {\bibinfo
  {journal} {JHEP}\ }\textbf {\bibinfo {volume} {06}},\ \bibinfo {pages} {077}
  (\bibinfo {year} {2015})},\ \Eprint {http://arxiv.org/abs/1412.3108}
  {arXiv:1412.3108 [hep-th]} \BibitemShut {NoStop}%
\bibitem [{\citenamefont {Moult}\ \emph {et~al.}(2019)\citenamefont {Moult},
  \citenamefont {Stewart},\ and\ \citenamefont {Vita}}]{Moult:2019mog}%
  \BibitemOpen
  \bibfield  {author} {\bibinfo {author} {\bibfnamefont {I.}~\bibnamefont
  {Moult}}, \bibinfo {author} {\bibfnamefont {I.~W.}\ \bibnamefont {Stewart}},
  \ and\ \bibinfo {author} {\bibfnamefont {G.}~\bibnamefont {Vita}},\ }\href
  {\doibase 10.1007/JHEP11(2019)153} {\bibfield  {journal} {\bibinfo  {journal}
  {JHEP}\ }\textbf {\bibinfo {volume} {11}},\ \bibinfo {pages} {153} (\bibinfo
  {year} {2019})},\ \Eprint {http://arxiv.org/abs/1905.07411} {arXiv:1905.07411
  [hep-ph]} \BibitemShut {NoStop}%
\bibitem [{\citenamefont {Beneke}\ \emph
  {et~al.}(2020{\natexlab{a}})\citenamefont {Beneke}, \citenamefont {Broggio},
  \citenamefont {Jaskiewicz},\ and\ \citenamefont {Vernazza}}]{Beneke:2019oqx}%
  \BibitemOpen
  \bibfield  {author} {\bibinfo {author} {\bibfnamefont {M.}~\bibnamefont
  {Beneke}}, \bibinfo {author} {\bibfnamefont {A.}~\bibnamefont {Broggio}},
  \bibinfo {author} {\bibfnamefont {S.}~\bibnamefont {Jaskiewicz}}, \ and\
  \bibinfo {author} {\bibfnamefont {L.}~\bibnamefont {Vernazza}},\ }\href
  {\doibase 10.1007/JHEP07(2020)078} {\bibfield  {journal} {\bibinfo  {journal}
  {JHEP}\ }\textbf {\bibinfo {volume} {07}},\ \bibinfo {pages} {078} (\bibinfo
  {year} {2020}{\natexlab{a}})},\ \Eprint {http://arxiv.org/abs/1912.01585}
  {arXiv:1912.01585 [hep-ph]} \BibitemShut {NoStop}%
\bibitem [{\citenamefont {Beneke}\ \emph
  {et~al.}(2018{\natexlab{a}})\citenamefont {Beneke}, \citenamefont {Garny},
  \citenamefont {Szafron},\ and\ \citenamefont {Wang}}]{Beneke:2017ztn}%
  \BibitemOpen
  \bibfield  {author} {\bibinfo {author} {\bibfnamefont {M.}~\bibnamefont
  {Beneke}}, \bibinfo {author} {\bibfnamefont {M.}~\bibnamefont {Garny}},
  \bibinfo {author} {\bibfnamefont {R.}~\bibnamefont {Szafron}}, \ and\
  \bibinfo {author} {\bibfnamefont {J.}~\bibnamefont {Wang}},\ }\href {\doibase
  10.1007/JHEP03(2018)001} {\bibfield  {journal} {\bibinfo  {journal} {JHEP}\
  }\textbf {\bibinfo {volume} {03}},\ \bibinfo {pages} {001} (\bibinfo {year}
  {2018}{\natexlab{a}})},\ \Eprint {http://arxiv.org/abs/1712.04416}
  {arXiv:1712.04416 [hep-ph]} \BibitemShut {NoStop}%
\bibitem [{\citenamefont {Beneke}\ \emph
  {et~al.}(2018{\natexlab{b}})\citenamefont {Beneke}, \citenamefont {Garny},
  \citenamefont {Szafron},\ and\ \citenamefont {Wang}}]{Beneke:2018rbh}%
  \BibitemOpen
  \bibfield  {author} {\bibinfo {author} {\bibfnamefont {M.}~\bibnamefont
  {Beneke}}, \bibinfo {author} {\bibfnamefont {M.}~\bibnamefont {Garny}},
  \bibinfo {author} {\bibfnamefont {R.}~\bibnamefont {Szafron}}, \ and\
  \bibinfo {author} {\bibfnamefont {J.}~\bibnamefont {Wang}},\ }\href {\doibase
  10.1007/JHEP11(2018)112} {\bibfield  {journal} {\bibinfo  {journal} {JHEP}\
  }\textbf {\bibinfo {volume} {11}},\ \bibinfo {pages} {112} (\bibinfo {year}
  {2018}{\natexlab{b}})},\ \Eprint {http://arxiv.org/abs/1808.04742}
  {arXiv:1808.04742 [hep-ph]} \BibitemShut {NoStop}%
\bibitem [{\citenamefont {Beneke}\ \emph
  {et~al.}(2019{\natexlab{b}})\citenamefont {Beneke}, \citenamefont {Garny},
  \citenamefont {Szafron},\ and\ \citenamefont {Wang}}]{Beneke:2019kgv}%
  \BibitemOpen
  \bibfield  {author} {\bibinfo {author} {\bibfnamefont {M.}~\bibnamefont
  {Beneke}}, \bibinfo {author} {\bibfnamefont {M.}~\bibnamefont {Garny}},
  \bibinfo {author} {\bibfnamefont {R.}~\bibnamefont {Szafron}}, \ and\
  \bibinfo {author} {\bibfnamefont {J.}~\bibnamefont {Wang}},\ }\href {\doibase
  10.1007/JHEP09(2019)101} {\bibfield  {journal} {\bibinfo  {journal} {JHEP}\
  }\textbf {\bibinfo {volume} {09}},\ \bibinfo {pages} {101} (\bibinfo {year}
  {2019}{\natexlab{b}})},\ \Eprint {http://arxiv.org/abs/1907.05463}
  {arXiv:1907.05463 [hep-ph]} \BibitemShut {NoStop}%
\bibitem [{\citenamefont {Moult}\ \emph
  {et~al.}(2017{\natexlab{a}})\citenamefont {Moult}, \citenamefont {Stewart},\
  and\ \citenamefont {Vita}}]{Moult:2017rpl}%
  \BibitemOpen
  \bibfield  {author} {\bibinfo {author} {\bibfnamefont {I.}~\bibnamefont
  {Moult}}, \bibinfo {author} {\bibfnamefont {I.~W.}\ \bibnamefont {Stewart}},
  \ and\ \bibinfo {author} {\bibfnamefont {G.}~\bibnamefont {Vita}},\ }\href
  {\doibase 10.1007/JHEP07(2017)067} {\bibfield  {journal} {\bibinfo  {journal}
  {JHEP}\ }\textbf {\bibinfo {volume} {07}},\ \bibinfo {pages} {067} (\bibinfo
  {year} {2017}{\natexlab{a}})},\ \Eprint {http://arxiv.org/abs/1703.03408}
  {arXiv:1703.03408 [hep-ph]} \BibitemShut {NoStop}%
\bibitem [{\citenamefont {Feige}\ \emph {et~al.}(2017)\citenamefont {Feige},
  \citenamefont {Kolodrubetz}, \citenamefont {Moult},\ and\ \citenamefont
  {Stewart}}]{Feige:2017zci}%
  \BibitemOpen
  \bibfield  {author} {\bibinfo {author} {\bibfnamefont {I.}~\bibnamefont
  {Feige}}, \bibinfo {author} {\bibfnamefont {D.~W.}\ \bibnamefont
  {Kolodrubetz}}, \bibinfo {author} {\bibfnamefont {I.}~\bibnamefont {Moult}},
  \ and\ \bibinfo {author} {\bibfnamefont {I.~W.}\ \bibnamefont {Stewart}},\
  }\href {\doibase 10.1007/JHEP11(2017)142} {\bibfield  {journal} {\bibinfo
  {journal} {JHEP}\ }\textbf {\bibinfo {volume} {11}},\ \bibinfo {pages} {142}
  (\bibinfo {year} {2017})},\ \Eprint {http://arxiv.org/abs/1703.03411}
  {arXiv:1703.03411 [hep-ph]} \BibitemShut {NoStop}%
\bibitem [{\citenamefont {Chang}\ \emph {et~al.}(2018)\citenamefont {Chang},
  \citenamefont {Stewart},\ and\ \citenamefont {Vita}}]{Chang:2017atu}%
  \BibitemOpen
  \bibfield  {author} {\bibinfo {author} {\bibfnamefont {C.-H.}\ \bibnamefont
  {Chang}}, \bibinfo {author} {\bibfnamefont {I.~W.}\ \bibnamefont {Stewart}},
  \ and\ \bibinfo {author} {\bibfnamefont {G.}~\bibnamefont {Vita}},\ }\href
  {\doibase 10.1007/JHEP04(2018)041} {\bibfield  {journal} {\bibinfo  {journal}
  {JHEP}\ }\textbf {\bibinfo {volume} {04}},\ \bibinfo {pages} {041} (\bibinfo
  {year} {2018})},\ \Eprint {http://arxiv.org/abs/1712.04343} {arXiv:1712.04343
  [hep-ph]} \BibitemShut {NoStop}%
\bibitem [{\citenamefont {Moult}\ \emph
  {et~al.}(2017{\natexlab{b}})\citenamefont {Moult}, \citenamefont {Rothen},
  \citenamefont {Stewart}, \citenamefont {Tackmann},\ and\ \citenamefont
  {Zhu}}]{Moult:2016fqy}%
  \BibitemOpen
  \bibfield  {author} {\bibinfo {author} {\bibfnamefont {I.}~\bibnamefont
  {Moult}}, \bibinfo {author} {\bibfnamefont {L.}~\bibnamefont {Rothen}},
  \bibinfo {author} {\bibfnamefont {I.~W.}\ \bibnamefont {Stewart}}, \bibinfo
  {author} {\bibfnamefont {F.~J.}\ \bibnamefont {Tackmann}}, \ and\ \bibinfo
  {author} {\bibfnamefont {H.~X.}\ \bibnamefont {Zhu}},\ }\href {\doibase
  10.1103/PhysRevD.95.074023} {\bibfield  {journal} {\bibinfo  {journal} {Phys.
  Rev. D}\ }\textbf {\bibinfo {volume} {95}},\ \bibinfo {pages} {074023}
  (\bibinfo {year} {2017}{\natexlab{b}})},\ \Eprint
  {http://arxiv.org/abs/1612.00450} {arXiv:1612.00450 [hep-ph]} \BibitemShut
  {NoStop}%
\bibitem [{\citenamefont {Boughezal}\ \emph {et~al.}(2017)\citenamefont
  {Boughezal}, \citenamefont {Liu},\ and\ \citenamefont
  {Petriello}}]{Boughezal:2016zws}%
  \BibitemOpen
  \bibfield  {author} {\bibinfo {author} {\bibfnamefont {R.}~\bibnamefont
  {Boughezal}}, \bibinfo {author} {\bibfnamefont {X.}~\bibnamefont {Liu}}, \
  and\ \bibinfo {author} {\bibfnamefont {F.}~\bibnamefont {Petriello}},\ }\href
  {\doibase 10.1007/JHEP03(2017)160} {\bibfield  {journal} {\bibinfo  {journal}
  {JHEP}\ }\textbf {\bibinfo {volume} {03}},\ \bibinfo {pages} {160} (\bibinfo
  {year} {2017})},\ \Eprint {http://arxiv.org/abs/1612.02911} {arXiv:1612.02911
  [hep-ph]} \BibitemShut {NoStop}%
\bibitem [{\citenamefont {Moult}\ \emph
  {et~al.}(2018{\natexlab{a}})\citenamefont {Moult}, \citenamefont {Rothen},
  \citenamefont {Stewart}, \citenamefont {Tackmann},\ and\ \citenamefont
  {Zhu}}]{Moult:2017jsg}%
  \BibitemOpen
  \bibfield  {author} {\bibinfo {author} {\bibfnamefont {I.}~\bibnamefont
  {Moult}}, \bibinfo {author} {\bibfnamefont {L.}~\bibnamefont {Rothen}},
  \bibinfo {author} {\bibfnamefont {I.~W.}\ \bibnamefont {Stewart}}, \bibinfo
  {author} {\bibfnamefont {F.~J.}\ \bibnamefont {Tackmann}}, \ and\ \bibinfo
  {author} {\bibfnamefont {H.~X.}\ \bibnamefont {Zhu}},\ }\href {\doibase
  10.1103/PhysRevD.97.014013} {\bibfield  {journal} {\bibinfo  {journal} {Phys.
  Rev. D}\ }\textbf {\bibinfo {volume} {97}},\ \bibinfo {pages} {014013}
  (\bibinfo {year} {2018}{\natexlab{a}})},\ \Eprint
  {http://arxiv.org/abs/1710.03227} {arXiv:1710.03227 [hep-ph]} \BibitemShut
  {NoStop}%
\bibitem [{\citenamefont {Boughezal}\ \emph {et~al.}(2018)\citenamefont
  {Boughezal}, \citenamefont {Isgr\`{o}},\ and\ \citenamefont
  {Petriello}}]{Boughezal:2018mvf}%
  \BibitemOpen
  \bibfield  {author} {\bibinfo {author} {\bibfnamefont {R.}~\bibnamefont
  {Boughezal}}, \bibinfo {author} {\bibfnamefont {A.}~\bibnamefont
  {Isgr\`{o}}}, \ and\ \bibinfo {author} {\bibfnamefont {F.}~\bibnamefont
  {Petriello}},\ }\href {\doibase 10.1103/PhysRevD.97.076006} {\bibfield
  {journal} {\bibinfo  {journal} {Phys. Rev. D}\ }\textbf {\bibinfo {volume}
  {97}},\ \bibinfo {pages} {076006} (\bibinfo {year} {2018})},\ \Eprint
  {http://arxiv.org/abs/1802.00456} {arXiv:1802.00456 [hep-ph]} \BibitemShut
  {NoStop}%
\bibitem [{\citenamefont {Ebert}\ \emph {et~al.}(2018)\citenamefont {Ebert},
  \citenamefont {Moult}, \citenamefont {Stewart}, \citenamefont {Tackmann},
  \citenamefont {Vita},\ and\ \citenamefont {Zhu}}]{Ebert:2018lzn}%
  \BibitemOpen
  \bibfield  {author} {\bibinfo {author} {\bibfnamefont {M.~A.}\ \bibnamefont
  {Ebert}}, \bibinfo {author} {\bibfnamefont {I.}~\bibnamefont {Moult}},
  \bibinfo {author} {\bibfnamefont {I.~W.}\ \bibnamefont {Stewart}}, \bibinfo
  {author} {\bibfnamefont {F.~J.}\ \bibnamefont {Tackmann}}, \bibinfo {author}
  {\bibfnamefont {G.}~\bibnamefont {Vita}}, \ and\ \bibinfo {author}
  {\bibfnamefont {H.~X.}\ \bibnamefont {Zhu}},\ }\href {\doibase
  10.1007/JHEP12(2018)084} {\bibfield  {journal} {\bibinfo  {journal} {JHEP}\
  }\textbf {\bibinfo {volume} {12}},\ \bibinfo {pages} {084} (\bibinfo {year}
  {2018})},\ \Eprint {http://arxiv.org/abs/1807.10764} {arXiv:1807.10764
  [hep-ph]} \BibitemShut {NoStop}%
\bibitem [{\citenamefont {Boughezal}\ \emph {et~al.}(2020)\citenamefont
  {Boughezal}, \citenamefont {Isgr\`{o}},\ and\ \citenamefont
  {Petriello}}]{Boughezal:2019ggi}%
  \BibitemOpen
  \bibfield  {author} {\bibinfo {author} {\bibfnamefont {R.}~\bibnamefont
  {Boughezal}}, \bibinfo {author} {\bibfnamefont {A.}~\bibnamefont
  {Isgr\`{o}}}, \ and\ \bibinfo {author} {\bibfnamefont {F.}~\bibnamefont
  {Petriello}},\ }\href {\doibase 10.1103/PhysRevD.101.016005} {\bibfield
  {journal} {\bibinfo  {journal} {Phys. Rev. D}\ }\textbf {\bibinfo {volume}
  {101}},\ \bibinfo {pages} {016005} (\bibinfo {year} {2020})},\ \Eprint
  {http://arxiv.org/abs/1907.12213} {arXiv:1907.12213 [hep-ph]} \BibitemShut
  {NoStop}%
\bibitem [{\citenamefont {Ebert}\ \emph {et~al.}(2019)\citenamefont {Ebert},
  \citenamefont {Moult}, \citenamefont {Stewart}, \citenamefont {Tackmann},
  \citenamefont {Vita},\ and\ \citenamefont {Zhu}}]{Ebert:2018gsn}%
  \BibitemOpen
  \bibfield  {author} {\bibinfo {author} {\bibfnamefont {M.~A.}\ \bibnamefont
  {Ebert}}, \bibinfo {author} {\bibfnamefont {I.}~\bibnamefont {Moult}},
  \bibinfo {author} {\bibfnamefont {I.~W.}\ \bibnamefont {Stewart}}, \bibinfo
  {author} {\bibfnamefont {F.~J.}\ \bibnamefont {Tackmann}}, \bibinfo {author}
  {\bibfnamefont {G.}~\bibnamefont {Vita}}, \ and\ \bibinfo {author}
  {\bibfnamefont {H.~X.}\ \bibnamefont {Zhu}},\ }\href {\doibase
  10.1007/JHEP04(2019)123} {\bibfield  {journal} {\bibinfo  {journal} {JHEP}\
  }\textbf {\bibinfo {volume} {04}},\ \bibinfo {pages} {123} (\bibinfo {year}
  {2019})},\ \Eprint {http://arxiv.org/abs/1812.08189} {arXiv:1812.08189
  [hep-ph]} \BibitemShut {NoStop}%
\bibitem [{\citenamefont {Moult}\ \emph
  {et~al.}(2018{\natexlab{b}})\citenamefont {Moult}, \citenamefont {Stewart},
  \citenamefont {Vita},\ and\ \citenamefont {Zhu}}]{Moult:2018jjd}%
  \BibitemOpen
  \bibfield  {author} {\bibinfo {author} {\bibfnamefont {I.}~\bibnamefont
  {Moult}}, \bibinfo {author} {\bibfnamefont {I.~W.}\ \bibnamefont {Stewart}},
  \bibinfo {author} {\bibfnamefont {G.}~\bibnamefont {Vita}}, \ and\ \bibinfo
  {author} {\bibfnamefont {H.~X.}\ \bibnamefont {Zhu}},\ }\href {\doibase
  10.1007/JHEP08(2018)013} {\bibfield  {journal} {\bibinfo  {journal} {JHEP}\
  }\textbf {\bibinfo {volume} {08}},\ \bibinfo {pages} {013} (\bibinfo {year}
  {2018}{\natexlab{b}})},\ \Eprint {http://arxiv.org/abs/1804.04665}
  {arXiv:1804.04665 [hep-ph]} \BibitemShut {NoStop}%
\bibitem [{\citenamefont {Beneke}\ \emph
  {et~al.}(2020{\natexlab{b}})\citenamefont {Beneke}, \citenamefont {Garny},
  \citenamefont {Jaskiewicz}, \citenamefont {Szafron}, \citenamefont
  {Vernazza},\ and\ \citenamefont {Wang}}]{Beneke:2019mua}%
  \BibitemOpen
  \bibfield  {author} {\bibinfo {author} {\bibfnamefont {M.}~\bibnamefont
  {Beneke}}, \bibinfo {author} {\bibfnamefont {M.}~\bibnamefont {Garny}},
  \bibinfo {author} {\bibfnamefont {S.}~\bibnamefont {Jaskiewicz}}, \bibinfo
  {author} {\bibfnamefont {R.}~\bibnamefont {Szafron}}, \bibinfo {author}
  {\bibfnamefont {L.}~\bibnamefont {Vernazza}}, \ and\ \bibinfo {author}
  {\bibfnamefont {J.}~\bibnamefont {Wang}},\ }\href {\doibase
  10.1007/JHEP01(2020)094} {\bibfield  {journal} {\bibinfo  {journal} {JHEP}\
  }\textbf {\bibinfo {volume} {01}},\ \bibinfo {pages} {094} (\bibinfo {year}
  {2020}{\natexlab{b}})},\ \Eprint {http://arxiv.org/abs/1910.12685}
  {arXiv:1910.12685 [hep-ph]} \BibitemShut {NoStop}%
\bibitem [{\citenamefont {Moult}\ \emph {et~al.}(2020)\citenamefont {Moult},
  \citenamefont {Stewart}, \citenamefont {Vita},\ and\ \citenamefont
  {Zhu}}]{Moult:2019uhz}%
  \BibitemOpen
  \bibfield  {author} {\bibinfo {author} {\bibfnamefont {I.}~\bibnamefont
  {Moult}}, \bibinfo {author} {\bibfnamefont {I.~W.}\ \bibnamefont {Stewart}},
  \bibinfo {author} {\bibfnamefont {G.}~\bibnamefont {Vita}}, \ and\ \bibinfo
  {author} {\bibfnamefont {H.~X.}\ \bibnamefont {Zhu}},\ }\href {\doibase
  10.1007/JHEP05(2020)089} {\bibfield  {journal} {\bibinfo  {journal} {JHEP}\
  }\textbf {\bibinfo {volume} {05}},\ \bibinfo {pages} {089} (\bibinfo {year}
  {2020})},\ \Eprint {http://arxiv.org/abs/1910.14038} {arXiv:1910.14038
  [hep-ph]} \BibitemShut {NoStop}%
\bibitem [{\citenamefont {Liu}\ and\ \citenamefont
  {Neubert}(2020{\natexlab{a}})}]{Liu:2019oav}%
  \BibitemOpen
  \bibfield  {author} {\bibinfo {author} {\bibfnamefont {Z.~L.}\ \bibnamefont
  {Liu}}\ and\ \bibinfo {author} {\bibfnamefont {M.}~\bibnamefont {Neubert}},\
  }\href {\doibase 10.1007/JHEP04(2020)033} {\bibfield  {journal} {\bibinfo
  {journal} {JHEP}\ }\textbf {\bibinfo {volume} {04}},\ \bibinfo {pages} {033}
  (\bibinfo {year} {2020}{\natexlab{a}})},\ \Eprint
  {http://arxiv.org/abs/1912.08818} {arXiv:1912.08818 [hep-ph]} \BibitemShut
  {NoStop}%
\bibitem [{\citenamefont {Wang}(2019)}]{Wang:2019mym}%
  \BibitemOpen
  \bibfield  {author} {\bibinfo {author} {\bibfnamefont {J.}~\bibnamefont
  {Wang}},\ }\href@noop {} {\  (\bibinfo {year} {2019})},\ \Eprint
  {http://arxiv.org/abs/1912.09920} {arXiv:1912.09920 [hep-ph]} \BibitemShut
  {NoStop}%
\bibitem [{\citenamefont {Liu}\ and\ \citenamefont
  {Neubert}(2020{\natexlab{b}})}]{Liu:2020ydl}%
  \BibitemOpen
  \bibfield  {author} {\bibinfo {author} {\bibfnamefont {Z.~L.}\ \bibnamefont
  {Liu}}\ and\ \bibinfo {author} {\bibfnamefont {M.}~\bibnamefont {Neubert}},\
  }\href@noop {} {\  (\bibinfo {year} {2020}{\natexlab{b}})},\ \Eprint
  {http://arxiv.org/abs/2003.03393} {arXiv:2003.03393 [hep-ph]} \BibitemShut
  {NoStop}%
\bibitem [{\citenamefont {Liu}\ \emph {et~al.}(2020{\natexlab{a}})\citenamefont
  {Liu}, \citenamefont {Mecaj}, \citenamefont {Neubert}, \citenamefont {Wang},\
  and\ \citenamefont {Fleming}}]{Liu:2020eqe}%
  \BibitemOpen
  \bibfield  {author} {\bibinfo {author} {\bibfnamefont {Z.~L.}\ \bibnamefont
  {Liu}}, \bibinfo {author} {\bibfnamefont {B.}~\bibnamefont {Mecaj}}, \bibinfo
  {author} {\bibfnamefont {M.}~\bibnamefont {Neubert}}, \bibinfo {author}
  {\bibfnamefont {X.}~\bibnamefont {Wang}}, \ and\ \bibinfo {author}
  {\bibfnamefont {S.}~\bibnamefont {Fleming}},\ }\href@noop {} {\  (\bibinfo
  {year} {2020}{\natexlab{a}})},\ \Eprint {http://arxiv.org/abs/2005.03013}
  {arXiv:2005.03013 [hep-ph]} \BibitemShut {NoStop}%
\bibitem [{\citenamefont {Beneke}\ \emph
  {et~al.}(2020{\natexlab{c}})\citenamefont {Beneke}, \citenamefont {Garny},
  \citenamefont {Jaskiewicz}, \citenamefont {Szafron}, \citenamefont
  {Vernazza},\ and\ \citenamefont {Wang}}]{Beneke:2020ibj}%
  \BibitemOpen
  \bibfield  {author} {\bibinfo {author} {\bibfnamefont {M.}~\bibnamefont
  {Beneke}}, \bibinfo {author} {\bibfnamefont {M.}~\bibnamefont {Garny}},
  \bibinfo {author} {\bibfnamefont {S.}~\bibnamefont {Jaskiewicz}}, \bibinfo
  {author} {\bibfnamefont {R.}~\bibnamefont {Szafron}}, \bibinfo {author}
  {\bibfnamefont {L.}~\bibnamefont {Vernazza}}, \ and\ \bibinfo {author}
  {\bibfnamefont {J.}~\bibnamefont {Wang}},\ }\href@noop {} {\  (\bibinfo
  {year} {2020}{\natexlab{c}})},\ \Eprint {http://arxiv.org/abs/2008.04943}
  {arXiv:2008.04943 [hep-ph]} \BibitemShut {NoStop}%
\bibitem [{\citenamefont {Liu}\ \emph {et~al.}(2020{\natexlab{b}})\citenamefont
  {Liu}, \citenamefont {Mecaj}, \citenamefont {Neubert},\ and\ \citenamefont
  {Wang}}]{Liu:2020tzd}%
  \BibitemOpen
  \bibfield  {author} {\bibinfo {author} {\bibfnamefont {Z.~L.}\ \bibnamefont
  {Liu}}, \bibinfo {author} {\bibfnamefont {B.}~\bibnamefont {Mecaj}}, \bibinfo
  {author} {\bibfnamefont {M.}~\bibnamefont {Neubert}}, \ and\ \bibinfo
  {author} {\bibfnamefont {X.}~\bibnamefont {Wang}},\ }\href@noop {} {\
  (\bibinfo {year} {2020}{\natexlab{b}})},\ \Eprint
  {http://arxiv.org/abs/2009.04456} {arXiv:2009.04456 [hep-ph]} \BibitemShut
  {NoStop}%
\bibitem [{\citenamefont {Laenen}\ \emph {et~al.}(2009)\citenamefont {Laenen},
  \citenamefont {Stavenga},\ and\ \citenamefont {White}}]{Laenen:2008gt}%
  \BibitemOpen
  \bibfield  {author} {\bibinfo {author} {\bibfnamefont {E.}~\bibnamefont
  {Laenen}}, \bibinfo {author} {\bibfnamefont {G.}~\bibnamefont {Stavenga}}, \
  and\ \bibinfo {author} {\bibfnamefont {C.~D.}\ \bibnamefont {White}},\ }\href
  {\doibase 10.1088/1126-6708/2009/03/054} {\bibfield  {journal} {\bibinfo
  {journal} {JHEP}\ }\textbf {\bibinfo {volume} {03}},\ \bibinfo {pages} {054}
  (\bibinfo {year} {2009})},\ \Eprint {http://arxiv.org/abs/0811.2067}
  {arXiv:0811.2067 [hep-ph]} \BibitemShut {NoStop}%
\bibitem [{\citenamefont {Gardi}\ \emph {et~al.}(2010)\citenamefont {Gardi},
  \citenamefont {Laenen}, \citenamefont {Stavenga},\ and\ \citenamefont
  {White}}]{Gardi:2010rn}%
  \BibitemOpen
  \bibfield  {author} {\bibinfo {author} {\bibfnamefont {E.}~\bibnamefont
  {Gardi}}, \bibinfo {author} {\bibfnamefont {E.}~\bibnamefont {Laenen}},
  \bibinfo {author} {\bibfnamefont {G.}~\bibnamefont {Stavenga}}, \ and\
  \bibinfo {author} {\bibfnamefont {C.~D.}\ \bibnamefont {White}},\ }\href
  {\doibase 10.1007/JHEP11(2010)155} {\bibfield  {journal} {\bibinfo  {journal}
  {JHEP}\ }\textbf {\bibinfo {volume} {1011}},\ \bibinfo {pages} {155}
  (\bibinfo {year} {2010})},\ \Eprint {http://arxiv.org/abs/1008.0098}
  {arXiv:1008.0098 [hep-ph]} \BibitemShut {NoStop}%
\bibitem [{\citenamefont {Beneke}\ and\ \citenamefont
  {Smirnov}(1998)}]{Beneke:1997zp}%
  \BibitemOpen
  \bibfield  {author} {\bibinfo {author} {\bibfnamefont {M.}~\bibnamefont
  {Beneke}}\ and\ \bibinfo {author} {\bibfnamefont {V.~A.}\ \bibnamefont
  {Smirnov}},\ }\href {\doibase 10.1016/S0550-3213(98)00138-2} {\bibfield
  {journal} {\bibinfo  {journal} {Nucl. Phys.}\ }\textbf {\bibinfo {volume}
  {B522}},\ \bibinfo {pages} {321} (\bibinfo {year} {1998})},\ \Eprint
  {http://arxiv.org/abs/hep-ph/9711391} {arXiv:hep-ph/9711391 [hep-ph]}
  \BibitemShut {NoStop}%
\bibitem [{\citenamefont {Smirnov}(2002)}]{Smirnov:2002pj}%
  \BibitemOpen
  \bibfield  {author} {\bibinfo {author} {\bibfnamefont {V.~A.}\ \bibnamefont
  {Smirnov}},\ }\href@noop {} {\bibfield  {journal} {\bibinfo  {journal}
  {Springer Tracts Mod.\ Phys.}\ }\textbf {\bibinfo {volume} {177}},\ \bibinfo
  {pages} {1} (\bibinfo {year} {2002})}\BibitemShut {NoStop}%
\bibitem [{\citenamefont {Bonocore}\ \emph
  {et~al.}(2015{\natexlab{b}})\citenamefont {Bonocore}, \citenamefont {Laenen},
  \citenamefont {Magnea}, \citenamefont {Vernazza},\ and\ \citenamefont
  {White}}]{Bonocore:2014wua}%
  \BibitemOpen
  \bibfield  {author} {\bibinfo {author} {\bibfnamefont {D.}~\bibnamefont
  {Bonocore}}, \bibinfo {author} {\bibfnamefont {E.}~\bibnamefont {Laenen}},
  \bibinfo {author} {\bibfnamefont {L.}~\bibnamefont {Magnea}}, \bibinfo
  {author} {\bibfnamefont {L.}~\bibnamefont {Vernazza}}, \ and\ \bibinfo
  {author} {\bibfnamefont {C.~D.}\ \bibnamefont {White}},\ }\href {\doibase
  10.1016/j.physletb.2015.02.008} {\bibfield  {journal} {\bibinfo  {journal}
  {Phys. Lett.}\ }\textbf {\bibinfo {volume} {B742}},\ \bibinfo {pages} {375}
  (\bibinfo {year} {2015}{\natexlab{b}})},\ \Eprint
  {http://arxiv.org/abs/1410.6406} {arXiv:1410.6406 [hep-ph]} \BibitemShut
  {NoStop}%
\bibitem [{\citenamefont {Bahjat-Abbas}\ \emph {et~al.}(2018)\citenamefont
  {Bahjat-Abbas}, \citenamefont {Sinninghe~Damst\'{e}}, \citenamefont
  {Vernazza},\ and\ \citenamefont {White}}]{Bahjat-Abbas:2018hpv}%
  \BibitemOpen
  \bibfield  {author} {\bibinfo {author} {\bibfnamefont {N.}~\bibnamefont
  {Bahjat-Abbas}}, \bibinfo {author} {\bibfnamefont {J.}~\bibnamefont
  {Sinninghe~Damst\'{e}}}, \bibinfo {author} {\bibfnamefont {L.}~\bibnamefont
  {Vernazza}}, \ and\ \bibinfo {author} {\bibfnamefont {C.~D.}\ \bibnamefont
  {White}},\ }\href {\doibase 10.1007/JHEP10(2018)144} {\bibfield  {journal}
  {\bibinfo  {journal} {JHEP}\ }\textbf {\bibinfo {volume} {10}},\ \bibinfo
  {pages} {144} (\bibinfo {year} {2018})},\ \Eprint
  {http://arxiv.org/abs/1807.09246} {arXiv:1807.09246 [hep-ph]} \BibitemShut
  {NoStop}%
\bibitem [{\citenamefont {Collins}(1989)}]{Collins:1989bt}%
  \BibitemOpen
  \bibfield  {author} {\bibinfo {author} {\bibfnamefont {J.~C.}\ \bibnamefont
  {Collins}},\ }\href@noop {} {\bibfield  {journal} {\bibinfo  {journal}
  {Adv.Ser.Direct.High Energy Phys.}\ }\textbf {\bibinfo {volume} {5}},\
  \bibinfo {pages} {573} (\bibinfo {year} {1989})},\ \Eprint
  {http://arxiv.org/abs/hep-ph/0312336} {arXiv:hep-ph/0312336 [hep-ph]}
  \BibitemShut {NoStop}%
\bibitem [{\citenamefont {Dixon}\ \emph {et~al.}(2008)\citenamefont {Dixon},
  \citenamefont {Magnea},\ and\ \citenamefont {Sterman}}]{Dixon:2008gr}%
  \BibitemOpen
  \bibfield  {author} {\bibinfo {author} {\bibfnamefont {L.~J.}\ \bibnamefont
  {Dixon}}, \bibinfo {author} {\bibfnamefont {L.}~\bibnamefont {Magnea}}, \
  and\ \bibinfo {author} {\bibfnamefont {G.~F.}\ \bibnamefont {Sterman}},\
  }\href {\doibase 10.1088/1126-6708/2008/08/022} {\bibfield  {journal}
  {\bibinfo  {journal} {JHEP}\ }\textbf {\bibinfo {volume} {0808}},\ \bibinfo
  {pages} {022} (\bibinfo {year} {2008})},\ \Eprint
  {http://arxiv.org/abs/0805.3515} {arXiv:0805.3515 [hep-ph]} \BibitemShut
  {NoStop}%
\bibitem [{\citenamefont {Landau}(1960)}]{Landau:1959fi}%
  \BibitemOpen
  \bibfield  {author} {\bibinfo {author} {\bibfnamefont {L.}~\bibnamefont
  {Landau}},\ }\href {\doibase 10.1016/B978-0-08-010586-4.50103-6} {\bibfield
  {journal} {\bibinfo  {journal} {Nucl. Phys.}\ }\textbf {\bibinfo {volume}
  {13}},\ \bibinfo {pages} {181} (\bibinfo {year} {1960})}\BibitemShut
  {NoStop}%
\bibitem [{\citenamefont {Coleman}\ and\ \citenamefont
  {Norton}(1965)}]{Coleman:1965xm}%
  \BibitemOpen
  \bibfield  {author} {\bibinfo {author} {\bibfnamefont {S.}~\bibnamefont
  {Coleman}}\ and\ \bibinfo {author} {\bibfnamefont {R.}~\bibnamefont
  {Norton}},\ }\href {\doibase 10.1007/BF02750472} {\bibfield  {journal}
  {\bibinfo  {journal} {Nuovo Cim.}\ }\textbf {\bibinfo {volume} {38}},\
  \bibinfo {pages} {438} (\bibinfo {year} {1965})}\BibitemShut {NoStop}%
\bibitem [{\citenamefont {Jantzen}(2011)}]{Jantzen:2011nz}%
  \BibitemOpen
  \bibfield  {author} {\bibinfo {author} {\bibfnamefont {B.}~\bibnamefont
  {Jantzen}},\ }\href {\doibase 10.1007/JHEP12(2011)076} {\bibfield  {journal}
  {\bibinfo  {journal} {JHEP}\ }\textbf {\bibinfo {volume} {12}},\ \bibinfo
  {pages} {076} (\bibinfo {year} {2011})},\ \Eprint
  {http://arxiv.org/abs/1111.2589} {arXiv:1111.2589 [hep-ph]} \BibitemShut
  {NoStop}%
\bibitem [{\citenamefont {Sterman}(1978)}]{Sterman:1978bi}%
  \BibitemOpen
  \bibfield  {author} {\bibinfo {author} {\bibfnamefont {G.~F.}\ \bibnamefont
  {Sterman}},\ }\href {\doibase 10.1103/PhysRevD.17.2773} {\bibfield  {journal}
  {\bibinfo  {journal} {Phys. Rev.}\ }\textbf {\bibinfo {volume} {D17}},\
  \bibinfo {pages} {2773} (\bibinfo {year} {1978})}\BibitemShut {NoStop}%
\bibitem [{\citenamefont {Akhoury}(1979)}]{Akhoury:1978vq}%
  \BibitemOpen
  \bibfield  {author} {\bibinfo {author} {\bibfnamefont {R.}~\bibnamefont
  {Akhoury}},\ }\href {\doibase 10.1103/PhysRevD.19.1250} {\bibfield  {journal}
  {\bibinfo  {journal} {Phys. Rev. D}\ }\textbf {\bibinfo {volume} {19}},\
  \bibinfo {pages} {1250} (\bibinfo {year} {1979})}\BibitemShut {NoStop}%
\bibitem [{\citenamefont {Bagan}\ \emph {et~al.}(1994)\citenamefont {Bagan},
  \citenamefont {Ball}, \citenamefont {Braun},\ and\ \citenamefont
  {Gosdzinsky}}]{Bagan:1994zd}%
  \BibitemOpen
  \bibfield  {author} {\bibinfo {author} {\bibfnamefont {E.}~\bibnamefont
  {Bagan}}, \bibinfo {author} {\bibfnamefont {P.}~\bibnamefont {Ball}},
  \bibinfo {author} {\bibfnamefont {V.~M.}\ \bibnamefont {Braun}}, \ and\
  \bibinfo {author} {\bibfnamefont {P.}~\bibnamefont {Gosdzinsky}},\ }\href
  {\doibase 10.1016/0550-3213(94)90591-6} {\bibfield  {journal} {\bibinfo
  {journal} {Nucl. Phys. B}\ }\textbf {\bibinfo {volume} {432}},\ \bibinfo
  {pages} {3} (\bibinfo {year} {1994})},\ \Eprint
  {http://arxiv.org/abs/hep-ph/9408306} {arXiv:hep-ph/9408306} \BibitemShut
  {NoStop}%
\bibitem [{\citenamefont {Aivazis}\ \emph {et~al.}(1994)\citenamefont
  {Aivazis}, \citenamefont {Collins}, \citenamefont {Olness},\ and\
  \citenamefont {Tung}}]{Aivazis:1993pi}%
  \BibitemOpen
  \bibfield  {author} {\bibinfo {author} {\bibfnamefont {M.}~\bibnamefont
  {Aivazis}}, \bibinfo {author} {\bibfnamefont {J.~C.}\ \bibnamefont
  {Collins}}, \bibinfo {author} {\bibfnamefont {F.~I.}\ \bibnamefont {Olness}},
  \ and\ \bibinfo {author} {\bibfnamefont {W.-K.}\ \bibnamefont {Tung}},\
  }\href {\doibase 10.1103/PhysRevD.50.3102} {\bibfield  {journal} {\bibinfo
  {journal} {Phys. Rev. D}\ }\textbf {\bibinfo {volume} {50}},\ \bibinfo
  {pages} {3102} (\bibinfo {year} {1994})},\ \Eprint
  {http://arxiv.org/abs/hep-ph/9312319} {arXiv:hep-ph/9312319} \BibitemShut
  {NoStop}%
\bibitem [{\citenamefont {Thorne}\ and\ \citenamefont
  {Roberts}(1998)}]{Thorne:1997ga}%
  \BibitemOpen
  \bibfield  {author} {\bibinfo {author} {\bibfnamefont {R.}~\bibnamefont
  {Thorne}}\ and\ \bibinfo {author} {\bibfnamefont {R.}~\bibnamefont
  {Roberts}},\ }\href {\doibase 10.1103/PhysRevD.57.6871} {\bibfield  {journal}
  {\bibinfo  {journal} {Phys. Rev. D}\ }\textbf {\bibinfo {volume} {57}},\
  \bibinfo {pages} {6871} (\bibinfo {year} {1998})},\ \Eprint
  {http://arxiv.org/abs/hep-ph/9709442} {arXiv:hep-ph/9709442} \BibitemShut
  {NoStop}%
\bibitem [{\citenamefont {Forte}\ \emph {et~al.}(2010)\citenamefont {Forte},
  \citenamefont {Laenen}, \citenamefont {Nason},\ and\ \citenamefont
  {Rojo}}]{Forte:2010ta}%
  \BibitemOpen
  \bibfield  {author} {\bibinfo {author} {\bibfnamefont {S.}~\bibnamefont
  {Forte}}, \bibinfo {author} {\bibfnamefont {E.}~\bibnamefont {Laenen}},
  \bibinfo {author} {\bibfnamefont {P.}~\bibnamefont {Nason}}, \ and\ \bibinfo
  {author} {\bibfnamefont {J.}~\bibnamefont {Rojo}},\ }\href {\doibase
  10.1016/j.nuclphysb.2010.03.014} {\bibfield  {journal} {\bibinfo  {journal}
  {Nucl. Phys. B}\ }\textbf {\bibinfo {volume} {834}},\ \bibinfo {pages} {116}
  (\bibinfo {year} {2010})},\ \Eprint {http://arxiv.org/abs/1001.2312}
  {arXiv:1001.2312 [hep-ph]} \BibitemShut {NoStop}%
\bibitem [{\citenamefont {Liu}\ and\ \citenamefont
  {Penin}(2017)}]{Liu:2017vkm}%
  \BibitemOpen
  \bibfield  {author} {\bibinfo {author} {\bibfnamefont {T.}~\bibnamefont
  {Liu}}\ and\ \bibinfo {author} {\bibfnamefont {A.~A.}\ \bibnamefont
  {Penin}},\ }\href {\doibase 10.1103/PhysRevLett.119.262001} {\bibfield
  {journal} {\bibinfo  {journal} {Phys. Rev. Lett.}\ }\textbf {\bibinfo
  {volume} {119}},\ \bibinfo {pages} {262001} (\bibinfo {year} {2017})},\
  \Eprint {http://arxiv.org/abs/1709.01092} {arXiv:1709.01092 [hep-ph]}
  \BibitemShut {NoStop}%
\bibitem [{\citenamefont {Beneke}\ \emph
  {et~al.}(2019{\natexlab{c}})\citenamefont {Beneke}, \citenamefont {Bobeth},\
  and\ \citenamefont {Szafron}}]{Beneke:2019slt}%
  \BibitemOpen
  \bibfield  {author} {\bibinfo {author} {\bibfnamefont {M.}~\bibnamefont
  {Beneke}}, \bibinfo {author} {\bibfnamefont {C.}~\bibnamefont {Bobeth}}, \
  and\ \bibinfo {author} {\bibfnamefont {R.}~\bibnamefont {Szafron}},\ }\href
  {\doibase 10.1007/JHEP10(2019)232} {\bibfield  {journal} {\bibinfo  {journal}
  {JHEP}\ }\textbf {\bibinfo {volume} {10}},\ \bibinfo {pages} {232} (\bibinfo
  {year} {2019}{\natexlab{c}})},\ \Eprint {http://arxiv.org/abs/1908.07011}
  {arXiv:1908.07011 [hep-ph]} \BibitemShut {NoStop}%
\bibitem [{\citenamefont {Pak}\ and\ \citenamefont
  {Smirnov}(2011)}]{Pak:2010pt}%
  \BibitemOpen
  \bibfield  {author} {\bibinfo {author} {\bibfnamefont {A.}~\bibnamefont
  {Pak}}\ and\ \bibinfo {author} {\bibfnamefont {A.}~\bibnamefont {Smirnov}},\
  }\href {\doibase 10.1140/epjc/s10052-011-1626-1} {\bibfield  {journal}
  {\bibinfo  {journal} {Eur. Phys. J. C}\ }\textbf {\bibinfo {volume} {71}},\
  \bibinfo {pages} {1626} (\bibinfo {year} {2011})},\ \Eprint
  {http://arxiv.org/abs/1011.4863} {arXiv:1011.4863 [hep-ph]} \BibitemShut
  {NoStop}%
\bibitem [{\citenamefont {Mertig}\ \emph {et~al.}(1991)\citenamefont {Mertig},
  \citenamefont {Bohm},\ and\ \citenamefont {Denner}}]{Mertig:1990an}%
  \BibitemOpen
  \bibfield  {author} {\bibinfo {author} {\bibfnamefont {R.}~\bibnamefont
  {Mertig}}, \bibinfo {author} {\bibfnamefont {M.}~\bibnamefont {Bohm}}, \ and\
  \bibinfo {author} {\bibfnamefont {A.}~\bibnamefont {Denner}},\ }\href
  {\doibase 10.1016/0010-4655(91)90130-D} {\bibfield  {journal} {\bibinfo
  {journal} {Comput. Phys. Commun.}\ }\textbf {\bibinfo {volume} {64}},\
  \bibinfo {pages} {345} (\bibinfo {year} {1991})}\BibitemShut {NoStop}%
\bibitem [{\citenamefont {Shtabovenko}\ \emph {et~al.}(2016)\citenamefont
  {Shtabovenko}, \citenamefont {Mertig},\ and\ \citenamefont
  {Orellana}}]{Shtabovenko:2016sxi}%
  \BibitemOpen
  \bibfield  {author} {\bibinfo {author} {\bibfnamefont {V.}~\bibnamefont
  {Shtabovenko}}, \bibinfo {author} {\bibfnamefont {R.}~\bibnamefont {Mertig}},
  \ and\ \bibinfo {author} {\bibfnamefont {F.}~\bibnamefont {Orellana}},\
  }\href {\doibase 10.1016/j.cpc.2016.06.008} {\bibfield  {journal} {\bibinfo
  {journal} {Comput. Phys. Commun.}\ }\textbf {\bibinfo {volume} {207}},\
  \bibinfo {pages} {432} (\bibinfo {year} {2016})},\ \Eprint
  {http://arxiv.org/abs/1601.01167} {arXiv:1601.01167 [hep-ph]} \BibitemShut
  {NoStop}%
\bibitem [{\citenamefont {Collins}\ \emph {et~al.}(1989)\citenamefont
  {Collins}, \citenamefont {Soper},\ and\ \citenamefont
  {Sterman}}]{Collins:1989gx}%
  \BibitemOpen
  \bibfield  {author} {\bibinfo {author} {\bibfnamefont {J.~C.}\ \bibnamefont
  {Collins}}, \bibinfo {author} {\bibfnamefont {D.~E.}\ \bibnamefont {Soper}},
  \ and\ \bibinfo {author} {\bibfnamefont {G.~F.}\ \bibnamefont {Sterman}},\
  }\enquote {\bibinfo {title} {{Factorization of Hard Processes in QCD}},}\ \
  (\bibinfo {year} {1989})\ pp.\ \bibinfo {pages} {1--91},\ \Eprint
  {http://arxiv.org/abs/hep-ph/0409313} {arXiv:hep-ph/0409313} \BibitemShut
  {NoStop}%
\bibitem [{\citenamefont {Huber}\ and\ \citenamefont
  {Maitre}(2006)}]{Huber:2005yg}%
  \BibitemOpen
  \bibfield  {author} {\bibinfo {author} {\bibfnamefont {T.}~\bibnamefont
  {Huber}}\ and\ \bibinfo {author} {\bibfnamefont {D.}~\bibnamefont {Maitre}},\
  }\href {\doibase 10.1016/j.cpc.2006.01.007} {\bibfield  {journal} {\bibinfo
  {journal} {Comput. Phys. Commun.}\ }\textbf {\bibinfo {volume} {175}},\
  \bibinfo {pages} {122} (\bibinfo {year} {2006})},\ \Eprint
  {http://arxiv.org/abs/hep-ph/0507094} {arXiv:hep-ph/0507094} \BibitemShut
  {NoStop}%
\bibitem [{\citenamefont {Huber}\ and\ \citenamefont
  {Maitre}(2008)}]{Huber:2007dx}%
  \BibitemOpen
  \bibfield  {author} {\bibinfo {author} {\bibfnamefont {T.}~\bibnamefont
  {Huber}}\ and\ \bibinfo {author} {\bibfnamefont {D.}~\bibnamefont {Maitre}},\
  }\href {\doibase 10.1016/j.cpc.2007.12.008} {\bibfield  {journal} {\bibinfo
  {journal} {Comput. Phys. Commun.}\ }\textbf {\bibinfo {volume} {178}},\
  \bibinfo {pages} {755} (\bibinfo {year} {2008})},\ \Eprint
  {http://arxiv.org/abs/0708.2443} {arXiv:0708.2443 [hep-ph]} \BibitemShut
  {NoStop}%
\end{thebibliography}%
\end{document}